\documentclass[12pt,a4paper]{article}

\usepackage{authblk}
\usepackage{amsmath,amssymb}
\usepackage{graphicx}
\usepackage{amsfonts}
\usepackage{epsfig}
\usepackage{times,pstricks,pst-node}
\usepackage{pst-all} 
\usepackage{tikz}
\usepackage{pgflibrarysnakes}
\usetikzlibrary[snakes]

\usetikzlibrary{arrows,shapes,positioning}
\usetikzlibrary{decorations.markings}
\tikzstyle arrowstyle=[scale=1]
\tikzstyle directed=[postaction={decorate,decoration={markings,
		mark=at position .65 with {\arrow[arrowstyle]{stealth}}}}]
\tikzstyle reverse directed=[postaction={decorate,decoration={markings,
		mark=at position .65 with {\arrowreversed[arrowstyle]{stealth};}}}]

\newcommand{\im}{\mathrm i}

\newcommand{\tr}{\operatorname{Tr}}

\newcommand{\ket}[1]{\left|#1\right\rangle}      
\newcommand{\bra}[1]{\left\langle #1\right|}     
\newcommand{\eq}{\begin{equation}}
\newcommand{\en}{\end{equation}}
\newcommand{\bear}{\begin{eqnarray}}
\newcommand{\ear}{\end{eqnarray}}

\title{Correlation functions of the integrable $SU(n)$ spin chain}

\author{G.A.P. Ribeiro\footnote{E-mail: pavan@df.ufscar.br} \ and A. Kl\"umper\footnote{E-mail: kluemper@uni-wuppertal.de}}
\affil{$^{*}$ Departamento de F\'{i}sica, Universidade Federal de S\~ao Carlos \\ S\~ao Carlos, SP 13565-905, Brazil
\\
$^{\dagger}$ Theoretische Physik, Bergische Universit\"at Wuppertal, \\ 42097
Wuppertal, Germany}
\date{}
  
\begin{document}

\maketitle
\thispagestyle{empty}

\begin{abstract}
We study the correlation functions of $SU(n)$, $n>2$, invariant spin chains in
the thermodynamic limit. We formulate a consistent framework for the
computation of short-range correlation functions via functional equations
which hold even at finite temperature. We give the explicit solution for two-
and three-site correlations for the $SU(3)$ case at zero temperature. The
correlators do not seem to be of factorizable form. From the two-sites result
we see that the correlation functions are given in terms of Hurwitz' zeta
function, which differs from the $SU(2)$ case where the correlations are
expressed in terms of Riemann's zeta function of odd arguments.
\end{abstract}
\newpage

\pagestyle{plain}
\pagenumbering{arabic}

\section{Introduction}

The study of correlation functions of integrable quantum spin chains has a very
long history \cite{BOOK-KBI,BOOK-JM}. There exist many results for the case of the
$SU(2)$ spin-$1/2$ chain
\cite{TAKA,JMMN92,JM96,KMT00,GAS05,BOKO01,BOOS05,BOOS2,BGKS06,DGHK07,KKMST09,AuKl12} and
its higher-spin
realizations \cite{BoWe94,Idzumi94,Kitanine01,DeMa10,GSS10,KNS013,RK2016}.

The correlation functions for the $SU(2)$ spin-$1/2$ case were realized to
be given in terms of Riemann's zeta function of odd arguments \cite{BOKO01}. 
Later higher-spin cases were studied and explicit results were obtained
where also zeta function values for even arguments appeared \cite{KNS013,RK2016}.
These results were obtained from solutions of functional equations for
suitably defined correlation functions. The derivation of the functional
equations is based on the Yang-Baxter equation, 
and crossing symmetry (valid in the $SU(2)$ case) for the $R$-matrix, see
\cite{AuKl12} for the finite temperature case.

Nevertheless, one still lacks a better understanding of the correlation
properties of models based on high rank algebras. In the $SU(n)$ case for
$n>2$ \cite{UIMIN,SUTHERLAND}, there is no result for correlation functions
and this stayed as a longstanding problem for decades. In this paper, we
devise a framework to tackle the problem of computing the short-range
correlations of the integrable $SU(n)$ spin chains. We also provide explicit
solutions for the first correlation functions for the $SU(3)$ case, where
already for the two-site case the solution is given in terms of Hurwitz' zeta
function (generalized zeta function). Besides that, we have indications of the
absence of factorization of the correlations in terms of two-point
correlations.

This paper is organized as follows. In section \ref{INTEGRA}, we introduce the
integrable Hamiltonians and their associated integrable structure. In section
\ref{density}, we introduce the density operator containing all correlation
data as well as a generalized density operator. In contrast to the standard
density operator, the generalized density operator allows for the derivation
of discrete functional equations. This and the analyticity properties of the
generalized density operator are presented in section \ref{functional}.
In section \ref{su3section}, we exemplify 
our approach for the case of $SU(3)$ spin chains and we present the zero
temperature solution for two- and three-site correlation functions for which
the use of a mixed density operator proves to be sufficient. In
section \ref{lack}, we present some evidence for the absence of factorization
of the correlations in terms of two-point correlations. Finally, our
conclusions are given in section \ref{conclusion}. Additional details are
given in the appendices.

\section{The integrable model}\label{INTEGRA}

The Hamiltonian of the integrable $SU(n)$ spin chain is given by \cite{UIMIN,SUTHERLAND},
\eq
H^{(n)}=\sum_{j=1}^L P_{j,j+1},
\label{hamiltonian}
\en 
where $P_{j,j+1}$ is the permutation operator and $L$ is the number of
sites. The Hilbert space is $V^{\otimes L}$ with local space $V=\mathbb{C}^n$.

For instance, in the case of $SU(3)$ spin chains the Hamiltonian can be written in terms of spin-$1$ matrices as follows,
\eq
H^{(3)}=\sum_{j=1}^L [ \vec{S}_{j}\cdot\vec{S}_{j+1} + (\vec{S}_{j}\cdot\vec{S}_{j+1})^2 ].
\en 

The integrable Hamiltonian (\ref{hamiltonian}) is obtained as the logarithmic
derivative of the row-to-row transfer matrix
\eq
T^{(n)}(\lambda)=\tr_{\cal A}{[ R^{(n,n)}_{{\cal A}L}(\lambda)\dots R^{(n,n)}_{{\cal A} 1}(\lambda)]},
\en 
where the $R$-matrix
$R^{(n,n)}_{ab}(\lambda)=P_{ab}\check{R}^{(n,n)}_{ab}(\lambda)$  acts
non-trivially on the indicated space $V_a\otimes V_b$ of the (long) tensor product 
where $V_a$ and $V_b$ are copies of the local space $V$. The
operation of $R^{(n,n)}_{ab}$ is co-variant under $SU(n)$ acting by the
product of two fundamental representations $[n]$. The representation $[n]$
is the irreducible representation of dimension $n$  denoted by a single box in
the Young-Tableaux notation. 
Later in applications we will associate spectral parameters $\lambda$, $\mu$
with the two local vector spaces the $R$-matrix acts on and the difference
$\lambda-\mu$ will enter as argument.
The rational solution of the Yang-Baxter equation can be written and depicted
as,
\eq
\begin{tikzpicture}[scale=1]
\draw (-3.75,0) node {$\check{R}_{12}^{(n,n)}(\lambda-\mu)= I_{12} + (\lambda-\mu ) P_{12}=$};

\tikzstyle directed=[postaction={decorate,decoration={markings,
		mark=at position .35 with {\arrow[arrowstyle]{stealth}}}}]
\tikzstyle reverse directed=[postaction={decorate,decoration={markings,
		mark=at position .35 with {\arrowreversed[arrowstyle]{stealth};}}}]
\draw (-0.5,0) [color=black, directed, thick, rounded corners=7pt]
+(0,0) -- +(1,0); 
\draw (-0.5,0) [color=black, directed, thick, rounded corners=7pt] +(0.5,-0.5) -- +(0.5,0.5);
\draw (0.85,0) node {$=$};
\draw (-0.25,0.25) node {$\lambda$};
\draw (0.2,-0.35) node {$\mu$};

\tikzstyle directed=[postaction={decorate,decoration={markings,
		mark=at position .65 with {\arrow[arrowstyle]{stealth}}}}]
\tikzstyle reverse directed=[postaction={decorate,decoration={markings,
		mark=at position .65 with {\arrowreversed[arrowstyle]{stealth};}}}]

\draw (1.5,0) [-,color=black, thick, rounded corners=8pt]
+(0,0) -- +(0.5,0) -- +(0.5,0.5);
\draw (1.5,0) [-,color=black,  thick, rounded corners=8pt]
+(0.5,-0.5) -- +(0.5,0) -- +(1,0);
\draw (3.5,0) node {$+(\lambda-\mu)$};
\draw (4.5,0) [color=black, thick, rounded corners=7pt]
+(0,0) -- +(1,0) +(0.5,0.05) -- +(0.5,0.5) +(0.5,-0.5) -- +(0.5,-0.05);
\draw (5.7,0) node {$,$};
\end{tikzpicture}
\label{Rmatrix}
\en where $P_{12}$ is the standard permutation operator such that
$P_{12}=\sum_{i,j,k,l=1}^n P_{ik}^{jl} \hat{e}_{ij}^{(1)}\otimes
\hat{e}_{kl}^{(2)}$ with $P_{ik}^{jl}=\delta_{il}\delta_{jk}$ and where
$\hat{e}_{ij}^{(a)}\in C_a^n$ are the standard $n\times n$ Weyl matrices
acting in the $a$-space. Likewise, the matrix elements of the identity matrix
are given as $I_{ik}^{jl}=\delta_{ij}\delta_{kl}$. 

Let us motivate the graphical depiction of algebraic quantities. In the
  main body of this paper we are going to study correlation functions which
  occur as ratios of certain (large) sums. The denominator will be the
  partition function of a certain classical vertex model on a square lattice
  or a minor modification thereof and the numerator will be a similar partition
  function of a slightly modified geometry with a few bonds cut and
  specifically chosen spin values at the open ends. The general rule for
  turning a graph into a number is like we are used from Feynman diagrams. We
  place spin variables on closed bonds, we evaluate all local objects for the
  given spin configuration and multiply these results, which are then summed
  over for all allowed spin configurations. In particular, a trace over a
  product of (transfer) matrices naturally turns into a (huge) sum over
  products of local objects. Very generally, graphs encode contractions of
  products of tensors.

Note that $R^{(n,n)}$ acts on $[n]\otimes [n]$, understood as $SU(n)$ module,
which is graphically indicated by arrows from left to right and from bottom to
top.  By use of the isomorphism of $\mbox{End}(W)$ and $W^*\otimes W$ for any linear
space $W$ and its dual $W^*$ we may alternatively view $R^{(n,n)}$ as a vector
in the tensor space $[\bar n]\otimes [\bar n]\otimes [n]\otimes [n]$, i.e. a multilinear map of
the type $[n]\times [n] \times [\bar n]\times [\bar n]\to \mathbb{C}$.

In order to further
illustrate, we show how to read the matrix elements of the $R$-matrix
(\ref{Rmatrix}) and the other operators in the graphical notation as follows,
\eq
\begin{tikzpicture}[scale=1]
\draw (-3.75,0) node {${[\check{R}^{(n,n)}(\lambda-\mu)]}_{ik}^{jl}= I_{ik}^{jl} + (\lambda-\mu ) P_{ik}^{jl}=$};

\tikzstyle directed=[postaction={decorate,decoration={markings,
		mark=at position .35 with {\arrow[arrowstyle]{stealth}}}}]
\tikzstyle reverse directed=[postaction={decorate,decoration={markings,
		mark=at position .35 with {\arrowreversed[arrowstyle]{stealth};}}}]

\draw (-0.15,0) [color=black, directed, thick, rounded corners=7pt]
+(0,0) -- +(1,0); 
\draw (-0.15,0) [color=black, directed, thick, rounded corners=7pt] +(0.5,-0.5) -- +(0.5,0.5);
\draw (-0.25,0) node {$i$};
\draw (0.35,0.8) node {$j$};
\draw (0.35,-0.8) node {$k$};
\draw (1.,0) node {$l$};

\draw (1.25,0) node {$=$};
\draw (0.1,0.25) node {$\lambda$};
\draw (0.6,-0.35) node {$\mu$};

\tikzstyle directed=[postaction={decorate,decoration={markings,
		mark=at position .65 with {\arrow[arrowstyle]{stealth}}}}]
\tikzstyle reverse directed=[postaction={decorate,decoration={markings,
		mark=at position .65 with {\arrowreversed[arrowstyle]{stealth};}}}]

\draw (1.68,0) node {$i$};
\draw (2.2,0.8) node {$j$};
\draw (2.2,-0.8) node {$k$};
\draw (2.9,0) node {$l$};
\draw (1.75,0) [-,color=black, thick, rounded corners=8pt]
+(0,0) -- +(0.5,0) -- +(0.5,0.5);
\draw (1.75,0) [-,color=black,  thick, rounded corners=8pt]
+(0.5,-0.5) -- +(0.5,0) -- +(1,0);
\draw (3.9,0) node {$+(\lambda-\mu)$};
\draw (4.85,0) node {$i$};
\draw (5.5,0.8) node {$j$};
\draw (5.5,-0.8) node {$k$};
\draw (6.1,0) node {$l$};
\draw (5,0) [color=black, thick, rounded corners=7pt]
+(0,0) -- +(1,0) +(0.5,0.05) -- +(0.5,0.5) +(0.5,-0.5) -- +(0.5,-0.05);
\draw (6.3,0) node {$.$};
\end{tikzpicture}
\en

The $R$-matrix with mixed representations of the fundamental $[n]$ and
anti-fundamental $[\bar n]$ representation of the $SU(n)$ can be written as follows,
\eq
\begin{tikzpicture}[scale=1]
\draw (-1.75,0) node {$\check{R}_{12}^{(n,\bar{n})}(\lambda-\mu)= E_{12} + (\lambda-\mu ) P_{12}=$};
\draw (1.5,0) [-,color=black, thick, rounded corners=7pt]
+(0,0) -- +(0.5,0) -- +(0.5,-0.5);
\draw (1.5,0) [-,color=black, thick, rounded corners=7pt]
+(0.5,0.5) -- +(0.5,0) -- +(1,0);
\draw (3.5,0) node {$+(\lambda-\mu)$};
\draw (4.5,0) [color=black, thick, rounded corners=7pt]
+(0,0) -- +(1,0) +(0.5,0.05) -- +(0.5,0.5) +(0.5,-0.5) -- +(0.5,-0.05);
\draw (5.7,0) node {$.$};
\end{tikzpicture}
\en 
where $E_{12}$ is the standard Temperley-Lieb operator such that
$E_{ik}^{jl}=\delta_{ik}\delta_{jl}$ and the anti-fundamental representation
is the other $n$ dimensional irreducible representation denoted as a column of
$n-1$ boxes in the Young-Tableaux notation.  
Note the reversed direction of the arrow on the vertical line.
For rational models, the
remaining combinations can be expressed in terms of the previous one such that
$\check{R}^{(\bar{n},n)}(\lambda)=\check{R}^{(n,\bar{n})}(\lambda)$ and
$\check{R}^{(\bar{n},\bar{n})}(\lambda)=\check{R}^{(n,n)}(\lambda)$ as
  linear operators on $V\otimes V$. Note that co-variance with respect to
  $SU(n)$ is guaranteed, i.e. $g\otimes g^*$ and $g^*\otimes g$ for any $g\in
  SU(n)$ commute with $\check{R}^{(\bar{n},n)}$ and of course with
  $\check{R}^{(n,\bar{n})}(\lambda)$.

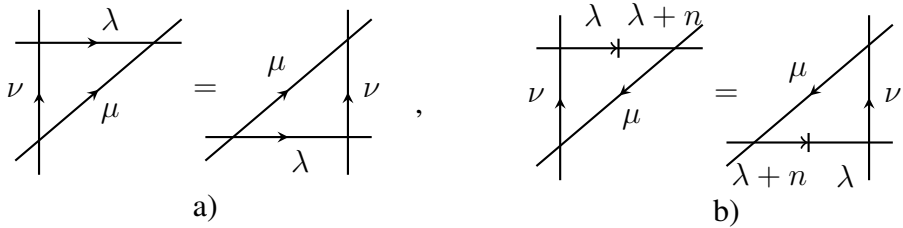
\begin{figure}[h]
	\begin{center}
		\begin{minipage}{0.5\linewidth}
		\begin{tikzpicture}[scale=1.25]
\tikzstyle directed=[postaction={decorate,decoration={markings,
		mark=at position .5 with {\arrow[arrowstyle]{stealth}}}}]

\draw (0,0) [-,color=black, thick,directed, rounded corners=8pt]+(-0.5,-0.25) -- +(1.25,1.25);
\draw (0,0) [-,color=black, thick,directed, rounded corners=8pt]+(-0.25,-0.4) -- +(-0.25,1.35);
\draw (0,0) [-,color=black, thick,directed, rounded corners=8pt]+(-0.5,1) -- +(1.25,1);

\draw (1.5	,0.5) node {$=$};

\draw (2,0) [-,color=black, thick,directed, rounded corners=8pt]+(-0.5,-0.25) -- +(1.25,1.25);
\draw (2,0) [-,color=black, thick,directed, rounded corners=8pt]+(1.,-0.4) -- +(1.,1.35);
\draw (2,0) [-,color=black, thick,directed, rounded corners=8pt]+(-0.5,0) -- +(1.25,0);

\tikzstyle directed=[postaction={decorate,decoration={markings,
		mark=at position .65 with {\arrow[arrowstyle]{stealth}}}}]

		\draw (0.5,0.25) node {$\mu$};
		\draw (0.5,1.25) node {$\lambda$};
		\draw (-0.5	,0.5) node {$\nu$};
		
		\draw (2.25,0.75) node {$\mu$};
		\draw (2.5,-0.25) node {$\lambda$};
		\draw (3.25,0.5) node {$\nu$};

		\draw (1.5,-0.75) node {a)};
		\draw (3.75,0.25) node {,};

		\end{tikzpicture}
		\end{minipage}%
		\begin{minipage}{0.5\linewidth}
		\begin{tikzpicture}[scale=1.25]
		\tikzstyle directed=[postaction={decorate,decoration={markings,
				mark=at position .5 with {\arrow[arrowstyle]{stealth}}}}]
		
		\draw (0,0) [-,color=black, thick,directed, rounded corners=8pt]+(1.25,1.25) -- +(-0.5,-0.25);
		\draw (0,0) [-,color=black, thick,directed, rounded corners=8pt]+(-0.25,-0.4) -- +(-0.25,1.35);
		
		\draw (1.5	,0.5) node {$=$};
		
		\draw (2,0) [-,color=black, thick,directed, rounded corners=8pt]+(1.25,1.25) -- +(-0.5,-0.25);
		\draw (2,0) [-,color=black, thick,directed, rounded corners=8pt]+(1.,-0.4) -- +(1.,1.35);

		\tikzstyle directed=[postaction={decorate,decoration={markings,
				mark=at position .5 with {\arrow[arrowstyle]{>|}}}}]
		\draw (0,0) [-,color=black, thick,directed, rounded corners=8pt]+(-0.5,1) -- +(1.25,1);
		\draw (2,0) [-,color=black, thick,directed, rounded corners=8pt]+(-0.5,0) -- +(1.25,0);

		\tikzstyle directed=[postaction={decorate,decoration={markings,
				mark=at position .65 with {\arrow[arrowstyle]{stealth}}}}]

		\draw (0.5,0.25) node {$\mu$};
		\draw (0.1,1.35) node {$\lambda$};
		\draw (0.85,1.35) node {$\lambda+n$};
		\draw (-0.5	,0.5) node {$\nu$};
		
		\draw (2.25,0.75) node {$\mu$};
		\draw (2.75,-0.35) node {$\lambda$};
		\draw (1.95,-0.35) node {$\lambda+n$};
		\draw (3.25,0.5) node {$\nu$};
		
		\draw (1.5,-0.75) node {b)};
		
		\end{tikzpicture}
		\end{minipage}

	\end{center}
	\caption{Graphical illustration of the Yang-Baxter equation 
            (where vertices from lower left to upper right correspond to
            $R$-matrices in (\ref{YB}) and (\ref{specialYB}) from right to
            left): a) the standard Yang-Baxter equation (\ref{YB}) for the
          fundamental representation $r_1=r_2=r_3=[n]$ (the $5$ remaining
          standard equations are obtained from a) by rotation); b) the special
          Yang-Baxter equation (\ref{specialYB}), where the shift in the
          argument of the $R$-matrix can be conveniently seen as a
          discontinuity of the spectral parameter along that line.}
	\label{YBeqs}
\end{figure}

With a grain of salt,
these four $R$-matrices are solution to the Yang-Baxter equation
\eq
\check{R}^{(r_1,r_2)}_{12}(\lambda-\mu) \check{R}^{(r_1,r_3)}_{23}(\lambda-\nu) \check{R}^{(r_2,r_3)}_{12}(\mu-\nu) =\check{R}^{(r_2,r_3)}_{23}(\mu-\nu) \check{R}^{(r_1,r_3)}_{12}(\lambda-\nu)  \check{R}^{(r_1,r_2)}_{23}(\lambda-\mu),
\label{YB}
\en
where $r_i \in \{n,\bar{n}\}$ for $i=1,2,3$. For
  having (\ref{YB}) literally for all combinations of $r_1$, $r_2$, $r_3$ we
  would have to introduce a shift by $n$ in the argument of for instance 
$\check{R}^{(n,\bar{n})}$ wherever it appears. However, we keep the definitions of the
  $R$-matrices as given above and have (\ref{YB}) for all $r_1$, $r_2$, $r_3$
  except for $n, \bar{n}, n$ and $\bar{n}, n, \bar{n}$.
In order to simplify our notation
it is convenient to list these equations in a group of six standard equations
as above (\ref{YB}) and a group of two special ones which have shifted arguments in the
intertwining matrix (see Figure \ref{YBeqs}). We explicitly write one of the
special Yang-Baxter equations,
\eq
\check{R}^{(n,\bar{n})}_{12}(\lambda-\mu+n) \check{R}^{(n,n)}_{23}(\lambda-\nu) \check{R}^{(\bar{n},n)}_{12}(\mu-\nu) =\check{R}^{(\bar{n},n)}_{23}(\mu-\nu) \check{R}^{(n,n)}_{12}(\lambda-\nu)  \check{R}^{(n,\bar{n})}_{23}(\lambda-\mu+n),
\label{specialYB}
\en and the other one is obtained by exchanging the representations $[n]$ and
$[\bar{n}]$. It is worth to note that in our graphical notation, e.g in Figure
\ref{YBeqs}, the lines upwards and to the right are associated to the
fundamental representation $[n]$ and conversely the lines downwards and to the
left are associated to the anti-fundamental representation $[\bar n]$.

The fundamental $R$-matrix has important properties,
\begin{align}
\check{R}^{(n,n)}_{12}(0)&= I, \qquad &\mbox{initial condition}, \label{regularity} \\
\check{R}^{(n,n)}_{12}(\lambda-\mu)\check{R}_{21}^{(n,n)}(\mu-\lambda) &= (1-(\lambda-\mu)^2)I, \qquad & \mbox{standard unitarity}, \label{unitarity}
\end{align}
where again $I$ is the $n^2\times n^2$ identity matrix. 
These relations hold literally also for $\check{R}^{(\bar n,\bar n)}$.
However, differently from
the $SU(2)$ case, the $SU(n)$ case for $n>2$ does not have crossing symmetry,
which makes this model special in the realm of integrable models. This is
because for $n>2$ the conjugate of the representation $[n]$, namely
$[\bar{n}]$, is inequivalent to $[n]$.

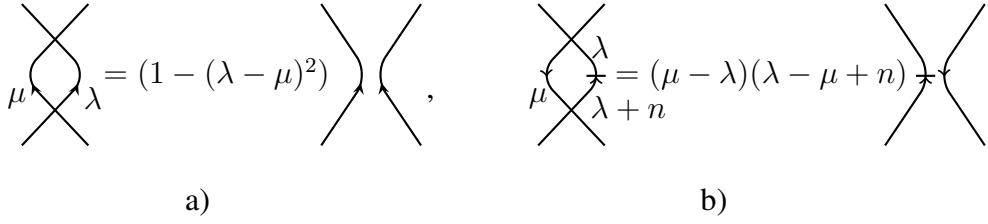
\begin{figure}[h]
	\begin{center}
		\begin{minipage}{0.5\linewidth}
			\begin{tikzpicture}[scale=1.25]
			\tikzstyle directed=[postaction={decorate,decoration={markings,
					mark=at position .45 with {\arrow[arrowstyle]{stealth}}}}]
			
			\draw (0,0.5) [-,color=black, thick, directed, rounded corners=8pt]+(-(0.35,-0.75) -- +(0.35,0)--+(-0.35,0.75);
			\draw (0,0.5) [-,color=black, thick,directed, rounded corners=8pt]+(0.35,-0.75) -- +(-0.35,0)--+(0.35,0.75);
			
			\draw (1.7,0.5) node {$=(1-(\lambda-\mu)^2)$};
			
			\draw (3.15,0.5) [-,color=black, thick, directed, rounded corners=8pt]+(-0.35,-0.75) -- +(0.15,0)--+(-0.35,0.75);
			\draw (3.5,0.5) [-,color=black, thick,directed, rounded corners=8pt]+(0.35,-0.75) -- +(-0.15,0)--+(0.35,0.75);

			\tikzstyle directed=[postaction={decorate,decoration={markings,
					mark=at position .65 with {\arrow[arrowstyle]{stealth}}}}]

			\draw (-0.4,0.25) node {$\mu$};
			\draw (0.4,0.25) node {$\lambda$};

			\draw (1.5,-0.85) node {a)};
			\draw (3.95,0.25) node {,};
			
			\end{tikzpicture}
		\end{minipage}%
		\begin{minipage}{0.5\linewidth}
			\begin{tikzpicture}[scale=1.25]
			\tikzstyle directed=[postaction={decorate,decoration={markings,
					mark=at position .5 with {\arrow[arrowstyle]{>|}}}}]
			\tikzstyle reverse directed=[postaction={decorate,decoration={markings,
					mark=at position .49 with {\arrowreversed[arrowstyle]{>};}}}]

			\draw (0,0.5) [-,color=black, thick, directed, rounded corners=8pt]+(-(0.35,-0.75) -- +(0.35,0)--+(-0.35,0.75);
			\draw (0,0.5) [-,color=black, thick, reverse directed, rounded corners=8pt]+(0.35,-0.75) -- +(-0.35,0)--+(0.35,0.75);
			
			\draw (2,0.5) node {$=(\mu-\lambda)(\lambda-\mu+n)$};
			
			\draw (3.65,0.5) [-,color=black, thick, directed, rounded corners=8pt]+(-0.35,-0.75) -- +(0.15,0)--+(-0.35,0.75);
			\draw (4,0.5) [-,color=black, thick, reverse directed, rounded corners=8pt]+(0.35,-0.75) -- +(-0.15,0)--+(0.35,0.75);

			\tikzstyle directed=[postaction={decorate,decoration={markings,
					mark=at position .65 with {\arrow[arrowstyle]{stealth}}}}]
			\tikzstyle reverse directed=[postaction={decorate,decoration={markings,
					mark=at position .65 with {\arrowreversed[arrowstyle]{stealth};}}}]

			\draw (-0.35,0.25) node {$\mu$};
			\draw (0.6,0.15) node {$\lambda+n$};
			\draw (0.3,0.8) node {$\lambda$};

			\draw (1.5,-0.85) node {b)};

			\end{tikzpicture}
		\end{minipage}
		
	\end{center}
	\caption{Graphical illustration of the unitarity  relations (two
            more are obtained by 180$^{\circ}$ rotations): a) the standard
          unitarity (\ref{unitarity}); b) the special unitarity
          (\ref{s-unitarity1}). Again we consider the spectral parameter to
          be discontinuous in order to describe the shift in the $R$-matrix,
          i.e. the spectral parameter value is $\lambda+n$ in the bottom part and
          $\lambda$ in the top part of the graph.}
	\label{Unitarities}
\end{figure}

In order to circumvent the difficulties which arise from the fact that the model
lacks the crossing symmetry, one has to add a few more ingredients to formulate
a consistent framework for the computation of correlation functions. 
The crucial observation is that one has to conveniently and largely on the same footing work with the
fundamental $[n]$ and anti-fundamental representation $[\bar{n}]$ of the
$SU(n)$. This is possible since as presented before, the Yang-Baxter equation
accommodates different representations in each vector space. Besides that, the
above $R$-matrices with mixed representations also have symmetry properties,
which we call special unitarity (see Figure \ref{Unitarities}b),
\bear
\check{R}^{(n,\bar{n})}_{12}(\lambda-\mu+n)\check{R}_{21}^{(\bar{n},n)}(\mu-\lambda) &=& (\mu-\lambda)(\lambda-\mu+n) I, \label{s-unitarity1}\\
 \check{R}^{(\bar{n},n)}_{12}(\lambda-\mu)\check{R}_{21}^{(n,\bar{n})}(\mu-\lambda+n) &=& (\lambda-\mu)(\mu-\lambda+n)I. \label{s-unitarity2}
\ear

Finally, in order to exploit the full $SU(n)$ symmetry we introduce relations
of suitable products of $n$ many $R$-matrices with the
completely antisymmetric state in $V^{\otimes n}$, i.e. the totally
  antisymmetric tensor $\epsilon$. For instance, in the $SU(3)$ case these relations read,
\eq
\begin{tikzpicture}[scale=1.7]
\draw (0.3,0.2) node {$\lambda$};
\draw (0.3,-0.3) node {$\lambda+1$};
\draw (0.3,-0.8) node {$\lambda+2$};

\draw (0,0) [<->,color=black, thick, rounded corners=7pt]
+(0,0) -- +(1.0,0) -- +(1.0,-1.) -- +(0.0,-1.0);
\draw (0,0) [->,color=black, thick, rounded corners=7pt]
+(1.0,-0.5) -- +(0,-0.5);
\draw (0.5,-1.3) node {$\mu$};
\draw (0.1,-0.5) [->,color=black, thick, rounded corners=7pt]
+(0.5,-1) -- +(0.5,1);
\draw (2.65,-0.5) node {$=(\lambda+2-\mu)(1-(\lambda-\mu)^2)$};
\draw (1.,-0.5)[fill=black]  circle (0.15ex);

\draw (4.5,0.2) node {$\lambda$};
\draw (4.5,-0.3) node {$\lambda+1$};
\draw (4.5,-0.8) node {$\lambda+2$};
\draw (4.2,0) [<->,color=black, thick, rounded corners=7pt]
+(0,0) -- +(1.0,0) -- +(1.0,-1.) -- +(0.0,-1.0);
\draw (4.2,0) [->,color=black, thick, rounded corners=7pt]
+(1.0,-0.5) -- +(0,-0.5);
\draw (5.3,-0.5) [->,color=black, thick, rounded corners=7pt]
+(0.5,-1) -- +(0.5,1);
\draw (5.4,-1.3) node {$\mu$};

\draw (5.2,-0.5)[fill=black]  circle (0.15ex);
\draw (6,-0.5) node {$,$};
\end{tikzpicture}
\label{sym-property1}
\en
and
\eq
\begin{tikzpicture}[scale=1.7]
\draw (0.3,0.2) node {$\lambda$};
\draw (0.3,-0.3) node {$\lambda+1$};
\draw (0.3,-0.8) node {$\lambda+2$};

\draw (0,0) [<->,color=black, thick, rounded corners=7pt]
+(0,0) -- +(1.0,0) -- +(1.0,-1.) -- +(0.0,-1.0);
\draw (0,0) [->,color=black, thick, rounded corners=7pt]
+(1.0,-0.5) -- +(0,-0.5);
\draw (0.5,-1.3) node {$\mu$};
\draw (0.1,-0.5) [->,color=black, thick, rounded corners=7pt]
+(0.5,1) -- +(0.5,-1);
\draw (2.6,-0.5) node {$=(\mu-\lambda)(1-(\lambda+2-\mu)^2)$};
\draw (1.,-0.5)[fill=black]  circle (0.15ex);

\draw (4.5,0.2) node {$\lambda$};
\draw (4.5,-0.3) node {$\lambda+1$};
\draw (4.5,-0.8) node {$\lambda+2$};
\draw (4.2,0) [<->,color=black, thick, rounded corners=7pt]
+(0,0) -- +(1.0,0) -- +(1.0,-1.) -- +(0.0,-1.0);
\draw (4.2,0) [->,color=black, thick, rounded corners=7pt]
+(1.0,-0.5) -- +(0,-0.5);
\draw (5.3,-0.5) [->,color=black, thick, rounded corners=7pt]
+(0.5,1) -- +(0.5,-1);
\draw (5.6,-1.3) node {$\mu$};
\draw (6,-0.5) node {$,$};
\draw (5.2,-0.5)[fill=black]  circle (0.15ex);
\end{tikzpicture}
\label{sym-property2}
\en
and the depicted objects are
\eq
\begin{minipage}{0.5\linewidth}
\begin{center}
	\begin{tikzpicture}[scale=1.25]
\draw (0,0.5) [-,color=black,  thick, rounded corners=7pt]
+(0,0) -- +(0.5,0) -- +(0.5,-1.) -- +(0.0,-1.0);
\draw (0,0.5) [-,color=black, thick, rounded corners=7pt]
+(0.5,-0.5) -- +(0,-0.5);
\draw (1.5,0.) node {$=\epsilon_{ijk},$};
\draw (0.,0)[fill=black]  circle (0.3ex);
		\draw (0.,-0.5)[fill=black]  circle (0.3ex);
		\draw (0.,0.5)[fill=black]  circle (0.3ex);
\draw (-0.25,-0.5) node {$i$};
\draw (-0.25,0) node {$j$};
\draw (-0.25,0.5) node {$k$};

\end{tikzpicture}
\end{center}
		\end{minipage}%
		\begin{minipage}{0.5\linewidth}
\begin{center}
\begin{tikzpicture}[scale=1.]
\draw (0,0) [-,color=black, thick, directed, rounded corners=8pt]+(-(0.,-0.75) --+(0.,0.75);
\draw (0,-1.1) node {$i$};
\draw (0,1.1) node {$j$};
\draw (0.75,0) node {$=\delta_{ij}$};
		\draw (0.,-0.75)[fill=black]  circle (0.3ex);
		\draw (0.,0.75)[fill=black]  circle (0.3ex);

\end{tikzpicture}
\end{center}
		\end{minipage}
\label{anti-symmetrizer}
\en
the fully anti-symmetric tensor (Levi-Civita tensor) and the Kronecker
delta. Besides that, a number of simple identities among the previous objects
are used throughout this work, e.g. tensor products and contractions
\bear
\begin{minipage}{0.5\linewidth}
	\begin{center}
		\begin{tikzpicture}[scale=1.25]
		\draw (0,0.5) [-,color=black,  thick, rounded corners=7pt]
		+(0,0) -- +(0.5,0) -- +(0.5,-1.) -- +(0.0,-1.0);
		\draw (0,0.5) [-,color=black,  thick, rounded corners=7pt]
		+(0,0) -- +(-0.5,0) -- +(-0.5,-1.) -- +(0.0,-1.0);
		\draw (0,0.5) [-,color=black, thick, rounded corners=7pt]
		+(0.5,-0.5) -- +(0,-0.5);
		\draw (0,0.5) [-,color=black, thick, rounded corners=7pt]
		+(-0.5,-0.5) -- +(0,-0.5);
		\draw (1.55,0.) node {$=\epsilon_{ijk} \epsilon^{ijk}=6,$};
		\draw (0.,0)[fill=black]  circle (0.3ex);
		\draw (0.,-0.5)[fill=black]  circle (0.3ex);
		\draw (0.,0.5)[fill=black]  circle (0.3ex);
		\draw (-0.,-0.7) node {$i$};
		\draw (-0.,-0.2) node {$j$};
		\draw (-0.,0.3) node {$k$};
		
		\end{tikzpicture}
	\end{center}
\end{minipage}%
\begin{minipage}{0.5\linewidth}
	\begin{center}
		\begin{tikzpicture}[scale=1.]
		\draw (0,0.5) node {$i$};
		\draw (1.6,0) node {$=\delta_{ii}=3$};
		\draw (0.,0)  circle (0.75);
		\draw (0.,0.75)[fill=black]  circle (0.3ex);
		
		\end{tikzpicture}
	\end{center}
\end{minipage}
\label{other1}\\
\begin{minipage}{0.45\linewidth}
	\begin{center}
		\begin{tikzpicture}[scale=1.2]
		\draw (0,0) [-,color=black, thick, rounded corners=7pt]
		+(0,0) -- +(0,0.5) -- +(-0.5,0.5) -- +(-0.5,0);
		\draw (0,0) [-,color=black,  thick, rounded corners=7pt]
		+(0,0) -- +(0,-0.5) -- +(-0.5,-0.5) -- +(-0.5,0);
		\draw (0,0) [-,color=black, thick, rounded corners=7pt]+(-0.25,0.5) -- +(-0.25,1);
		\draw (0,0) [-,color=black, thick, rounded corners=7pt]+(-0.25,-0.5) -- +(-0.25,-1);
		\draw (0.,0)[fill=black]  circle (0.3ex);
		\draw (-0.5,0)[fill=black]  circle (0.3ex);
		\draw (-0.25,1)[fill=black]  circle (0.3ex);
		\draw (-0.25,-1)[fill=black]  circle (0.3ex);
		\draw (-0.15,-0.8) node {$i$};
		\draw (-0.7,-0.) node {$j$};
		\draw (0.2,0.) node {$k$};
		\draw (-0.15,0.8) node {$l$};
		\draw (0.75,0.) node {$=2$};

\draw (1.2,0) [-,color=black, thick, directed, rounded corners=8pt]+(-(0.,-0.75) --+(0.,0.75);
\draw (1.2,-1.1) node {$i$};
\draw (1.2,1.1) node {$l$};
\draw (1.2,-0.75)[fill=black]  circle (0.3ex);
\draw (1.2,0.75)[fill=black]  circle (0.3ex);

\draw (2.45,0.) node {$=\epsilon_{ijk} \epsilon^{jkl}=2 \delta_{il},$};
		
		\end{tikzpicture}
	\end{center}
\end{minipage}%
\begin{minipage}{0.55\linewidth}
	\begin{center}
		\begin{tikzpicture}[scale=1.2]
		\draw (-0.5,0) [-,color=black,  thick, rounded corners=7pt]
		+(0,0.5) -- +(0,0) -- +(0.5,0) -- +(0.5,0.5);
		\draw (-0.5,-0.5) [-,color=black,  thick, rounded corners=7pt]
		+(0,-0.5) -- +(0,0) -- +(0.5,0) -- +(0.5,-0.5);
		\draw (0,-1.0) [-,color=black, thick, rounded corners=7pt]+(-0.25,0.5) -- +(-0.25,1);

		\draw (0.,-1)[fill=black]  circle (0.3ex);
		\draw (-0.5,0.5)[fill=black]  circle (0.3ex);
		\draw (0,0.5)[fill=black]  circle (0.3ex);
		\draw (-0.5,-1)[fill=black]  circle (0.3ex);
		\draw (-0.25,-0.25)[fill=black]  circle (0.3ex);

		\draw (-0.5,-1.25) node {$i$};
		\draw (-0.0,-1.25) node {$j$};
		\draw (-0.5,0.8) node {$l$};
		\draw (-0.45,-0.25) node {$k$};
		\draw (-0.,0.8) node {$m$};
		\draw (0.25,-0.3) node {$=$};

\draw (0.65,0) [-,color=black, thick, directed, rounded corners=8pt]+(-(0.,-1) --+(0.,0.5);
\draw (0.65,-1.25) node {$i$};
\draw (0.65,0.75) node {$l$};
\draw (0.65,-1)[fill=black]  circle (0.3ex);
\draw (0.65,0.5)[fill=black]  circle (0.3ex);

\draw (0.9,0) [-,color=black, thick, directed, rounded corners=8pt]+(-(0.,-1) --+(0.,0.5);
\draw (0.9,-1.25) node {$j$};
\draw (0.9,0.75) node {$m$};
\draw (0.9,-1)[fill=black]  circle (0.3ex);
\draw (0.9,0.5)[fill=black]  circle (0.3ex);

\draw (1.15,-0.3) node {$-$};

\draw (1.4,0) [-,color=black, thick, directed, rounded corners=8pt]+(-(0.,-1)--+(0.11,-0.25);
\draw (1.4,0) [-,color=black, thick, rounded corners=8pt]+((0.15,-0.15)--+(0.25,0.5);

\draw (1.4,0) [-,color=black, thick, directed, rounded corners=8pt]+((0.25,-1)--+(0.,0.5);
\draw (1.4,-1.25) node {$i$};
\draw (1.4,0.75) node {$l$};
\draw (1.4,-1)[fill=black]  circle (0.3ex);
\draw (1.4,0.5)[fill=black]  circle (0.3ex);

\draw (1.65,-1.25) node {$j$};
\draw (1.65,0.75) node {$m$};
\draw (1.65,-1)[fill=black]  circle (0.3ex);
\draw (1.65,0.5)[fill=black]  circle (0.3ex);

\draw (3.65,-0.3) node {$=\epsilon_{ijk} \epsilon^{klm}=\delta_{il}\delta_{jm}-\delta_{im}\delta_{jl}.$};
		
		\end{tikzpicture}
	\end{center}
\end{minipage}
\label{other2}
\ear

\section{Density matrices}\label{density}

\begin{figure}[h]
\begin{tikzpicture}[scale=1.6]
\draw (0,0) [->,color=black, thick, rounded corners=7pt] +(0.25,0)--(0.25,3);

\draw (0,0) [->,color=black, thick, rounded corners=7pt] +(0.8,1.75)--(0.8,3);
\draw (0,0) [-,color=black, thick, rounded corners=7pt] +(0.8,0)--(0.8,1.25);

\draw (0,0) [->,color=black, thick, rounded corners=7pt] +(1.4,1.75)--(1.4,3);
\draw (0,0) [-,color=black, thick, rounded corners=7pt] +(1.4,0)--(1.4,1.25);

\draw (0,0) [->,color=black, thick, rounded corners=7pt] +(2.05,1.75)--(2.05,3);
\draw (0,0) [-,color=black, thick, rounded corners=7pt] +(2.05,0)--(2.05,1.25);

\draw (0,0) [->,color=black, thick, rounded corners=7pt] +(2.65,0)--(2.65,3);

\draw (0.25,-0.25) node {$0$};
\draw (0.8,-0.25) node {$\lambda_1$};
\draw (1.4,-0.25) node {$\lambda_2$};
\draw (1.75	,-0.25) node {$\dots$};
\draw (2.05,-0.25) node {$\lambda_m$};
\draw (2.65,-0.25) node {$0$};

\draw (0.8,1.25)[fill=black]  circle (0.15ex);
\draw (0.8,1.75)[fill=black]  circle (0.15ex);

\draw (1.4,1.25)[fill=black]  circle (0.15ex);
\draw (1.4,1.75)[fill=black]  circle (0.15ex);

\draw (2.05,1.25)[fill=black]  circle (0.15ex);
\draw (2.05,1.75)[fill=black]  circle (0.15ex);

\draw (-0.25,0) [->,color=black, thick, rounded corners=7pt] +(0,0.5)--(3.5,0.5);
\draw (-0.25,0) [->,color=black, thick, rounded corners=7pt] +(0,1.0)--(3.5,1);

\draw (-0.25,0) [->,color=black, thick, rounded corners=7pt] +(0,2.5)--(3.5,2.5);
\draw (-0.25,0) [->,color=black, thick, rounded corners=7pt] +(0,2.0)--(3.5,2.0);

\draw (-0.4,0.5) node {$\dots$};
\draw (-0.4,1.0) node {$\dots$};
\draw (-0.4,2.5) node {$\dots$};
\draw (-0.4,2.0) node {$\dots$};

\draw (3.7,0.5) node {$\dots$};
\draw (3.7,1.0) node {$\dots$};
\draw (3.7,2.5) node {$\dots$};
\draw (3.7,2.0) node {$\dots$};

\draw (-1.5,3.) node {a)};

\draw (-1.7,1.5) node {$D_m(\lambda_1,\dots,\lambda_m)=$};

\draw (3.1,0.6) node {$u_1$};
\draw (3.1,1.1) node {$u_2$};
\draw (3.1,1.6) node {$\vdots$};
\draw (3.1,2.6) node {$u_{N}$};
\draw (3.1,2.1) node {$u_{N-1}$};

\end{tikzpicture}
\begin{tikzpicture}[scale=1.6]
\draw (0,0) [->,color=black, thick, rounded corners=7pt] +(0.8,1.75)--(0.8,3);
\draw (0,0) [-,color=black, thick, rounded corners=7pt] +(0.8,0)--(0.8,1.25);

\draw (0,0) [->,color=black, thick, rounded corners=7pt] +(1.4,1.75)--(1.4,3);
\draw (0,0) [-,color=black, thick, rounded corners=7pt] +(1.4,0)--(1.4,1.25);

\draw (0,0) [->,color=black, thick, rounded corners=7pt] +(2.05,1.75)--(2.05,3);
\draw (0,0) [-,color=black, thick, rounded corners=7pt] +(2.05,0)--(2.05,1.25);

\draw (0.8,-0.25) node {$\lambda_1$};
\draw (1.4,-0.25) node {$\lambda_2$};
\draw (1.75	,-0.25) node {$\dots$};
\draw (2.05,-0.25) node {$\lambda_m$};

\draw (0.8,1.25)[fill=black]  circle (0.15ex);
\draw (0.8,1.75)[fill=black]  circle (0.15ex);

\draw (1.4,1.25)[fill=black]  circle (0.15ex);
\draw (1.4,1.75)[fill=black]  circle (0.15ex);

\draw (2.05,1.25)[fill=black]  circle (0.15ex);
\draw (2.05,1.75)[fill=black]  circle (0.15ex);

\draw (-0.25,0) [->,color=black, thick, rounded corners=7pt] +(0.5,0.5)--(3.,0.5);
\draw (-0.25,0) [->,color=black, thick, rounded corners=7pt] +(0.5,1.0)--(3.,1);

\draw (-0.25,0) [->,color=black, thick, rounded corners=7pt] +(0.5,2.5)--(3.,2.5);
\draw (-0.25,0) [->,color=black, thick, rounded corners=7pt] +(0.5,2.0)--(3.,2.0);

\draw (-1.5,3.) node {b)};

\draw (-1.7,1.5) node {$D_m(\lambda_1,\dots,\lambda_m)=$};

\draw (2.5,0.6) node {$u_1$};
\draw (2.5,1.1) node {$u_2$};
\draw (2.5,1.6) node {$\vdots$};
\draw (2.5,2.6) node {$u_{N}$};
\draw (2.5,2.1) node {$u_{N-1}$};

\draw (0,1.5) node {$\Phi_L$};
\draw (0,0) [-,color=black, thick, rounded corners=7pt] +(0.25,0)--(0.25,3)--(-0.3,1.5)--(0.25,0);
\draw (3.25,1.5) node {$\Phi_R$};
\draw (0,0) [-,color=black, thick, rounded corners=7pt] +(3.,0)--(3.,3)--(3.55,1.5)--(3.,0);

\end{tikzpicture}
\caption{Graphical illustration of the un-normalized density operator
  $D_m(\lambda_1,\dots,\lambda_m)$:  a) an infinite
  cylinder with $N$ infinitely long horizontal lines carrying spectral
  parameters $u_j$ and $m$ open bonds associated to the spectral parameters
  $\lambda_1,\dots,\lambda_m$; b) the infinitely many column-to-column
  transfer matrices to the left and to the right are replaced by the boundary
  states they project onto.}
\label{figDmatrix}
\end{figure}
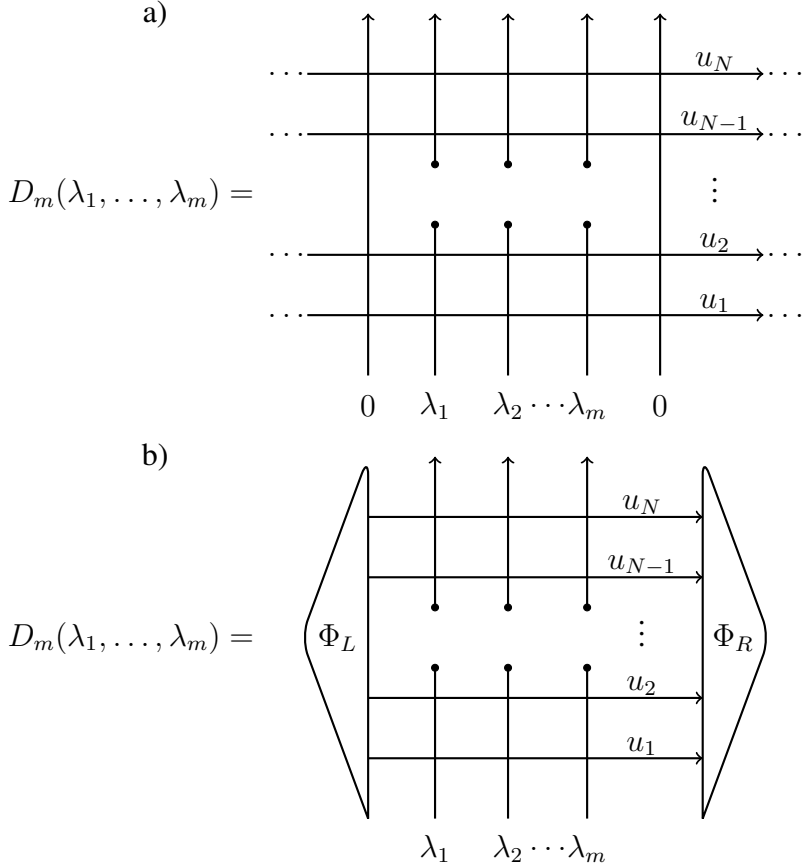

The framework for calculating thermal correlation functions of
integrable Hamiltonians was introduced in \cite{GoKlSe04} and has been
applied to the case of integrable $SU(2)$ spin chains several times, see
e.g.~\cite{GSS10,KNS013,RK2016}.  This approach makes use of the usual
inhomogeneous reduced density operator, see Figure \ref{figDmatrix}a, in the
thermodynamic limit $L \to \infty$, however with finite Trotter number $N$
\cite{GoKlSe04}. This formulation can be naturally extended to the case of
$SU(n)$ spin chains. As the infinitely many column-to-column transfer matrices
on the left (right) project onto the leading eigenstate $\bra{\Phi_L}$
($\ket{\Phi_R}$) we obtain the compact form
\eq
D_m(\lambda_1,\cdots,\lambda_m)=\frac{  \bra{\Phi_L}{\cal T}^{(n)}_1(\lambda_1)\cdots {\cal T}^{(n)}_m(\lambda_m) \ket{\Phi_R}}{ \Lambda_0^{(n)}(\lambda_1)\cdots \Lambda_0^{(n)}(\lambda_m) },
\label{i-dm}
\en
where ${\cal T}^{(n)}_j(x)$ is the usual $j$-th monodromy matrix ${\cal
  T}^{(n)}_j(x)=R^{(n,n)}_{j,N}(x-u_N)\dots $ $R^{(n,n)}_{j,2}(x-u_2) R^{(n,n)}_{j,1}(x-u_1)$ associated
to the quantum transfer matrix for the $SU(n)$ quantum spin chains $t_j^{QTM}(x)=\tr[{\cal T}_j^{(n)}(x)]$, $\Phi_L$
and $\Phi_R$ represent the left and right leading eigenstates of the quantum transfer
matrix and $\Lambda_0^{(n)}(x)$ is the leading eigenvalue. 
For instance, the matrix element ${D_m}_{11\cdots1}^{11\cdots1}=\tr{\left[\hat{e}^{(1)}_{11}\hat{e}^{(2)}_{11}\cdots \hat{e}^{(m)}_{11}D_m\right]}$ (which in Figure \ref{figDmatrix} corresponds to assign $1$ to all the indices sitting at the black dots) is the standard emptiness formation probability $P_m(\lambda_1,\cdots,\lambda_m)$ \cite{BOOK-KBI,EFP}.

The physically interesting result is typically obtained from the above reduced
density operator (\ref{i-dm}) by taking the homogeneous limit $\lambda_j \to 0$
and the Trotter limit $N\to\infty$. However, we take advantage of the
dependence on arbitrary $\lambda\in\mathbb{C}$. 
For the density operator of the $SU(2)$ case a set
of discrete functional equations can be derived by use of the usual
integrability structure plus the crossing symmetry \cite{AuKl12}. This and
transparent analyticity properties
allow for the complete determination of the reduced density operator at
finite temperature and for an alternative proof of factorization of the
correlation functions in terms of sums over products of nearest-neighbor
correlators.
 
\begin{figure}
\begin{tikzpicture}[scale=1.6]

\draw (-1.5,4.5) node {a)};

\draw (0,0) [->,color=black, thick, rounded corners=7pt] +(0.25,0)--(0.25,4.7);
\draw (0,0) [->,color=black, thick, rounded corners=7pt] +(0.8,0)--(0.8,4.7);

\draw (0,0) [<-,color=black, thick, rounded corners=7pt] +(-0.25,1.0)--(1.3,1.0);
\draw (0,0) [<-,color=black, thick, rounded corners=7pt] +(-0.25,1.4)--(1.3,1.4);
\draw (0,0) [<-,color=black, thick, rounded corners=7pt] +(-0.25,1.8)--(1.3,1.8);

\draw (0,0) [<-,color=black, thick, rounded corners=7pt] +(-0.25,2.2)--(1.3,2.2);
\draw (0,0) [<-,color=black, thick, rounded corners=7pt] +(-0.25,2.6)--(1.3,2.6);
\draw (0,0) [<-,color=black, thick, rounded corners=7pt] +(-0.25,3.0)--(1.3,3.0);

\draw (0,0) [<-,color=black, thick, rounded corners=7pt] +(-0.25,3.4)--(1.3,3.4);
\draw (0,0) [<-,color=black, thick, rounded corners=7pt] +(-0.25,3.8)--(1.3,3.8);
\draw (0,0) [<-,color=black, thick, rounded corners=7pt] +(-0.25,4.2)--(1.3,4.2);

\draw (-0.4,1.0) node {$\dots$};
\draw (-0.4,1.4) node {$\dots$};
\draw (-0.4,1.8) node {$\dots$};
\draw (-0.4,2.2) node {$\dots$};
\draw (-0.4,2.6) node {$\dots$};
\draw (-0.4,3.0) node {$\dots$};
\draw (-0.4,3.4) node {$\dots$};
\draw (-0.4,3.8) node {$\dots$};
\draw (-0.4,4.2) node {$\dots$};

\draw (-0.4,0.25) node {$\dots$};
\draw (-0.4,0.6) node {$\dots$};
\draw (-0.4,4.55) node {$\dots$};

\draw (3.95,0.25) node {$\dots$};
\draw (3.95,0.6) node {$\dots$};
\draw (3.95,4.55) node {$\dots$};

\draw (0,0) [->,color=black, thick, rounded corners=7pt] +(2.6,0)--(2.6,4.7);
\draw (0,0) [->,color=black, thick, rounded corners=7pt] +(3.05,0)--(3.05,4.7);

\draw (1.97,0.98) node {$\lambda_m+n-1$};
\draw (1.97,1.23) node {$\vdots$};
\draw (1.7,1.4) node {$\lambda_m+1$};
\draw (1.5,1.8) node {$\lambda_m$};

\draw (1.0,2.06) node {$\vdots$};

\draw (1.95,2.18) node {$\lambda_2+n-1$};
\draw (1.95,2.43) node {$\vdots$};
\draw (1.7,2.6) node {$\lambda_2+1$};
\draw (1.5,3.0) node {$\lambda_2$};

\draw (1.95,3.38) node {$\lambda_1+n-1$};
\draw (1.95,3.63) node {$\vdots$};
\draw (1.7,3.8) node {$\lambda_1+1$};
\draw (1.5,4.2) node {$\lambda_1$};

\draw (0.25,-0.15) node {$0$};
\draw (0.8,-0.15) node {$0$};
\draw (2.6,-0.15) node {$0$};
\draw (3.05,-0.15) node {$0$};

\draw (-0.25,0) [->,color=black, thick, rounded corners=7pt] +(0,0.25)--(3.75,0.25);
\draw (-0.25,0) [->,color=black, thick, rounded corners=7pt] +(0,0.6)--(3.75,0.6);

\draw (-0.25,0) [->,color=black, thick, rounded corners=7pt] +(0,4.55)--(3.75,4.55);

\draw (3.4,0.35) node {$u_1$};
\draw (3.4,0.75) node {$u_2$};
\draw (3.4,1.1) node {$\vdots$};
\draw (3.4,4.35) node {$u_{N}$};

\draw (1.3,1)[fill=black]  circle (0.15ex);
\draw (1.3,1.4)[fill=black]  circle (0.15ex);
\draw (1.3,1.8)[fill=black]  circle (0.15ex);

\draw (1.3,2.2)[fill=black]  circle (0.15ex);
\draw (1.3,2.6)[fill=black]  circle (0.15ex);
\draw (1.3,3)[fill=black]  circle (0.15ex);

\draw (1.3,3.4)[fill=black]  circle (0.15ex);
\draw (1.3,3.8)[fill=black]  circle (0.15ex);
\draw (1.3,4.2)[fill=black]  circle (0.15ex);

\draw (-1.8,2.5) node {${\mathbb D}_m(\lambda_1,\dots,\lambda_m)=$};

\end{tikzpicture}
\begin{tikzpicture}[scale=1.65]
\draw (-1.,4.5) node {b)};

\draw (0,0) [<-,color=black, thick, rounded corners=7pt] +(0.5,1.0)--(1.3,1.0);
\draw (0,0) [<-,color=black, thick, rounded corners=7pt] +(0.5,1.4)--(1.3,1.4);
\draw (0,0) [<-,color=black, thick, rounded corners=7pt] +(0.5,1.8)--(1.3,1.8);

\draw (0,0) [<-,color=black, thick, rounded corners=7pt] +(0.5,2.2)--(1.3,2.2);
\draw (0,0) [<-,color=black, thick, rounded corners=7pt] +(0.5,2.6)--(1.3,2.6);
\draw (0,0) [<-,color=black, thick, rounded corners=7pt] +(0.5,3.0)--(1.3,3.0);

\draw (0,0) [<-,color=black, thick, rounded corners=7pt] +(0.5,3.4)--(1.3,3.4);
\draw (0,0) [<-,color=black, thick, rounded corners=7pt] +(0.5,3.8)--(1.3,3.8);
\draw (0,0) [<-,color=black, thick, rounded corners=7pt] +(0.5,4.2)--(1.3,4.2);

\draw (1.97,0.98) node {$\lambda_m+n-1$};
\draw (1.97,1.23) node {$\vdots$};
\draw (1.7,1.4) node {$\lambda_m+1$};
\draw (1.5,1.8) node {$\lambda_m$};

\draw (1.0,2.06) node {$\vdots$};

\draw (1.95,2.18) node {$\lambda_2+n-1$};
\draw (1.95,2.43) node {$\vdots$};
\draw (1.7,2.6) node {$\lambda_2+1$};
\draw (1.5,3.0) node {$\lambda_2$};

\draw (1.95,3.38) node {$\lambda_1+n-1$};
\draw (1.95,3.63) node {$\vdots$};
\draw (1.7,3.8) node {$\lambda_1+1$};
\draw (1.5,4.2) node {$\lambda_1$};

\draw (-0.25,0) [->,color=black, thick, rounded corners=7pt] +(0.75,0.25)--(3.25,0.25);
\draw (-0.25,0) [->,color=black, thick, rounded corners=7pt] +(0.75,0.6)--(3.25,0.6);

\draw (-0.25,0) [->,color=black, thick, rounded corners=7pt] +(0.75,4.55)--(3.25,4.55);

\draw (0.25,2.5) node {$\widetilde{\Phi}_L$};
\draw (0,0) [-,color=black, thick, rounded corners=7pt] +(0.5,0)--(0.5,4.85)--(-0.05,2.5)--(0.5,0);

\draw (3.5,2.5) node {$\Phi_R$};
\draw (0,0) [-,color=black, thick, rounded corners=7pt] +(3.25,0)--(3.25,4.85)--(3.8,2.5)--(3.25,0);

\draw (2.9,0.35) node {$u_1$};
\draw (2.9,0.75) node {$u_2$};
\draw (2.9,1.1) node {$\vdots$};
\draw (2.9,4.35) node {$u_{N}$};

\draw (1.3,1)[fill=black]  circle (0.15ex);
\draw (1.3,1.4)[fill=black]  circle (0.15ex);
\draw (1.3,1.8)[fill=black]  circle (0.15ex);

\draw (1.3,2.2)[fill=black]  circle (0.15ex);
\draw (1.3,2.6)[fill=black]  circle (0.15ex);
\draw (1.3,3)[fill=black]  circle (0.15ex);

\draw (1.3,3.4)[fill=black]  circle (0.15ex);
\draw (1.3,3.8)[fill=black]  circle (0.15ex);
\draw (1.3,4.2)[fill=black]  circle (0.15ex);

\draw (-1.3,2.5) node {${\mathbb D}_m(\lambda_1,\dots,\lambda_m)=$};

\end{tikzpicture}
\caption{Graphical illustration of the un-normalized generalized density
  operator ${\mathbb D}_m(\lambda_1,\dots,\lambda_m)$: a) the
  infinite cylinder with $N$ infinitely long horizontal lines carrying
  spectral parameters $u_j$ and $m$ bunches of $n$ semi-infinite lines with
  spectral parameters $\{\lambda_j,\lambda_j+1,\dots,\lambda_j+n-1 \}$ for
  $j=1,\dots, m$;  b) the infinitely many column-to-column
  transfer matrices to the left and to the right are replaced by the boundary
  states they project onto.}
\label{figBunch}
\end{figure}
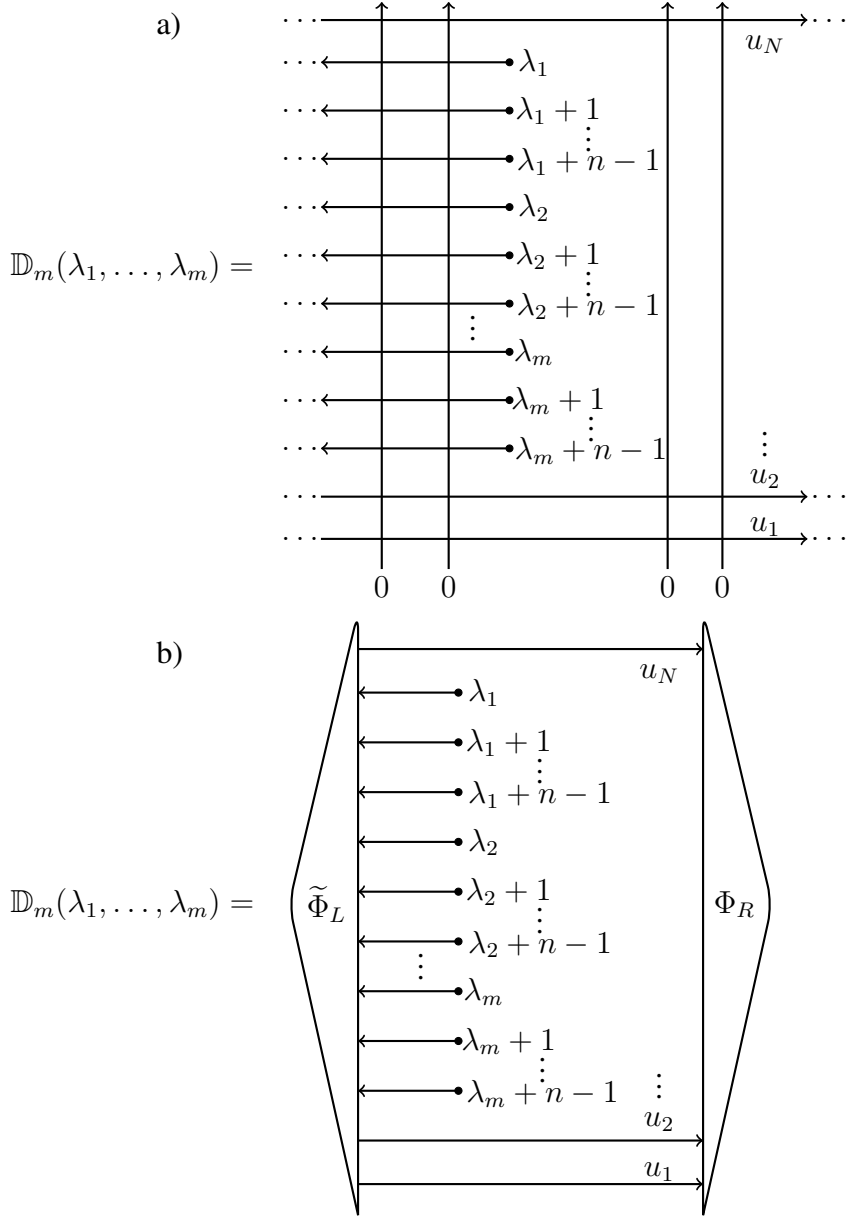

Unfortunately, for the case of $SU(n)$ spin chains with $n>2$ we
do not have crossing symmetry and hence adopting the line of reasoning of the
$SU(2)$ case does not result in a closed set of functional
equations for $D_m$. However, we can derive an analogous set of discrete functional
equations provided we consider a slightly more general density operator we
denote by ${\mathbb D}_m(\lambda_1,\dots,\lambda_m)$. 
Like the usual density operator, the generalized density
operator ${\mathbb D}_m$ is defined on a horizontal infinite cylinder with $N$ horizontal
lines, carrying spectral parameters $u_j$, however with additional semi-infinite rows as
depicted in Figure \ref{figBunch}a. Alternatively, this generalized
correlator can be written with boundary states
where now $\widetilde{\Phi}_L$ is the leading eigenstate of the modified
quantum transfer matrix acting on a tensor product of $N+m\cdot n$ copies of
$V$ (see Figure \ref{figBunch}b). 
The monodromy matrix is given by
$\widetilde{\cal
  T}_j^{(n)}(x)={\cal T}_j^{(n)}(x) \cdot \prod_{\alpha=1}^m
\prod_{\beta=1}^{n}
R^{(n,\bar{n})}_{j,N+(\alpha-1)n+\beta}(x-(\lambda_{\alpha}+\beta-1))$.

Besides the correlations contained in $D_m$, 
the generalized density operator ${\mathbb D}_m$ also contains other correlation functions,
like those contained in variants of $D_m$ with just anti-fundamental
representations or any mixture of
fundamental and anti-fundamental representations in the spaces indexed by 1 to
$m$.

Note that $D_m$ may be viewed as a vector in the tensor space $[\bar
  n]^{\otimes m}\otimes [n]^{\otimes m}$, i.e. a multilinear map $[n]^m\times
[\bar n]^m\to \mathbb{C}$, and likewise ${\mathbb D}_m$ is a vector in the
tensor space $[\bar n]^{\otimes mn}$, i.e. a multilinear map $[n]^{mn}\to
\mathbb{C}$.

We have to show that
\begin{itemize}
\item the generalized density operator ${\mathbb D}_m$ contains the physically
  interesting correlations (and more),
\item it admits a closed set of functional equations,
\item it has controlled analyticity properties (which are not obvious from the
  definition).
\end{itemize}
Before turning to the proofs we like to mention the normalization of the
  generalized density operator. For this we take the action of ${\mathbb
    D}_m(\lambda_1,\dots,\lambda_m)$ on $m$ completely antisymmetric states
  (the $SU(n)$ singlets) in each bunch of $n$ basis states. We like to mention
  the useful reduction property of ${\mathbb D}_m$ applied to just $k$
  antisymmetric states resulting into a density matrix ${\mathbb D}_{m-k}$,
  see Appendix A.

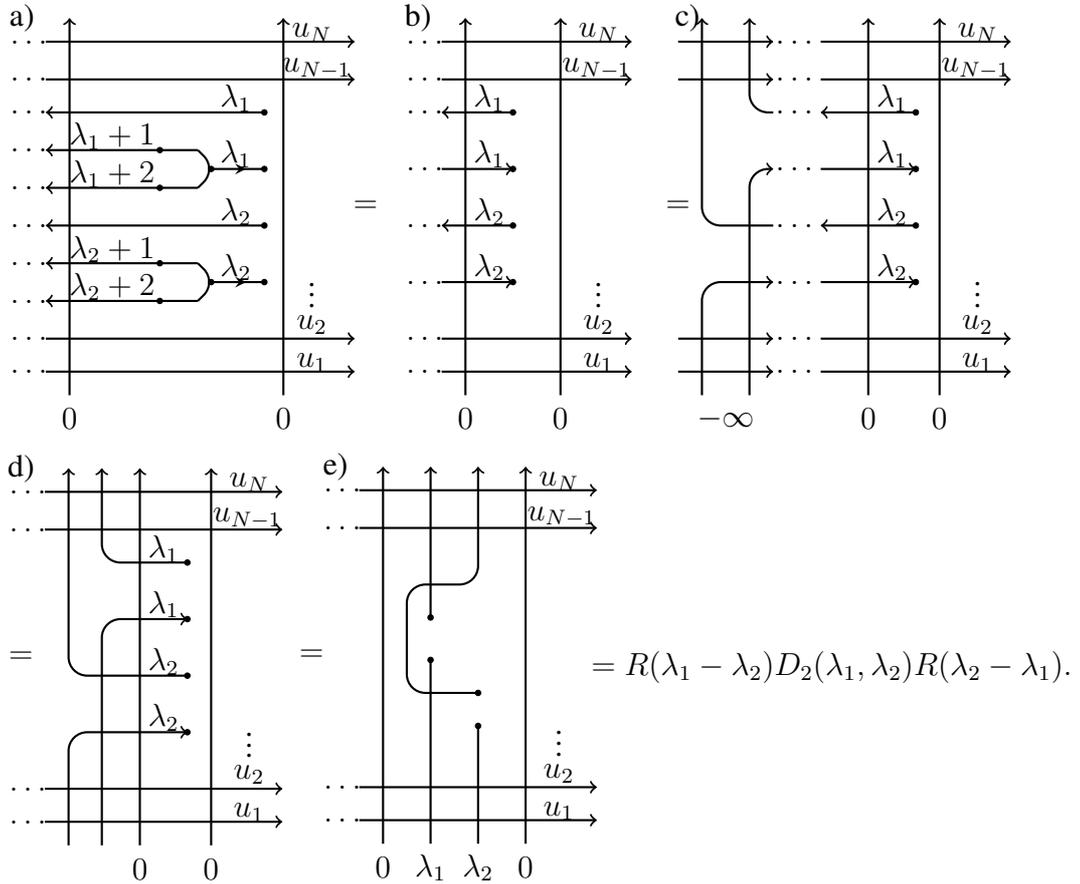
\begin{figure}[h]
	\begin{center}
\begin{minipage}{0.33\linewidth}
		\begin{tikzpicture}[scale=1.25]
\draw (-0.25,4.) node {a)};
		\draw (0,0) [->,color=black, thick, rounded corners=7pt] +(0.25,0)--(0.25,4);
		
		\draw (0,0) [<-,color=black, thick, rounded corners=7pt] +(0,1.0)--(1.6,1.0);
		\draw (0,0) [<-,color=black, thick, rounded corners=7pt] +(0,1.4)--(1.6,1.4);
		\draw (0,0) [<-,color=black, thick, rounded corners=7pt] +(0,1.8)--(2.3,1.8);
		
		\draw (0,0) [<-,color=black, thick, rounded corners=7pt] +(0,2.2)--(1.6,2.2);
		\draw (0,0) [<-,color=black, thick, rounded corners=7pt] +(0,2.6)--(1.6,2.6);
		\draw (0,0) [<-,color=black, thick, rounded corners=7pt] +(0,3.0)--(2.3,3.0);
		
		\draw (0,0) [-,color=black, directed, thick, rounded corners=7pt] +(1.75,1.2)--(2.3,1.2);
		\draw (0,0) [-,color=black, thick, rounded corners=7pt] +(1.6,1.0) -- +(1.8,1.2) -- +(1.6,1.4);
		\draw (1.74,1.2)[fill=black]  circle (0.15ex);
		\draw (2.3,1.2)[fill=black]  circle (0.15ex);
		\draw (2,1.36) node {$\lambda_2$};
		
		\draw (0,0) [-,color=black, directed, thick, rounded corners=7pt] +(1.75,2.4)--(2.3,2.4);
		\draw (0,0) [-,color=black, thick, rounded corners=7pt] +(1.6,2.2) -- +(1.8,2.4) -- +(1.6,2.6);
		\draw (1.74,2.4)[fill=black]  circle (0.15ex);
		\draw (2.3,2.4)[fill=black]  circle (0.15ex);
		\draw (2,2.56) node {$\lambda_1$};

		\draw (-0.15,0.25) node {$\dots$};
		\draw (-0.15,0.6) node {$\dots$};
		\draw (-0.15,3.35) node {$\dots$};
		\draw (-0.15,3.75) node {$\dots$};
		\draw (-0.15,1) node {$\dots$};
		\draw (-0.15,1.4) node {$\dots$};
		\draw (-0.15,1.8) node {$\dots$};
		\draw (-0.15,2.2) node {$\dots$};
		\draw (-0.15,2.6) node {$\dots$};
		\draw (-0.15,3.0) node {$\dots$};

		\draw (0,0) [->,color=black, thick, rounded corners=7pt] +(2.5,0)--(2.5,4);
		
		\draw (0.7,1.15) node {$\lambda_2+2$};
		\draw (0.7,1.55) node {$\lambda_2+1$};
		\draw (2,1.95) node {$\lambda_2$};

		\draw (0.7,2.35) node {$\lambda_1+2$};
		\draw (0.7,2.75) node {$\lambda_1+1$};
		\draw (2.,3.15) node {$\lambda_1$};		
		
		\draw (0.25,-0.25) node {$0$};
		\draw (2.5,-0.25) node {$0$};
				
		\draw (-0.25,0) [->,color=black, thick, rounded corners=7pt] +(0.25,0.25)--(3.25,0.25);
		\draw (-0.25,0) [->,color=black, thick, rounded corners=7pt] +(0.25,0.6)--(3.25,0.6);
		
		\draw (-0.25,0) [->,color=black, thick, rounded corners=7pt] +(0.25,3.35)--(3.25,3.35);
		\draw (-0.25,0) [->,color=black, thick, rounded corners=7pt] +(0.25,3.75)--(3.25,3.75);
		
		\draw (2.8,0.35) node {$u_1$};
		\draw (2.8,0.75) node {$u_2$};
		\draw (2.8,1.15) node {$\vdots$};
		\draw (2.8,3.85) node {$u_{N}$};
		\draw (2.85,3.47) node {$u_{N-1}$};
		\draw (1.2,1)[fill=black]  circle (0.15ex);
		\draw (1.2,1.4)[fill=black]  circle (0.15ex);
		\draw (2.3,1.8)[fill=black]  circle (0.15ex);
		
		\draw (1.2,2.2)[fill=black]  circle (0.15ex);
		\draw (1.2,2.6)[fill=black]  circle (0.15ex);
		\draw (2.3,3)[fill=black]  circle (0.15ex);
				
		\end{tikzpicture}
\end{minipage}%
\begin{minipage}{0.3\linewidth}
		\begin{tikzpicture}[scale=1.25]
\draw (-0.25,4.) node {b)};
		\draw (0,0) [->,color=black, thick, rounded corners=7pt] +(0.25,0)--(0.25,4);

		\draw (0,0) [<-,color=black, thick, rounded corners=7pt] +(0,1.8)--(0.75,1.8);

		\draw (0,0) [<-,color=black, thick, rounded corners=7pt] +(0,3.0)--(0.75,3.0);
		
		\draw (0,0) [->,color=black,  thick, rounded corners=7pt] +(0,1.2)--(0.75,1.2);

		\draw (-0.15,0.25) node {$\dots$};
		\draw (-0.15,0.6) node {$\dots$};
		\draw (-0.15,3.35) node {$\dots$};
		\draw (-0.15,3.75) node {$\dots$};

		\draw (-0.15,1.2) node {$\dots$};
		\draw (-0.15,1.8) node {$\dots$};
		\draw (-0.15,2.4) node {$\dots$};
		\draw (-0.15,3.0) node {$\dots$};

		\draw (0.75,1.2)[fill=black]  circle (0.15ex);
		\draw (0.5,1.36) node {$\lambda_2$};
		
		\draw (0,0) [->,color=black,  thick, rounded corners=7pt] +(0,2.4)--(0.75,2.4);

		\draw (0.75,2.4)[fill=black]  circle (0.15ex);
		\draw (0.5,2.56) node {$\lambda_1$};
		
		\draw (0,0) [->,color=black, thick, rounded corners=7pt] +(1.25,0)--(1.25,4);
		
		\draw (0.5,1.95) node {$\lambda_2$};

		\draw (0.5,3.15) node {$\lambda_1$};		
		
		\draw (0.25,-0.25) node {$0$};
		\draw (1.25,-0.25) node {$0$};

		\draw (-0.25,0) [->,color=black, thick, rounded corners=7pt] +(0.25,0.25)--(2,0.25);
		\draw (-0.25,0) [->,color=black, thick, rounded corners=7pt] +(0.25,0.6)--(2,0.6);
		
		\draw (-0.25,0) [->,color=black, thick, rounded corners=7pt] +(0.25,3.35)--(2,3.35);
		\draw (-0.25,0) [->,color=black, thick, rounded corners=7pt] +(0.25,3.75)--(2,3.75);
				
		\draw (1.65,0.35) node {$u_1$};
		\draw (1.65,0.75) node {$u_2$};
		\draw (1.65,1.15) node {$\vdots$};
		\draw (1.65,3.85) node {$u_{N}$};
		\draw (1.63,3.47) node {$u_{N-1}$};
		\draw (0.75,1.8)[fill=black]  circle (0.15ex);
		
		\draw (0.75,3)[fill=black]  circle (0.15ex);
		
		\draw (-0.8,2) node {$=$};
		
		\end{tikzpicture}
\end{minipage}%
\begin{minipage}{0.3666\linewidth}
	\begin{tikzpicture}[scale=1.25]
\draw (-0.9,4.) node {c)};
	\draw (0,0) [->,color=black, thick, rounded corners=7pt] +(1.,0)--(1.,4);
	
	\draw (0,0) [<-,color=black, thick, rounded corners=7pt] +(-0.75,4)--(-0.75,1.8)--(0,1.8);
	\draw (0,0) [->,color=black,  thick, rounded corners=7pt] +(-0.75,0)--(-0.75,1.2)--(0,1.2);
	
	\draw (0,0) [<-,color=black, thick, rounded corners=7pt] +(-0.25,4)--(-0.25,3.0)--(0,3.0);
	\draw (0,0) [->,color=black,  thick, rounded corners=7pt] +(-0.25,0)--(-0.25,2.4)--(0,2.4);

	\draw (0,0) [<-,color=black, thick, rounded corners=7pt] +(0.5,1.8)--(1.5,1.8);
	\draw (0,0) [->,color=black,  thick, rounded corners=7pt] +(0.5,1.2)--(1.5,1.2);
	
	\draw (0,0) [<-,color=black, thick, rounded corners=7pt] +(0.5,3.0)--(1.5,3.0);
	\draw (0,0) [->,color=black,  thick, rounded corners=7pt] +(0.5,2.4)--(1.5,2.4);

	\draw (0.25,0.25) node {$\dots$};
	\draw (0.25,0.6) node {$\dots$};

	\draw (0.25,1.2) node {$\dots$};
	\draw (0.25,1.8) node {$\dots$};
	\draw (0.25,2.4) node {$\dots$};
	\draw (0.25,3.0) node {$\dots$};

	\draw (0.25,3.35) node {$\dots$};
	\draw (0.25,3.75) node {$\dots$};

	\draw (1.5,1.2)[fill=black]  circle (0.15ex);
	\draw (1.5,1.8)[fill=black]  circle (0.15ex);	
	\draw (1.5,2.4)[fill=black]  circle (0.15ex);
	\draw (1.5,3)[fill=black]  circle (0.15ex);

	\draw (1.25,1.36) node {$\lambda_2$};
	\draw (1.25,2.56) node {$\lambda_1$};
	\draw (1.25,1.95) node {$\lambda_2$};
	\draw (1.25,3.15) node {$\lambda_1$};		
	
	\draw (0,0) [->,color=black, thick, rounded corners=7pt] +(1.75,0)--(1.75,4);

	\draw (1.0,-0.25) node {$0$};
	\draw (1.75,-0.25) node {$0$};
	\draw (-0.5,-0.25) node {$-\infty$};

	\draw (0.25,0) [->,color=black, thick, rounded corners=7pt] +(0.25,0.25)--(2.5,0.25);
	\draw (0.25,0) [->,color=black, thick, rounded corners=7pt] +(0.25,0.6)--(2.5,0.6);
	
	\draw (0.25,0) [->,color=black, thick, rounded corners=7pt] +(0.25,3.35)--(2.5,3.35);
	\draw (0.25,0) [->,color=black, thick, rounded corners=7pt] +(0.25,3.75)--(2.5,3.75);

	\draw (0.,0) [->,color=black, thick, rounded corners=7pt] +(-1.,0.25)--(0.,0.25);
	\draw (0.,0) [->,color=black, thick, rounded corners=7pt] +(-1,0.6)--(0,0.6);
	
	\draw (0.,0) [->,color=black, thick, rounded corners=7pt] +(-1.,3.35)--(0,3.35);
	\draw (0.,0) [->,color=black, thick, rounded corners=7pt] +(-1.,3.75)--(0,3.75);

	\draw (2.15,0.35) node {$u_1$};
	\draw (2.15,0.75) node {$u_2$};
	\draw (2.15,1.15) node {$\vdots$};
	\draw (2.15,3.85) node {$u_{N}$};
	\draw (2.13,3.47) node {$u_{N-1}$};
	
	\draw (-1.,2) node {$=$};
	
	\end{tikzpicture}
\end{minipage}
\begin{minipage}{0.28\linewidth}
	\begin{tikzpicture}[scale=1.25]
\draw (-0.25,4.) node {d)};
	\draw (0,0) [->,color=black, thick, rounded corners=7pt] +(1.,0)--(1.,4);
	
	\draw (0,0) [<-,color=black, thick, rounded corners=7pt] +(0.25,4)--(0.25,1.8)--(1.5,1.8);
	\draw (0,0) [->,color=black,  thick, rounded corners=7pt] +(0.25,0)--(0.25,1.2)--(1.5,1.2);
	
	\draw (0,0) [<-,color=black, thick, rounded corners=7pt] +(0.6,4)--(0.6,3.0)--(1.5,3.0);
	\draw (0,0) [->,color=black,  thick, rounded corners=7pt] +(0.6,0)--(0.6,2.4)--(1.5,2.4);

	\draw (-0.15,0.25) node {$\dots$};
	\draw (-0.15,0.6) node {$\dots$};
	\draw (-0.15,3.35) node {$\dots$};
	\draw (-0.15,3.75) node {$\dots$};

	\draw (1.5,1.2)[fill=black]  circle (0.15ex);
	\draw (1.5,1.8)[fill=black]  circle (0.15ex);	
	\draw (1.5,2.4)[fill=black]  circle (0.15ex);
	\draw (1.5,3)[fill=black]  circle (0.15ex);
	
	\draw (1.25,1.36) node {$\lambda_2$};
	\draw (1.25,2.56) node {$\lambda_1$};
	\draw (1.25,1.95) node {$\lambda_2$};
	\draw (1.25,3.15) node {$\lambda_1$};		
	
	\draw (0,0) [->,color=black, thick, rounded corners=7pt] +(1.75,0)--(1.75,4);

	\draw (1.0,-0.25) node {$0$};
	\draw (1.75,-0.25) node {$0$};

	\draw (-0.25,0) [->,color=black, thick, rounded corners=7pt] +(0.25,0.25)--(2.5,0.25);
	\draw (-0.25,0) [->,color=black, thick, rounded corners=7pt] +(0.25,0.6)--(2.5,0.6);
	
	\draw (-0.25,0) [->,color=black, thick, rounded corners=7pt] +(0.25,3.35)--(2.5,3.35);
	\draw (-0.25,0) [->,color=black, thick, rounded corners=7pt] +(0.25,3.75)--(2.5,3.75);
	
	\draw (2.15,0.35) node {$u_1$};
	\draw (2.15,0.75) node {$u_2$};
	\draw (2.15,1.15) node {$\vdots$};
	\draw (2.15,3.85) node {$u_{N}$};
	\draw (2.13,3.47) node {$u_{N-1}$};
	
	\draw (-0.25,2) node {$=$};
	
	\end{tikzpicture}
\end{minipage}%
\begin{minipage}{0.28\linewidth}
	\begin{tikzpicture}[scale=1.25]
\draw (-0.25,4.) node {e)};
	\draw (0,0) [->,color=black, thick, rounded corners=7pt] +(0.25,0)--(0.25,4);
	
	\draw (0,0) [->,color=black, thick, rounded corners=7pt] +(0.75,2.4)--(0.75,4);
	\draw (0,0) [-,color=black, thick, rounded corners=7pt] +(0.75,0.0)--(0.75,1.95);
	\draw (0.75,2.4)[fill=black]  circle (0.15ex);
	\draw (0.75,1.95)[fill=black]  circle (0.15ex);
	
		\draw (-0.15,0.25) node {$\dots$};
		\draw (-0.15,0.6) node {$\dots$};
		\draw (-0.15,3.35) node {$\dots$};
		\draw (-0.15,3.75) node {$\dots$};

	\draw (0,0) [-,color=black,  thick, rounded corners=7pt] +(1.25,0)--(1.25,1.25);
	\draw (0,0) [->,color=black,  thick, rounded corners=7pt] +(1.25,1.6)--(0.5,1.6)--(0.5,2.75)-- (1.25,2.75)--(1.25,4);
	\draw (1.25,1.25)[fill=black]  circle (0.15ex);
	\draw (1.25,1.6)[fill=black]  circle (0.15ex);

	\draw (0.75,-0.25) node {$\lambda_1$};
	\draw (1.25,-0.25) node {$\lambda_2$};

	\draw (0.25,-0.25) node {$0$};
	\draw (1.75,-0.25) node {$0$};
	
	\draw (0,0) [->,color=black, thick, rounded corners=7pt] +(1.75,0)--(1.75,4);
	
	\draw (-0.25,0) [->,color=black, thick, rounded corners=7pt] +(0.25,0.25)--(2.5,0.25);
	\draw (-0.25,0) [->,color=black, thick, rounded corners=7pt] +(0.25,0.6)--(2.5,0.6);
	
	\draw (-0.25,0) [->,color=black, thick, rounded corners=7pt] +(0.25,3.35)--(2.5,3.35);
	\draw (-0.25,0) [->,color=black, thick, rounded corners=7pt] +(0.25,3.75)--(2.5,3.75);
	
	\draw (2.1,0.35) node {$u_1$};
	\draw (2.1,0.75) node {$u_2$};
	\draw (2.1,1.15) node {$\vdots$};
	\draw (2.1,3.85) node {$u_{N}$};
	\draw (2.13,3.47) node {$u_{N-1}$};
	
	\draw (-0.5,2) node {$=$};
	
	\end{tikzpicture}
\end{minipage}%
\begin{minipage}{0.44\linewidth}
	\begin{tikzpicture}[scale=0.9]

\draw (-1.75,2.25) node {$= R(\lambda_1-\lambda_2) D_2(\lambda_1,\lambda_2) R(\lambda_2-\lambda_1).$};
	
	\end{tikzpicture}
\end{minipage}

\caption{Graphical illustration of the reduction of the generalized density
  operator ${\mathbb D}_2(\lambda_1,\lambda_2)$ to the density operator
  $D_2(\lambda_1,\lambda_2)$ for $SU(3)$. 
}
\label{DD3-D3}
\end{center}
\end{figure}

Next we turn to the embedding of $D_m$ in ${\mathbb D}_m$. For this we apply
anti-symmetrizations to the lower (upper) $n-1$ lines of a
bunch of $n$ lines in ${\mathbb D}_m$ resulting in a line carrying the conjugate
representation. For the $SU(3)$ case we have:
\bear
\begin{minipage}{0.5\linewidth}
\begin{center}
	\begin{tikzpicture}[scale=1.25]
	\draw (0,0) [->,color=black, thick, rounded corners=7pt] +(0.25,0.5)--(0.25,2.5);
	\draw (0,0) [->,color=black, thick, rounded corners=7pt] +(0.5,0.5)--(0.5,2.5);

		\draw (0,0) [<-,color=black, thick, rounded corners=7pt] +(0,1.0)--(1.6,1.0);
		\draw (0,0) [<-,color=black, thick, rounded corners=7pt] +(0,1.4)--(1.6,1.4);
		\draw (0,0) [<-,color=black, thick, rounded corners=7pt] +(0,1.8)--(2.3,1.8);

		\draw (0,0) [-,color=black, directed, thick, rounded corners=7pt] +(1.75,1.2)--(2.3,1.2);
		\draw (0,0) [-,color=black, thick, rounded corners=7pt] +(1.6,1.0) -- +(1.8,1.2) -- +(1.6,1.4);
		\draw (2.3,1.8)[fill=black]  circle (0.15ex);
		\draw (2.3,1.2)[fill=black]  circle (0.15ex);

		\draw (1.15,1.2) node {$\lambda+2$};
		\draw (1.15,1.6) node {$\lambda+1$};
		\draw (0.85,2.) node {$\lambda$};		
    	\draw (3,1.5) node {$=$};
	\end{tikzpicture}
\end{center}
\end{minipage}%
		\begin{minipage}{0.5\linewidth}
\begin{center}
		\begin{tikzpicture}[scale=1.25]
		\draw (0,0) [->,color=black, thick, rounded corners=7pt] +(0.25,0.5)--(0.25,2.5);
		\draw (0,0) [->,color=black, thick, rounded corners=7pt] +(0.5,0.5)--(0.5,2.5);
 
 		\draw (0,0) [<-,color=black, thick, rounded corners=7pt] +(0,1.8)--(2.3,1.8);
		
		\draw (0,0) [->,color=black, thick, rounded corners=7pt] +(0,1.2)--(2.3,1.2);
		\draw (2.3,1.8)[fill=black]  circle (0.15ex);
		\draw (2.3,1.2)[fill=black]  circle (0.15ex);
		
		\draw (1.15,1.45) node {$\lambda$};
		\draw (1.15,2.) node {$\lambda$};		
	\draw (2.75,1.5) node {$,$};
	\end{tikzpicture}
\end{center}
\end{minipage}
\label{fus1}\\		
\begin{minipage}{0.5\linewidth}
\begin{center}
	\begin{tikzpicture}[scale=1.25]
	\draw (0,0) [->,color=black, thick, rounded corners=7pt] +(0.25,0.5)--(0.25,2.5);
	\draw (0,0) [->,color=black, thick, rounded corners=7pt] +(0.5,0.5)--(0.5,2.5);
	
	\draw (0,0) [<-,color=black, thick, rounded corners=7pt] +(0,1.0)--(2.3,1.0);
	\draw (0,0) [<-,color=black, thick, rounded corners=7pt] +(0,1.4)--(1.6,1.4);
	\draw (0,0) [<-,color=black, thick, rounded corners=7pt] +(0,1.8)--(1.6,1.8);
	
	\draw (0,0) [-,color=black, directed, thick, rounded corners=7pt] +(1.75,1.6)--(2.3,1.6);
	\draw (0,0) [-,color=black, thick, rounded corners=7pt] +(1.6,1.4) -- +(1.8,1.6) -- +(1.6,1.8);
	\draw (2.3,1.6)[fill=black]  circle (0.15ex);
	\draw (2.3,1.0)[fill=black]  circle (0.15ex);
	
	\draw (1.15,1.2) node {$\lambda+2$};
	\draw (1.15,1.6) node {$\lambda+1$};
	\draw (0.85,2.) node {$\lambda$};		
	\draw (3,1.5) node {$=$};
	\end{tikzpicture}
\end{center}
\end{minipage}%
\begin{minipage}{0.5\linewidth}
\begin{center}
	\begin{tikzpicture}[scale=1.25]
	\draw (0,0) [->,color=black, thick, rounded corners=7pt] +(0.25,0.5)--(0.25,2.5);
	\draw (0,0) [->,color=black, thick, rounded corners=7pt] +(0.5,0.5)--(0.5,2.5);
	
	\draw (0,0) [->,color=black, thick, rounded corners=7pt] +(0,1.8)--(2.3,1.8);
	
	\draw (0,0) [<-,color=black, thick, rounded corners=7pt] +(0,1.2)--(2.3,1.2);
	\draw (2.3,1.8)[fill=black]  circle (0.15ex);
	\draw (2.3,1.2)[fill=black]  circle (0.15ex);

	\draw (1.15,1.45) node {$\lambda+2$};
	\draw (1.15,2.) node {$\lambda-1$};		
	\draw (2.75,1.5) node {$.$};
	\end{tikzpicture}
\end{center}
\end{minipage}
\label{fus2}
\ear 
Let us consider the anti-symmetrization of the lower $n-1$ lines.
{In Figure \ref{DD3-D3} we depict the simplest case of two-point
  ($m=2$) correlations for $SU(3)$. First, the antisymmetrizers are carried to
  the very left by virtue of (\ref{fus1}), see Figure \ref{DD3-D3} a) and b).}
We modify the lattice at the far left by bending the upper horizontal line
with arrow pointing to the left upwards and bending the lower horizontal line
with arrow pointing to the right downwards, finally connecting the two ends
{(carrying the same spectral parameter)} by exploiting the
periodic boundary condition in vertical direction, Figure \ref{DD3-D3}
c). This manipulation of the far left boundary may introduce a factor which
however is independent of the spins on the open bonds inside the
lattice. Finally, we use the Yang-Baxter equation and unitarity to move the
closed loops at the far left as simple vertical lines to the center of the
lattice, {Figure \ref{DD3-D3} d) and e). The resulting object is
  equal to the density operator ${D}_2(\lambda_1,\lambda_2)$ under the action
  of two $R$-matrices.}

Quite generally, by the above described procedure of anti-symmetrization of the
lower $n-1$ lines of each bunch of $n$ lines we have the reduction of the
operator ${\mathbb D}_m(\lambda_1,\dots,\lambda_m)$ to the usual density
operator $D_m(\lambda_1,\dots,\lambda_m)$ with additional action of $m(m-1)$
$R$-matrices.

Just as a short remark we want to point out that the procedure of
anti-symme\-tri\-zation of the {\em upper} $n-1$ lines of a bunch of $n$ lines and
subsequent use of Yang-Baxter and (special) unitarity leads to a vertical line
with conjugate representation of 
$SU(n)$
and carrying the spectral parameter
$\lambda+n-1$.

It is worth reminding that the physically interesting object we want to compute
is precisely the full density operator $D_m$. However, in order to formulate
consistent functional equations we have to work in the more general setting of
${\mathbb D}_m$ and at the end of the calculation to project onto the
physically relevant subspace. The derivation of functional equations and
analyticity properties for ${\mathbb D}_m$ is the subject of the next section.

\section{Discrete functional equations and analyticity}\label{functional}

\begin{figure}
	\begin{center}
		\begin{minipage}{0.5\linewidth}
			\begin{tikzpicture}[scale=1.15]
\draw (-1.25,4.) node {a)};

\draw (0,0) [->,color=black, thick, rounded corners=7pt] +(-1,0.4)--(-1,4);

\draw (0,0) [-,color=black, thick, rounded corners=9pt] +(-1.25,3)--(-1.5,3.2)--(-1.25,3.35);

\draw (0,0) [<-,color=black, thick, rounded corners=7pt] +(-1.25,1.0)--(-0.75,1.0);
\draw (0,0) [<-,color=black, thick, rounded corners=7pt] +(-1.25,1.4)--(-0.75,1.4);
\draw (0,0) [<-,color=black, thick, rounded corners=7pt] +(-1.25,1.8)--(-0.75,1.8);
			
\draw (0,0) [<-,color=black, thick, rounded corners=7pt] +(-1.25,2.2)--(-0.75,2.2);
\draw (0,0) [<-,color=black, thick, rounded corners=7pt] +(-1.25,2.6)--(-0.75,2.6);
\draw (0,0) [<-,color=black, thick, rounded corners=7pt] +(-1.25,3.0)--(-0.75,3.0);

\draw (0,0) [->,color=black, thick, rounded corners=7pt] +(-1.25,0.6)--(-0.75,0.6);
		
\draw (0,0) [->,color=black, thick, rounded corners=7pt] +(-1.25,3.35)--(-0.75,3.35);
\draw (0,0) [->,color=black, thick, rounded corners=7pt] +(-1.25,3.75)--(-0.75,3.75);

\draw (-1.1,0.25) node {$-\infty$};

\draw (-0.5,0.6) node {$\dots$};
\draw (-0.5,3.35) node {$\dots$};
\draw (-0.5,3.75) node {$\dots$};
\draw (-0.5,2.2) node {$\dots$};
\draw (-0.5,2.6) node {$\dots$};
\draw (-0.5,3.0) node {$\dots$};
\draw (-0.5,1.0) node {$\dots$};
\draw (-0.5,1.4) node {$\dots$};
\draw (-0.5,1.8) node {$\dots$};

			\draw (0,0) [->,color=black, thick, rounded corners=7pt] +(0.25,0.4)--(0.25,4);
			
			\draw (0,0) [<-,color=black, thick, rounded corners=7pt] +(-0.25,1.0)--(0.5,1.0);
			\draw (0,0) [<-,color=black, thick, rounded corners=7pt] +(-0.25,1.4)--(0.5,1.4);
			\draw (0,0) [<-,color=black, thick, rounded corners=7pt] +(-0.25,1.8)--(0.5,1.8);
			
			\draw (0,0) [<-,color=black, thick, rounded corners=7pt] +(-0.25,2.2)--(0.5,2.2);
			\draw (0,0) [<-,color=black, thick, rounded corners=7pt] +(-0.25,2.6)--(0.5,2.6);
			\draw (0,0) [<-,color=black, thick, rounded corners=7pt] +(-0.25,3.0)--(0.5,3.0);

			\draw (0,0) [->,color=black, thick, rounded corners=7pt] +(2.1,0.4)--(2.1,4);
			
			\draw (1.,1.0) node {$\lambda_2+2$};
			\draw (1.,1.4) node {$\lambda_2+1$};
			\draw (0.75,1.8) node {$\lambda_2$};

			\draw (1.,2.2) node {$\lambda_1+2$};
			\draw (1.,2.6) node {$\lambda_1+1$};
			\draw (0.75,3.0) node {$\lambda_1$};
				
			\draw (-0.25,0) [->,color=black, thick, rounded corners=7pt] +(0,0.6)--(2.75,0.6);
			
			\draw (-0.25,0) [->,color=black, thick, rounded corners=7pt] +(0,3.35)--(2.75,3.35);
			\draw (-0.25,0) [->,color=black, thick, rounded corners=7pt] +(0,3.75)--(2.75,3.75);
						
			\draw (2.4,0.45) node {$u_1$};
			\draw (2.4,3.6) node {$u_{N}$};
			\draw (2.4,3.2) node {$u_{i}$};
			
			\draw (0.5,1)[fill=black]  circle (0.15ex);
			\draw (0.5,1.4)[fill=black]  circle (0.15ex);
			\draw (0.5,1.8)[fill=black]  circle (0.15ex);
			
			\draw (0.5,2.2)[fill=black]  circle (0.15ex);
			\draw (0.5,2.6)[fill=black]  circle (0.15ex);
			\draw (0.5,3)[fill=black]  circle (0.15ex);

			\draw (3.54,2.2) node {$\stackrel{\lambda_1=u_i}{=}$};
			
			\end{tikzpicture}
		\end{minipage}%
		\begin{minipage}{0.5\linewidth}
			\begin{tikzpicture}[scale=1.15]
\draw (-0.25,4.) node {b)};
			
			\draw (0,0) [->,color=black, thick, rounded corners=7pt] +(0.25,0.4)--(0.25,4);
			
			\draw (0,0) [<-,color=black, thick, rounded corners=7pt] +(-0.25,1.0)--(0.5,1.0);
			\draw (0,0) [<-,color=black, thick, rounded corners=7pt] +(-0.25,1.4)--(0.5,1.4);
			\draw (0,0) [<-,color=black, thick, rounded corners=7pt] +(-0.25,1.8)--(0.5,1.8);
			
			\draw (0,0) [<-,color=black, thick, rounded corners=7pt] +(-0.25,2.2)--(0.5,2.2);
			\draw (0,0) [<-,color=black, thick, rounded corners=7pt] +(-0.25,2.6)--(0.5,2.6);
			\draw (0,0) [<-,color=black, thick, rounded corners=7pt] +(-0.25,3.0)--(0.5,3.0);

			\draw (0,0) [->,color=black, thick, rounded corners=7pt] +(2.1,0.4)--(2.1,4);
			
			\draw (1.,1.0) node {$\lambda_2+2$};
			\draw (1.,1.4) node {$\lambda_2+1$};
			\draw (0.75,1.8) node {$\lambda_2$};

			\draw (1.,2.2) node {$\lambda_1+2$};
			\draw (1.,2.6) node {$\lambda_1+1$};
			\draw (0.75,3.0) node {$\lambda_1$};
						
			\draw (-0.25,0) [->,color=black, thick, rounded corners=7pt] +(0,0.6)--(2.75,0.6);
			
			\draw (-0.25,0) [->,color=black, thick, rounded corners=7pt] +(0,3.35)--(2.75,3.35);
			\draw (-0.25,0) [->,color=black, thick, rounded corners=7pt] +(0,3.75)--(2.75,3.75);
			
			\draw (2.4,0.45) node {$u_1$};
			\draw (2.4,3.6) node {$u_{N}$};
			\draw (2.4,3.2) node {$u_{i}$};
			
			\draw (0.5,1)[fill=black]  circle (0.15ex);
			\draw (0.5,1.4)[fill=black]  circle (0.15ex);
			\draw (0.5,1.8)[fill=black]  circle (0.15ex);
			
			\draw (0.5,2.2)[fill=black]  circle (0.15ex);
			\draw (0.5,2.6)[fill=black]  circle (0.15ex);
			\draw (0.5,3)[fill=black]  circle (0.15ex);
			
			\draw (0,0) [-,color=black, thick, rounded corners=9pt] +(-0.25,3)--(-0.42,3.2)--(-0.25,3.35);
						
			\end{tikzpicture}
		\end{minipage}
		\begin{minipage}{0.5\linewidth}
			\begin{tikzpicture}[scale=1.15]
\draw (-0.5,3.75) node {c)};
			
			\draw (0,0) [->,color=black, thick, rounded corners=7pt] +(0.25,0.4)--(0.25,3.8);
			
			\draw (0,0) [<-,color=black, thick, rounded corners=7pt] +(-0.25,1.0)--(0.5,1.0);
			\draw (0,0) [<-,color=black, thick, rounded corners=7pt] +(-0.25,1.4)--(0.5,1.4);
			\draw (0,0) [<-,color=black, thick, rounded corners=7pt] +(-0.25,1.8)--(0.5,1.8);
			
			\draw (0,0) [<-,color=black, thick, rounded corners=7pt] +(-0.25,2.2)--(0.5,2.2);
			\draw (0,0) [<-,color=black, thick, rounded corners=7pt] +(-0.25,2.6)--(0.5,2.6);
			\draw (0,0) [->,color=black, thick, rounded corners=7pt] +(0.5,3.0)--(2.75,3.0);

			\draw (0,0) [->,color=black, thick, rounded corners=7pt] +(2.1,0.4)--(2.1,3.8);
			
			\draw (1.,1.0) node {$\lambda_2+2$};
			\draw (1.,1.4) node {$\lambda_2+1$};
			\draw (0.75,1.8) node {$\lambda_2$};
			
			\draw (1.,2.2) node {$\lambda_1+2$};
			\draw (1.,2.6) node {$\lambda_1+1$};
			\draw (0.75,3.2) node {$\lambda_1$};
			
			\draw (-0.25,0) [->,color=black, thick, rounded corners=7pt] +(0,0.6)--(2.75,0.6);
			
			\draw (-0.25,0) [->,color=black, thick, rounded corners=7pt] +(0,3.45)--(2.75,3.45);

			\draw (2.4,0.45) node {$u_1$};
			\draw (2.4,3.6) node {$u_{N}$};
			
			\draw (0.5,1)[fill=black]  circle (0.15ex);
			\draw (0.5,1.4)[fill=black]  circle (0.15ex);
			\draw (0.5,1.8)[fill=black]  circle (0.15ex);
			
			\draw (0.5,2.2)[fill=black]  circle (0.15ex);
			\draw (0.5,2.6)[fill=black]  circle (0.15ex);
			\draw (0.5,3)[fill=black]  circle (0.15ex);

			\draw (-0.75,2.2) node {$=$};

			\draw (3.54,2.2) node {$\stackrel{\lambda_1=u_i}{=}$};
			
			\end{tikzpicture}
		\end{minipage}%
		\begin{minipage}{0.5\linewidth}
			\begin{tikzpicture}[scale=1.15]
\draw (-0.,3.75) node {d)};
			
			\draw (0,0) [->,color=black, thick, rounded corners=7pt] +(0.25,0.4)--(0.25,3.8);
			
			\draw (0,0) [<-,color=black, thick, rounded corners=7pt] +(-0.25,1.0)--(0.5,1.0);
			\draw (0,0) [<-,color=black, thick, rounded corners=7pt] +(-0.25,1.4)--(0.5,1.4);
			\draw (0,0) [<-,color=black, thick, rounded corners=7pt] +(-0.25,1.8)--(0.5,1.8);
			
			\draw (0,0) [<-,color=black, thick, rounded corners=7pt] +(-0.25,2.2)--(0.5,2.2);
			\draw (0,0) [<-,color=black, thick, rounded corners=7pt] +(-0.25,2.6)--(0.5,2.6);
			\draw (0,0) [->|,color=black, thick, rounded corners=7pt] +(0.5,3.0)--(2.35,3.0)--(2.35,2);
			\draw (0,0) [->|,color=black, thick, rounded corners=7pt] +(2.35,2)--(2.35,1.35);
			\draw (0,0) [->,color=black, thick, rounded corners=7pt] +(2.35,1.35)--(2.35,0.85)--(2.75,0.85);

			\draw (0,0) [->,color=black, thick, rounded corners=7pt] +(2.1,0.4)--(2.1,3.8);

			\draw (1.,1.0) node {$\lambda_2+2$};
			\draw (1.,1.4) node {$\lambda_2+1$};
			\draw (0.75,1.8) node {$\lambda_2$};
			
			\draw (1.,2.2) node {$\lambda_1+2$};
			\draw (1.,2.6) node {$\lambda_1+1$};
			\draw (2.57,2.35) node {$\lambda_1$};
			\draw (2.85,1.65) node {$\lambda_1+3$};
			\draw (2.57,1.1) node {$\lambda_1$};
			
			\draw (-0.25,0) [->,color=black, thick, rounded corners=7pt] +(0,0.6)--(2.75,0.6);
			
			\draw (-0.25,0) [->,color=black, thick, rounded corners=7pt] +(0,3.45)--(2.75,3.45);

			\draw (2.4,0.45) node {$u_1$};
			\draw (2.4,3.6) node {$u_{N}$};
			
			\draw (0.5,1)[fill=black]  circle (0.15ex);
			\draw (0.5,1.4)[fill=black]  circle (0.15ex);
			\draw (0.5,1.8)[fill=black]  circle (0.15ex);
			
			\draw (0.5,2.2)[fill=black]  circle (0.15ex);
			\draw (0.5,2.6)[fill=black]  circle (0.15ex);
			\draw (0.5,3)[fill=black]  circle (0.15ex);
			
			\end{tikzpicture}
		\end{minipage}
		\begin{minipage}{0.5\linewidth}
			\begin{tikzpicture}[scale=1.15]
\draw (-0.75,3.8) node {e)};
			
			\draw (0,0) [->,color=black, thick, rounded corners=7pt] +(0.25,-0.25)--(0.25,3.8);
			
			\draw (0,0) [<-,color=black, thick, rounded corners=7pt] +(-0.25,1.0)--(0.5,1.0);
			\draw (0,0) [<-,color=black, thick, rounded corners=7pt] +(-0.25,1.4)--(0.5,1.4);
			\draw (0,0) [<-,color=black, thick, rounded corners=7pt] +(-0.25,1.8)--(0.5,1.8);
			
			\draw (0,0) [<-,color=black, thick, rounded corners=7pt] +(-0.25,2.2)--(0.5,2.2);
			\draw (0,0) [<-,color=black, thick, rounded corners=7pt] +(-0.25,2.6)--(0.5,2.6);
			\draw (0,0) [->|,color=black, thick, rounded corners=7pt] +(0.5,3.0)--(2.35,3.0)--(2.35,1.85);
			\draw (0,0) [->|,color=black, thick, rounded corners=7pt] +(2.35,1.85)--(2.35,0.7)--(-0.25,0.7);			
			\draw (0,0) [->,color=black, thick, rounded corners=7pt] +(-0.25,0.35)--(2.75,0.35);

			\draw (0,0) [-,color=black, thick, rounded corners=9pt] +(-0.25,0.35)--(-0.42,0.55)--(-0.25,0.7);

			\draw (0,0) [->,color=black, thick, rounded corners=7pt] +(2.1,-0.25)--(2.1,3.8);
			
			\draw (1.,1.0) node {$\lambda_2+2$};
			\draw (1.,1.4) node {$\lambda_2+1$};
			\draw (0.75,1.8) node {$\lambda_2$};
			
			\draw (1.,2.2) node {$\lambda_1+2$};
			\draw (1.,2.6) node {$\lambda_1+1$};
			\draw (2.55,2.5) node {$\lambda_1$};
			\draw (2.8,1.25) node {$\lambda_1+3$};
			\draw (2.4,0.2) node {$\lambda_1$};

			\draw (-0.25,0) [->,color=black, thick, rounded corners=7pt] +(0,0.)--(2.75,0.);
			
			\draw (-0.25,0) [->,color=black, thick, rounded corners=7pt] +(0,3.65)--(2.75,3.65);
					
			\draw (2.4,-0.15) node {$u_1$};
			\draw (2.4,3.5) node {$u_{N}$};
			
			\draw (0.5,1)[fill=black]  circle (0.15ex);
			\draw (0.5,1.4)[fill=black]  circle (0.15ex);
			\draw (0.5,1.8)[fill=black]  circle (0.15ex);
			
			\draw (0.5,2.2)[fill=black]  circle (0.15ex);
			\draw (0.5,2.6)[fill=black]  circle (0.15ex);
			\draw (0.5,3)[fill=black]  circle (0.15ex);
			
			\draw (3.54,2.2) node {$\stackrel{\lambda_1=u_i}{=}$};
			
			\draw (-0.75,2.1) node {$=$};
			
			\end{tikzpicture}
		\end{minipage}%
		\begin{minipage}{0.5\linewidth}
			\begin{tikzpicture}[scale=1.15]
\draw (-1.65,3.65) node {f)};
			
			\draw (0,0) [->,color=black, thick, rounded corners=7pt] +(0.,-0.25)--(0.,3.8);
			
			\draw (0,0) [<-,color=black, thick, rounded corners=7pt] +(-0.25,1.0)--(0.5,1.0);
			\draw (0,0) [<-,color=black, thick, rounded corners=7pt] +(-0.25,1.4)--(0.5,1.4);
			\draw (0,0) [<-,color=black, thick, rounded corners=7pt] +(-0.25,0.6)--(0.5,0.6);
			
			\draw (0,0) [<-,color=black, thick, rounded corners=7pt] +(-0.25,2.2)--(0.5,2.2);
			\draw (0,0) [<-,color=black, thick, rounded corners=7pt] +(-0.25,2.6)--(0.5,2.6);
			
			\draw (0,0) [->|,color=black, thick, rounded corners=7pt] +(0.5,3.0) -- (1.85,3.0)--(1.85,1.75);
			\draw (0,0) [->,color=black, thick, rounded corners=7pt] +(1.85,1.75)--(1.85,0.4)--(0.25,0.4)--(0.25,1.8)--(-0.25,1.8);

			\draw (-0.5,0.6) node {$\cdots$};
			\draw (-0.5,1.) node {$\cdots$};
			\draw (-0.5,1.4) node {$\cdots$};
			\draw (-0.5,1.8) node {$\cdots$};
			\draw (-0.5,2.2) node {$\cdots$};
			\draw (-0.5,2.6) node {$\cdots$};

			\draw (0,0) [->,color=black, thick, rounded corners=7pt] +(-1.,-0.25)--(-1.,3.8);

			\draw (0,0) [->|,color=black, thick, rounded corners=7pt] +(-0.75,1.8) -- (-1.25,1.8)--(-1.25,0.25)--(-0.75,0.25);

			\draw (0,0) [<-,color=black, thick, rounded corners=7pt] +(-0.75,3.65) -- (-1.5,3.65);
			\draw (0,0) [<-,color=black, thick, rounded corners=7pt] +(-0.75,-0.1) -- (-1.5,-0.1);

			\draw (0,0) [->,color=black, thick, rounded corners=7pt] +(-0.75,2.2) -- (-1.5,2.2);
			\draw (0,0) [->,color=black, thick, rounded corners=7pt] +(-0.75,2.6) -- (-1.5,2.6);
			\draw (0,0) [->,color=black, thick, rounded corners=7pt] +(-0.75,1.4) -- (-1.5,1.4);
			\draw (0,0) [->,color=black, thick, rounded corners=7pt] +(-0.75,1.) -- (-1.5,1.);
			\draw (0,0) [->,color=black, thick, rounded corners=7pt] +(-0.75,0.6) -- (-1.5,0.6);

			\draw (0,0) [->,color=black, thick, rounded corners=7pt] +(2.1,-0.25)--(2.1,3.8);
			
			\draw (1.,0.6) node {$\lambda_2+2$};
			\draw (1.,1.0) node {$\lambda_2+1$};
			\draw (0.75,1.4) node {$\lambda_2$};

			\draw (1.,1.8) node {$\lambda_1+3$};
			\draw (1.,2.2) node {$\lambda_1+2$};
			\draw (1.,2.6) node {$\lambda_1+1$};
			
			\draw (0.8,3.15) node {$\lambda_1$};

			\draw (-0.5,0.25) node {$\cdots$};
			\draw (-0.5,-0.1) node {$\cdots$};
			\draw (-0.5,3.65) node {$\cdots$};
			
			\draw (-0.25,0) [->,color=black, thick, rounded corners=7pt] +(0,0.25)--(2.75,0.25);
			\draw (-0.25,0) [->,color=black, thick, rounded corners=7pt] +(0,-0.1)--(2.75,-0.1);
			\draw (-0.25,0) [->,color=black, thick, rounded corners=7pt] +(0,3.65)--(2.75,3.65);
			
			\draw (2.4,-0.25) node {$u_1$};
			\draw (2.4,0.1) node {$\lambda_1$};
			\draw (2.4,3.5) node {$u_{N}$};
			\draw (-1.15,-0.35) node {$-\infty$};
			
			\draw (0.5,0.6)[fill=black]  circle (0.15ex);
			\draw (0.5,1)[fill=black]  circle (0.15ex);
			\draw (0.5,1.4)[fill=black]  circle (0.15ex);
			
			\draw (0.5,2.2)[fill=black]  circle (0.15ex);
			\draw (0.5,2.6)[fill=black]  circle (0.15ex);
			\draw (0.5,3)[fill=black]  circle (0.15ex);
			
			\end{tikzpicture}
		\end{minipage}
		
				\begin{minipage}{0.5\linewidth}
					\begin{tikzpicture}[scale=1.15]
					\draw (-0.25,3.85) node {g)};
			\draw (-0.75,2.1) node {$=$};
					
					\draw (0,0) [->,color=black, thick, rounded corners=7pt] +(0.,0.)--(0.,3.8);
					
					\draw (0,0) [<-,color=black, thick, rounded corners=7pt] +(-0.25,1.0)--(0.5,1.0);
					\draw (0,0) [<-,color=black, thick, rounded corners=7pt] +(-0.25,1.4)--(0.5,1.4);
					\draw (0,0) [<-,color=black, thick, rounded corners=7pt] +(-0.25,0.6)--(0.5,0.6);
					
					\draw (0,0) [<-,color=black, thick, rounded corners=7pt] +(-0.25,2.2)--(0.5,2.2);
					\draw (0,0) [<-,color=black, thick, rounded corners=7pt] +(-0.25,2.6)--(0.5,2.6);
					
					\draw (0,0) [->|,color=black, thick, rounded corners=7pt] +(0.5,3.0) -- (1.85,3.0)--(1.85,1.75);
					\draw (0,0) [->,color=black, thick, rounded corners=7pt] +(1.85,1.75)--(1.85,0.4)--(0.25,0.4)--(0.25,1.8)--(-0.25,1.8);

					\draw (0,0) [->,color=black, thick, rounded corners=7pt] +(2.1,-0.)--(2.1,3.8);
					
					\draw (1.,0.6) node {$\lambda_2+2$};
					\draw (1.,1.0) node {$\lambda_2+1$};
					\draw (0.75,1.4) node {$\lambda_2$};

					\draw (1.,1.8) node {$\lambda_1+3$};
					\draw (1.,2.2) node {$\lambda_1+2$};
					\draw (1.,2.6) node {$\lambda_1+1$};
					
					\draw (0.8,3.15) node {$\lambda_1$};
					
					\draw (-0.25,0) [->,color=black, thick, rounded corners=7pt] +(0,0.25)--(2.75,0.25);
					
					\draw (-0.25,0) [->,color=black, thick, rounded corners=7pt] +(0,3.35)--(2.75,3.35);
					\draw (-0.25,0) [->,color=black, thick, rounded corners=7pt] +(0,3.65)--(2.75,3.65);
					
					\draw (2.4,0.10) node {$u_1$};
					\draw (2.4,3.5) node {$u_{N}$};
					\draw (2.4,3.2) node {$u_{i}$};
					
					\draw (0.5,0.6)[fill=black]  circle (0.15ex);
					\draw (0.5,1)[fill=black]  circle (0.15ex);
					\draw (0.5,1.4)[fill=black]  circle (0.15ex);
					
					\draw (0.5,2.2)[fill=black]  circle (0.15ex);
					\draw (0.5,2.6)[fill=black]  circle (0.15ex);
					\draw (0.5,3)[fill=black]  circle (0.15ex);
					
					\end{tikzpicture}
				\end{minipage}
\caption{Outline of the derivation of the functional equation for the
  two-sites ($m=2$) generalized density operator for the $SU(3)$ case.}
\label{derivationfunc-eqs}
\end{center}
\end{figure}
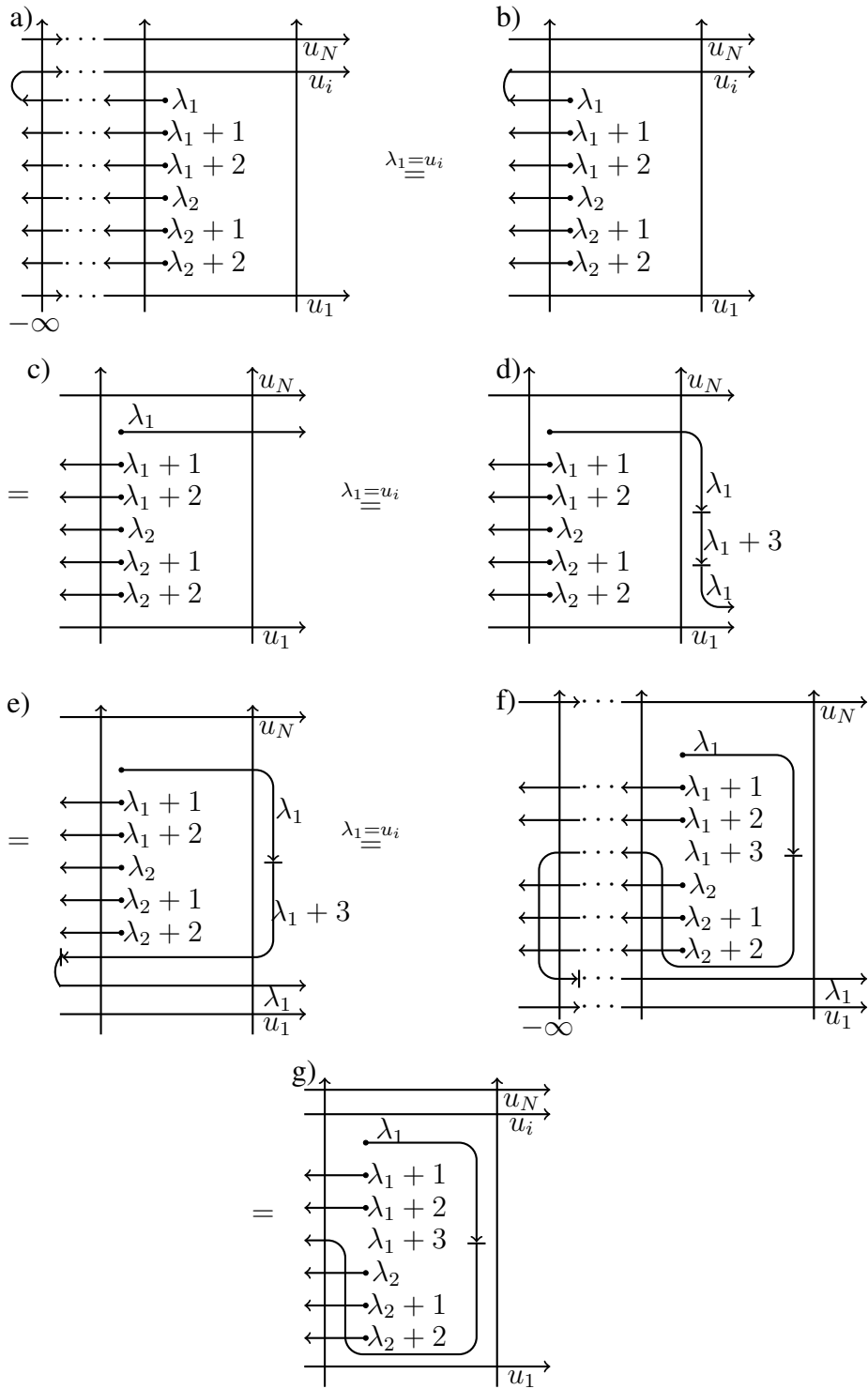

In order to derive closed functional equations for the correlators of the
$SU(n)$ quantum spin chain, we explore the consequences of setting the value
of for instance $\lambda_1$ equal to one of the spectral parameters $u_i$ on
the horizontal lines.
We illustrate a sequence of
manipulations in Figure \ref{derivationfunc-eqs} for the case
$m=2$ of $SU(3)$.

Having $\lambda_1=u_i$ allows us to connect the left going semi-infinite
line carrying $\lambda_1$ with the right-going line carrying $u_i$, Figure
\ref{derivationfunc-eqs}a, and to use the unitarity property
(\ref{unitarity}) for moving the link towards the right, Figure
\ref{derivationfunc-eqs} b) and c). 
Note that operation a) may change the partition function by some factor which
however is independent of the spins on the interior open bonds.
{Next,} we use special unitarity
(\ref{s-unitarity1}) to move the line around and back to the left, Figure
\ref{derivationfunc-eqs} d) and e). 

{Then we use standard unitarity and widen the narrow loop as shown
  in Figure \ref{derivationfunc-eqs} f).  This introduces the action of
  additional $R$-matrices at the open ends in the middle part of the lattice
  and at the far left, Figure \ref{derivationfunc-eqs} f). The boundary part
  at $-\infty$ is dropped, the horizontal line carrying $\lambda_1$ is moved
  by use of periodic boundary condition in vertical direction as $u_i$ line
  upwards. Finally we obtain the original density operator with the spectral
  parameter $\lambda_1$ shifted by $1$ with the action of $R$-matrices upon
  it, Figure \ref{derivationfunc-eqs} g).}

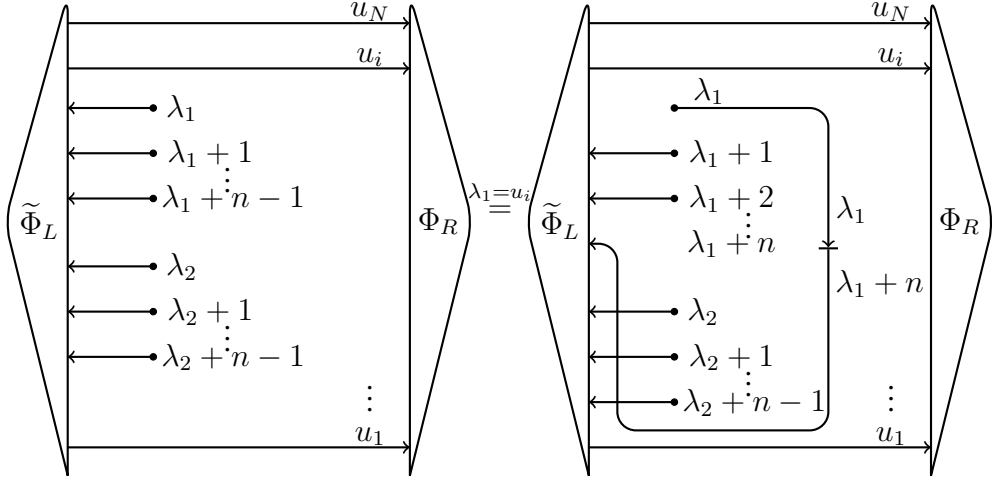
\begin{figure}[h]
	\begin{center}
		\begin{minipage}{0.5\linewidth}
		\begin{tikzpicture}[scale=1.5]
		
		\draw (0,0) [<-,color=black, thick, rounded corners=7pt] +(-0.25,0.8)--(0.5,0.8);
		\draw (0,0) [<-,color=black, thick, rounded corners=7pt] +(-0.25,1.2)--(0.5,1.2);
		\draw (0,0) [<-,color=black, thick, rounded corners=7pt] +(-0.25,1.6)--(0.5,1.6);
		
		\draw (0,0) [<-,color=black, thick, rounded corners=7pt] +(-0.25,2.2)--(0.5,2.2);
		\draw (0,0) [<-,color=black, thick, rounded corners=7pt] +(-0.25,2.6)--(0.5,2.6);
		\draw (0,0) [<-,color=black, thick, rounded corners=7pt] +(-0.25,3.0)--(0.5,3.0);
	
		\draw (1.2,0.8) node {$\lambda_2+n-1$};
		\draw (1.15,1.03) node {$\vdots$};
		\draw (1.,1.2) node {$\lambda_2+1$};
		\draw (0.75,1.6) node {$\lambda_2$};
		
		\draw (1.2,2.2) node {$\lambda_1+n-1$};
		\draw (1.15,2.43) node {$\vdots$};
		\draw (1.,2.6) node {$\lambda_1+1$};
		\draw (0.75,3.0) node {$\lambda_1$};
		
		\draw (-0.25,0) [->,color=black, thick, rounded corners=7pt] +(0,0.)--(2.75,0.);
		
		\draw (-0.25,0) [->,color=black, thick, rounded corners=7pt] +(0,3.35)--(2.75,3.35);
		\draw (-0.25,0) [->,color=black, thick, rounded corners=7pt] +(0,3.75)--(2.75,3.75);
		
		\draw (-0.5,2) node {$\widetilde{\Phi}_L$};
		\draw (0,0) [-,color=black, thick, rounded corners=7pt] +(-0.25,-0.25)--(-0.25,4)--(-0.8,2)--(-0.25,-0.25);

		\draw (3,2) node {$\Phi_R$};
		\draw (0,0) [-,color=black, thick, rounded corners=7pt] +(2.75,-0.25)--(2.75,4)--(3.3,2)--(2.75,-0.25);
		
		\draw (2.4,0.1) node {$u_1$};
		\draw (2.4,0.5) node {$\vdots$};
		\draw (2.4,3.85) node {$u_{N}$};
		\draw (2.4,3.45) node {$u_{i}$};

		\draw (0.5,0.8)[fill=black]  circle (0.15ex);
		\draw (0.5,1.2)[fill=black]  circle (0.15ex);
		\draw (0.5,1.6)[fill=black]  circle (0.15ex);
		
		\draw (0.5,2.2)[fill=black]  circle (0.15ex);
		\draw (0.5,2.6)[fill=black]  circle (0.15ex);
		\draw (0.5,3)[fill=black]  circle (0.15ex);
		
		\draw (3.54,2.2) node {$\stackrel{\lambda_1=u_i}{=}$};
			
\end{tikzpicture}
\end{minipage}%
		\begin{minipage}{0.5\linewidth}
		\begin{tikzpicture}[scale=1.5]
		
		\draw (0,0) [<-,color=black, thick, rounded corners=7pt] +(-0.25,0.8)--(0.5,0.8);
		\draw (0,0) [<-,color=black, thick, rounded corners=7pt] +(-0.25,1.2)--(0.5,1.2);
		\draw (0,0) [<-,color=black, thick, rounded corners=7pt] +(-0.25,0.4)--(0.5,0.4);
		
		\draw (0,0) [<-,color=black, thick, rounded corners=7pt] +(-0.25,2.2)--(0.5,2.2);
		\draw (0,0) [<-,color=black, thick, rounded corners=7pt] +(-0.25,2.6)--(0.5,2.6);

		\draw (0,0) [->|,color=black, thick, rounded corners=7pt] +(0.5,3.0) -- (1.85,3.0)--(1.85,1.75);
		\draw (0,0) [->,color=black, thick, rounded corners=7pt] +(1.85,1.75)--(1.85,0.15)--(-0,0.15)--(0,1.8)--(-0.25,1.8);
		
		\draw (1.2,0.4) node {$\lambda_2+n-1$};
		\draw (1.15,0.63) node {$\vdots$};
		\draw (1.,0.8) node {$\lambda_2+1$};
		\draw (0.75,1.2) node {$\lambda_2$};

		\draw (2.05,2.1) node {$\lambda_1$};
		\draw (2.3,1.45) node {$\lambda_1+n$};

		\draw (1.,1.8) node {$\lambda_1+n$};
		\draw (1.15,2.03) node {$\vdots$};
		\draw (1.,2.2) node {$\lambda_1+2$};
		\draw (1.,2.6) node {$\lambda_1+1$};
		\draw (0.8,3.15) node {$\lambda_1$};

		\draw (-0.25,0) [->,color=black, thick, rounded corners=7pt] +(0,0.)--(2.75,0.);
		
		\draw (-0.25,0) [->,color=black, thick, rounded corners=7pt] +(0,3.35)--(2.75,3.35);
		\draw (-0.25,0) [->,color=black, thick, rounded corners=7pt] +(0,3.75)--(2.75,3.75);
		
		\draw (-0.5,2) node {$\widetilde{\Phi}_L$};
		\draw (0,0) [-,color=black, thick, rounded corners=7pt] +(-0.25,-0.25)--(-0.25,4)--(-0.8,2)--(-0.25,-0.25);

		\draw (3,2) node {$\Phi_R$};
		\draw (0,0) [-,color=black, thick, rounded corners=7pt] +(2.75,-0.25)--(2.75,4)--(3.3,2)--(2.75,-0.25);
		
		\draw (2.4,0.1) node {$u_1$};
		\draw (2.4,0.5) node {$\vdots$};
		\draw (2.4,3.85) node {$u_{N}$};
		\draw (2.4,3.45) node {$u_{i}$};
		
		\draw (0.5,0.4)[fill=black]  circle (0.15ex);
		\draw (0.5,0.8)[fill=black]  circle (0.15ex);
		\draw (0.5,1.2)[fill=black]  circle (0.15ex);
		
		\draw (0.5,2.2)[fill=black]  circle (0.15ex);
		\draw (0.5,2.6)[fill=black]  circle (0.15ex);
		\draw (0.5,3)[fill=black]  circle (0.15ex);

		\end{tikzpicture}
		\end{minipage}
\caption{Graphical illustration of the functional equations for two-sites
($m=2$) for the generalized density operator ${\mathbb D}_2(\lambda_1,\lambda_2)$ valid for $\lambda_1=u_i$. Note that the spectral parameter on the manipulated line is $\lambda_1$ on the left hand side, and $\lambda_1+n$ on the right hand side.}
\label{func-eqs}
\end{center}
\end{figure}

In summary, for arbitrary $m$ and $SU(n)$ we
derive a functional equation for the generalized density operator ${\mathbb
  D}_m(\lambda_1,\dots,\lambda_m)$ given by,
\eq
{\mathbb A}_m(\lambda_1,\dots,\lambda_m)[{\mathbb D}_m(\lambda_1+1,\lambda_2,\dots,\lambda_m)] = {\mathbb D}_m(\lambda_1,\lambda_2,\dots,\lambda_m), \qquad \lambda_1 = u_i,
\label{qkzsun}
\en
where the linear ${\mathbb A}_m(\lambda_1,\dots,\lambda_m)$ operator is a
product of $(m-1)n$ $R$-matrices given as
\bear
{\mathbb A}_m(\lambda_1,\dots,\lambda_m)^{{i}_1 \cdots {i}_{nm}}_{\bar{i}_1 \cdots \bar{i}_{nm}}&:=&
\prod_{k=1}^{m-1} \prod_{j=1+(k-1)n}^{ k n} \check{R}^{(n,n)}(\lambda_1-\lambda_{m-k+1}+j)_{\alpha_j, \bar{i}_j}^{i_j, \alpha_{j-1}} \nonumber \\
&\times& \delta_{\alpha_0}^{i_{nm}} \delta^{\alpha_{n(m-1)}}_{\bar{i}_{n(m-1)+1}} \delta^{i_{n(m-1)+1}}_{\bar{i}_{n(m-1)+2}} \cdots \delta^{i_{nm-1}}_{\bar{i}_{nm}},
\ear
where we assume summation over repeated indices $\alpha_j$.
The action of ${\mathbb A}_2(\lambda_1,\lambda_2)$ for arbitrary $SU(n)$ is illustrated in Figure
\ref{func-eqs}.

Next we turn to the analytical properties of ${\mathbb D}_m(\lambda_1,\dots,\lambda_m)$. By definition, an infinite number of
vertices carrying the parameters $\lambda_j$ enter which may result in an
uncontrolled analytical dependence. Here we are going to show that fortunately
this is not so.

\begin{figure}
	\begin{center}
		\begin{minipage}{0.5\linewidth}
			\begin{tikzpicture}[scale=1.7]
			
			
			\draw (0,0) [->,color=black, thick , rounded corners=7pt] +(-0.5,1.6)--(-0.5,0);
			\draw (0,0) [-,color=black, directed, thick , rounded corners=7pt] +(-1.1,0)--(-1.1,1.9);
			\draw (0,0) [-,color=black, reverse directed, thick, rounded corners=7pt] +(-1.1,1.9)--(-0.5,1.9);
			\draw (0,0) [->,color=black, thick, rounded corners=7pt] +(-0.5,2.2)--(-1.1,2.55)--(-1.1,3);
			\draw (0,0) [-,color=black,  directed, thick, rounded corners=7pt] +(-0.5,3)--(-0.5,2.55)--(-1.1,1.9);
			
			\draw (0,0) [->,color=black, thick , rounded corners=7pt] +(-1.5,0.7)--(-1.5,0);
			\draw (0,0) [-,color=black, directed, thick , rounded corners=7pt] +(-2.1,0)--(-2.1,1.);
			\draw (0,0) [-,color=black, reverse directed, thick, rounded corners=7pt] +(-2.1,1.)--(-1.5,1.);
			\draw (0,0) [->,color=black, thick, rounded corners=7pt] +(-1.5,1.3)--(-2.1,1.65)--(-2.1,3);
			\draw (0,0) [-,color=black,  directed, thick, rounded corners=7pt] +(-1.5,3)--(-1.5,1.65)--(-2.1,1.);
			
			\draw (-0.27,0.7) node {$\dots$};
			\draw (-0.27,1.) node {$\dots$};
			\draw (-0.27,1.3) node {$\dots$};
			
			\draw (-0.27,1.6) node {$\dots$};
			\draw (-0.27,1.9) node {$\dots$};
			\draw (-0.27,2.2) node {$\dots$};
			
			\draw (0,0) [->,color=black, thick, rounded corners=7pt] +(-2.25,0.25)--(-0.4,0.25);
			\draw (0,0) [->,color=black, thick, rounded corners=7pt] +(-2.25,2.75)--(-0.4,2.75);
			
			\draw (0,0) [-,color=black, thick, rounded corners=7pt] +(-1.5,0.7)--(-0.45,0.7);
			\draw (0,0) [-,color=black, thick, rounded corners=7pt] +(-1.5,1.)--(-0.45,1.);
			\draw (0,0) [-,color=black, thick, rounded corners=7pt] +(-1.5,1.3)--(-0.45,1.3);

			\draw (-0.22,0.25) node {$\dots$};
			\draw (-0.22,2.75) node {$\dots$};
			
			\draw (-0.45,-0.25) node {$\lambda_1+2$};
			\draw (-0.85,1.75) node {{\tiny $\lambda_1+1$}};
			\draw (-1.05,-0.25) node {$\lambda_1$};
			
			\draw (-1.55,-0.25) node {$\lambda_2+2$};
			\draw (-1.85,0.85) node {{\tiny $\lambda_2+1$}};
			\draw (-2.1,-0.25) node {$\lambda_2$};

			\draw (0,0) [<-,color=black, thick, rounded corners=7pt] +(-0.1,0.7)--(1.,0.7);
			\draw (0,0) [<-,color=black, thick, rounded corners=7pt] +(-0.1,1.)--(1.,1.);
			\draw (0,0) [<-,color=black, thick, rounded corners=7pt] +(-0.1,1.3)--(1.,1.3);
			\draw (1.,0.7)[fill=black]  circle (0.15ex);
			\draw (1.,1.)[fill=black]  circle (0.15ex);
			\draw (1.,1.3)[fill=black]  circle (0.15ex);
			\draw (0.65,0.82) node {$\lambda_2+2$};
			\draw (0.65,1.12) node {$\lambda_2+1$};
			\draw (0.48,1.42) node {$\lambda_2$};

			\draw (0,0) [<-,color=black, thick, rounded corners=7pt] +(-0.1,1.6)--(1.,1.6);
			\draw (0,0) [<-,color=black, thick, rounded corners=7pt] +(-0.1,1.9)--(1.,1.9);
			\draw (0,0) [<-,color=black, thick, rounded corners=7pt] +(-0.1,2.2)--(1.,2.2);
			\draw (1.,1.6)[fill=black]  circle (0.15ex);
			\draw (1.,1.9)[fill=black]  circle (0.15ex);
			\draw (1.,2.2)[fill=black]  circle (0.15ex);
			\draw (0.65,1.72) node {$\lambda_1+2$};
			\draw (0.65,2.02) node {$\lambda_1+1$};
			\draw (0.48,2.32) node {$\lambda_1$};
			
			\draw (0,0) [->,color=black, thick, rounded corners=7pt] +(0.25,0)--(0.25,3);
						
			\draw (0,0) [->,color=black, thick, rounded corners=7pt] +(1.25,0)--(1.25,3);
			
			\draw (0.25,-0.25) node {$0$};
			\draw (1.25,-0.25) node {$0$};
						
			\draw (0,0) [->,color=black, thick, rounded corners=7pt] +(-0.05,0.25)--(1.65,0.25);
			
			\draw (0,0) [->,color=black, thick, rounded corners=7pt] +(-0.05,2.75)--(1.65,2.75);

			\draw (1.45,0.1) node {$u_1$};
			\draw (1.45,0.45) node {$\vdots$};
			\draw (1.45,2.6) node {$u_{N}$};
			
			\end{tikzpicture}
		\end{minipage}%
		\begin{minipage}{0.5\linewidth}
			\begin{tikzpicture}[scale=1.7]
			
			\draw (0,0) [->,color=black, thick , rounded corners=7pt] +(0.25,1.6)--(0.25,0);
			\draw (0,0) [-,color=black, directed, thick , rounded corners=7pt] +(-0.35,0)--(-0.35,1.9);
			\draw (0,0) [-,color=black, reverse directed, thick, rounded corners=7pt] +(-0.35,1.9)--(0.25,1.9);
			\draw (0,0) [->,color=black, thick, rounded corners=7pt] +(0.25,2.2)--(-0.35,2.55)--(-0.35,3);
			\draw (0,0) [-,color=black,  directed, thick, rounded corners=7pt] +(0.25,3)--(0.25,2.55)--(-0.35,1.9);
			
			\draw (0,0) [->,color=black, thick , rounded corners=7pt] +(-0.75,0.7)--(-0.75,0);
			\draw (0,0) [-,color=black, directed, thick , rounded corners=7pt] +(-1.35,0)--(-1.35,1.);
			\draw (0,0) [-,color=black, reverse directed, thick, rounded corners=7pt] +(-1.35,1.)--(-0.75,1.);
			\draw (0,0) [->,color=black, thick, rounded corners=7pt] +(-0.75,1.3)--(-1.35,1.65)--(-1.35,3);
			\draw (0,0) [-,color=black,  directed, thick, rounded corners=7pt] +(-0.75,3)--(-0.75,1.65)--(-1.35,1.);

			\draw (0.3,-0.25) node {$\lambda_1+2$};
			\draw (-0.1,1.75) node {{\tiny $\lambda_1+1$}};
			\draw (-0.3,-0.25) node {$\lambda_1$};
			
			\draw (-0.8,-0.25) node {$\lambda_2+2$};
			\draw (-1.1,0.85) node {{\tiny $\lambda_2+1$}};
			\draw (-1.35,-0.25) node {$\lambda_2$};

			\draw (0,0) [-,color=black, thick, rounded corners=7pt] +(-0.75,0.7)--(1.,0.7);
			\draw (0,0) [-,color=black, thick, rounded corners=7pt] +(-0.75,1.)--(1.,1.);
			\draw (0,0) [-,color=black, thick, rounded corners=7pt] +(-0.75,1.3)--(1.,1.3);
			\draw (1.,0.7)[fill=black]  circle (0.15ex);
			\draw (1.,1.)[fill=black]  circle (0.15ex);
			\draw (1.,1.3)[fill=black]  circle (0.15ex);
			\draw (0.65,0.82) node {$\lambda_2+2$};
			\draw (0.65,1.12) node {$\lambda_2+1$};
			\draw (0.48,1.42) node {$\lambda_2$};

			\draw (0,0) [-,color=black, thick, rounded corners=7pt] +(0.25,1.6)--(1.,1.6);
			\draw (0,0) [-,color=black, thick, rounded corners=7pt] +(0.25,1.9)--(1.,1.9);
			\draw (0,0) [-,color=black, thick, rounded corners=7pt] +(0.25,2.2)--(1.,2.2);
			\draw (1.,1.6)[fill=black]  circle (0.15ex);
			\draw (1.,1.9)[fill=black]  circle (0.15ex);
			\draw (1.,2.2)[fill=black]  circle (0.15ex);
			\draw (0.65,1.72) node {$\lambda_1+2$};
			\draw (0.65,2.02) node {$\lambda_1+1$};
			\draw (0.48,2.32) node {$\lambda_1$};
			
			\draw (0,0) [->,color=black, thick, rounded corners=7pt] +(-1.65,0)--(-1.65,3);
			
			\draw (0,0) [->,color=black, thick, rounded corners=7pt] +(1.25,0)--(1.25,3);
			
			\draw (-1.65,-0.25) node {$0$};
			
			\draw (1.25,-0.25) node {$0$};

			\draw (0,0) [->,color=black, thick, rounded corners=7pt] +(-1.75,0.25)--(1.65,0.25);
			
			\draw (0,0) [->,color=black, thick, rounded corners=7pt] +(-1.75,2.75)--(1.65,2.75);

			\draw (1.45,0.1) node {$u_1$};
			\draw (1.45,0.45) node {$\vdots$};
			\draw (1.45,2.6) node {$u_{N}$};
			
			\draw (-2,1.5) node {$=$};

			\end{tikzpicture}
			
		\end{minipage}
	\end{center}
	\caption{The generalized density operator ${\mathbb D}_m(\lambda_1,\dots,\lambda_m)$ for the
		case $SU(3)$ and $m=2$, i.e. two bunches of semi-infinite lines.  The
		semi-infinite lines can be rearranged without changing the correlations in the
		form of two vertical lines with spectral parameters $\lambda_i$ and
		$\lambda_i+2$, however with different arrow directions resp.~representations.}
	\label{analyticity}
\end{figure}
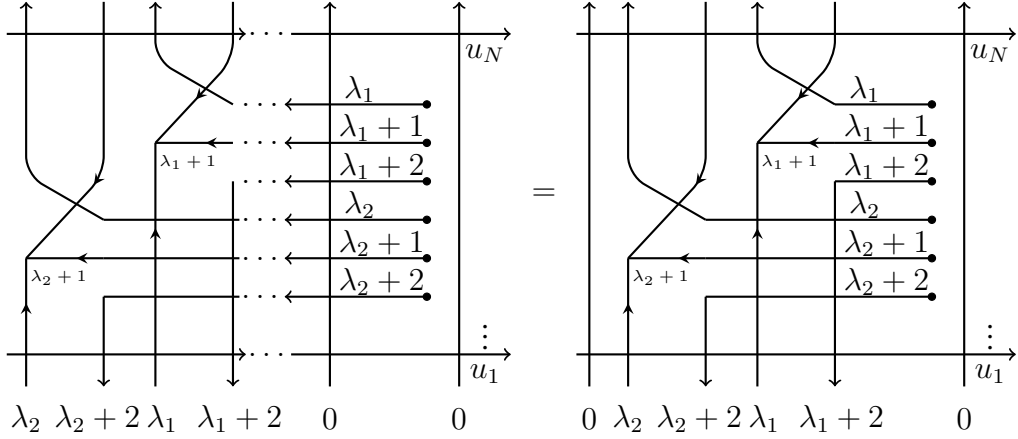

In order to represent the density operator in a way that the analyticity
properties become transparent, we attach at the far left boundary the operator
defined graphically on the left hand side of Figure \ref{analyticity}. This
operator can be moved inside the lattice by use of the 180$^{\circ}$ rotated
version of (\ref{sym-property1}-\ref{sym-property2}), as well as unitarity and
special unitarity, see Figure \ref{analyticity}.

This manipulation makes the analytical structure of the generalized density
operator ${\mathbb D}_m(\lambda_1,\dots,\lambda_m)$ clear. For the
$SU(3)$ case illustrated in Figure \ref{analyticity} the numerator
of the (unnormalized)
matrix elements must be a multivariate polynomial of degree $2N$ in the
variables $\lambda_j$ times an $N$ independent number of linear factors of the
type $\lambda_i-\lambda_j+const$. 
Therefore, the matrix elements of the generalized density
operator after normalization are of the kind,
\eq
\frac{P(\lambda_1,\dots,\lambda_m)}{\prod_{i=1}^m\Lambda_0^{(3)}(\lambda_i)\Lambda_0^{(\bar
    3)}(\lambda_i+2) \cdot \prod_{i<j}\Phi(\lambda_i-\lambda_j)},
\label{analyt-prop}
\en
where $P(\lambda_1,\dots,\lambda_m)$ is a multivariate polynomial of degree
up to $2N + 6(m-1)m/2$ in each variable,
$\Phi(\delta):=(4-\delta^2)(1-\delta^2)^2$ from the intersection
  of three semi-infinite lines with two vertical lines
and $\Lambda_0^{(3)}(\lambda)$ is the leading
eigenvalue of the quantum transfer matrix with fundamental representation in
the auxiliary space and $\Lambda_0^{(\bar 3)}(\lambda)$ is the leading
eigenvalue of the quantum transfer matrix with anti-fundamental representation
in the auxiliary space. The normalization in the denominator is obtained by use of
	(\ref{sym-property1}-\ref{sym-property2}) to move the lower anti-symmetrizer
	to the left, which generates the $\Phi(\delta)$ function. Finally, by use of
	properties (\ref{other1}-\ref{other2}) we obtain two decoupled up- and down-
	going lines with spectral parameters $\lambda_i$ and $\lambda_i+2$, which are
	associated with the corresponding leading eigenvalues.

In the next section, we are going to discuss the solution of the
above functional equations for two and three-sites density operators for the
case of the $SU(3)$ spin chain.

\section{SU(3) spin chain}\label{su3section}

In order to compute the two ($m=2$) and three ($m=3$) sites density operator
$\mathbb D_m(\lambda_1,\dots,\lambda_m)$ we have to propose a suitable
ansatz. This is usually done by choosing a certain number of linearly
independent operators (states) as a basis and working out the resulting
equations for the expansion coefficients. However, the number of these
operators is determined by the dimension of the singlet subspace of the total
space of density operators on $m$ sites referred to as $\dim(m)$, which for
the two and three-sites case are $5$ and $42$, respectively. Although the
two-sites case can still be treated, the high number of required independent
operators for three-sites makes the problem very hard to treat.

\begin{figure}[h]
		\begin{tikzpicture}[scale=1.5]
		\draw (0,0) [<-,color=black, thick, rounded corners=7pt] +(-0.25,1.2)--(1.1,1.2);
		\draw (0,0) [<-,color=black, thick, rounded corners=7pt] +(-0.25,1.5)--(1.1,1.5);
		\draw (0,0) [<-,color=black, thick, rounded corners=7pt] +(-0.25,1.8)--(1.1,1.8);
		\draw (1.1,1.2)[fill=black]  circle (0.15ex);
		\draw (1.1,1.5)[fill=black]  circle (0.15ex);
		\draw (1.1,1.8)[fill=black]  circle (0.15ex);
		\draw (0.65,1.35) node {$\lambda_1+2$};
		\draw (0.65,1.65) node {$\lambda_1+1$};
		\draw (0.45,1.95) node {$\lambda_1$};
		
		\draw (0,0) [->,color=black, thick, rounded corners=7pt] +(0.25,0.25)--(0.25,2.75);
		
		\draw (0,0) [->,color=black, thick, rounded corners=7pt] +(1.6,1.65)--(1.6,2.75);
		\draw (0,0) [-,color=black, thick, rounded corners=7pt] +(1.6,0.25)--(1.6,1.35);
		
		\draw (0,0) [->,color=black, thick, rounded corners=7pt] +(2.15,1.65)--(2.15,2.75);
		\draw (0,0) [-,color=black, thick, rounded corners=7pt] +(2.15,0.25)--(2.15,1.35);
		
		\draw (0,0) [->,color=black, thick, rounded corners=7pt] +(2.65,0.25)--(2.65,2.75);
		
		\draw (0.25,0.) node {$0$};
		
		\draw (1.6,0.) node {$\lambda_2$};
		\draw (1.9,0.) node {$\dots$};
		\draw (2.2,0.) node {$\lambda_m$};
		\draw (2.65,0.) node {$0$};

		\draw (1.6,1.35)[fill=black]  circle (0.15ex);
		\draw (1.6,1.65)[fill=black]  circle (0.15ex);
		
		\draw (2.15,1.35)[fill=black]  circle (0.15ex);
		\draw (2.15,1.65)[fill=black]  circle (0.15ex);

		\draw (-0.25,0) [->,color=black, thick, rounded corners=7pt] +(0,0.5)--(3.2,0.5);
		
		\draw (-0.25,0) [->,color=black, thick, rounded corners=7pt] +(0,2.5)--(3.2,2.5);

		\draw (-1.9,2.5) node {a)};
		\draw (-1.9,1.5) node {${\mathfrak D}_m(\lambda_1,\dots,\lambda_m)=$};

		\draw (-0.4,0.5) node {$\dots$};
		\draw (-0.4,2.5) node {$\dots$};
		\draw (3.4,0.5) node {$\dots$};
		\draw (3.4,2.5) node {$\dots$};

		\draw (-0.4,1.2) node {$\dots$};
		\draw (-0.4,1.5) node {$\dots$};
		\draw (-0.4,1.8) node {$\dots$};

		\draw (2.9,0.6) node {$u_1$};
		\draw (2.9,0.95) node {$\vdots$};
		\draw (2.9,2.6) node {$u_{N}$};

		\end{tikzpicture}
		\begin{tikzpicture}[scale=1.5]
		
		\draw (0,0) [<-,color=black, thick, rounded corners=7pt] +(-0.25,1.2)--(1.1,1.2);
		\draw (0,0) [<-,color=black, thick, rounded corners=7pt] +(-0.25,1.5)--(1.1,1.5);
		\draw (0,0) [<-,color=black, thick, rounded corners=7pt] +(-0.25,1.8)--(1.1,1.8);
		\draw (1.1,1.2)[fill=black]  circle (0.15ex);
		\draw (1.1,1.5)[fill=black]  circle (0.15ex);
		\draw (1.1,1.8)[fill=black]  circle (0.15ex);
		\draw (0.65,1.35) node {$\lambda_1+2$};
		\draw (0.65,1.65) node {$\lambda_1+1$};
		\draw (0.45,1.95) node {$\lambda_1$};
				
		\draw (0,0) [->,color=black, thick, rounded corners=7pt] +(1.6,1.65)--(1.6,2.75);
		\draw (0,0) [-,color=black, thick, rounded corners=7pt] +(1.6,0.25)--(1.6,1.35);
		
		\draw (0,0) [->,color=black, thick, rounded corners=7pt] +(2.15,1.65)--(2.15,2.75);
		\draw (0,0) [-,color=black, thick, rounded corners=7pt] +(2.15,0.25)--(2.15,1.35);
		
		\draw (1.6,0) node {$\lambda_2$};
		\draw (1.9,0) node {$\dots$};
		\draw (2.2,0) node {$\lambda_m$};

		\draw (1.6,1.35)[fill=black]  circle (0.15ex);
		\draw (1.6,1.65)[fill=black]  circle (0.15ex);
		
		\draw (2.15,1.35)[fill=black]  circle (0.15ex);
		\draw (2.15,1.65)[fill=black]  circle (0.15ex);

		\draw (-0.25,0) [->,color=black, thick, rounded corners=7pt] +(0,0.5)--(3.25,0.5);
		
		\draw (-0.25,0) [->,color=black, thick, rounded corners=7pt] +(0,2.5)--(3.25,2.5);

		\draw (-1.9,2.5) node {b)};
		\draw (-1.9,1.5) node {${\mathfrak D}_m(\lambda_1,\dots,\lambda_m)=$};
		
		\draw (2.9,0.6) node {$u_1$};
		\draw (2.9,0.95) node {$\vdots$};
		\draw (2.9,2.6) node {$u_{N}$};
		
		\draw (-0.5,1.5) node {$\widetilde{\Phi}_L$};
		\draw (0,0) [-,color=black, thick, rounded corners=7pt] +(-0.25,0.25)--(-0.25,2.75)--(-0.8,1.5)--(-0.25,0.25);
		\draw (3.5,1.5) node {$\Phi_R$};
		\draw (0,0) [-,color=black, thick, rounded corners=7pt] +(3.25,0.25)--(3.25,2.75)--(3.8,1.5)--(3.25,0.25);
		
		\end{tikzpicture}
\caption{Graphical illustration of the un-normalized mixed density
operator ${\mathfrak D}_m(\lambda_1,\dots,\lambda_m)$ for the $SU(3)$
case. a) We have an infinite cylinder with $N$ infinitely long
horizontal lines carrying spectral parameters $u_j$, $1$ bunch of
three semi-infinite lines carrying spectral parameters $\{\lambda_1,\lambda_1+1,\lambda_1+2 \}$ and $m-1$ vertically open bonds associated to the spectral parameters $\lambda_2,\dots,\lambda_m$; b) the infinitely many column-to-column transfer matrices to the left and right replaced by the boundary states they project onto.}
\label{mixDmatrix}
\end{figure}
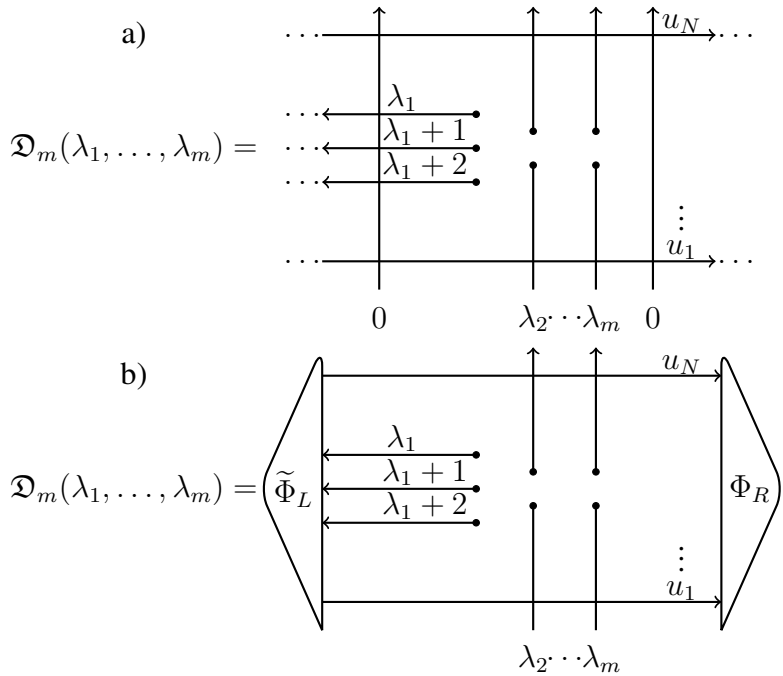

A crucial observation to turn the three-sites case feasible is to reduce the
number of bunches of semi-infinite horizontal lines to the minimal
possible.  For the problem at hand we found that
this can be done by working with only one bunch of
semi-infinite horizontal lines and the remaining $m-1$ bunches reduced by
partial anti-symmetrization to $m-1$
vertically open bonds resulting in a mixed density operator denoted by
${\mathfrak D}_m(\lambda_1,\dots,\lambda_m)$ (see Figure \ref{mixDmatrix}).

This
strategy reduces the dimension of the singlet subspace for the two and
three-sites case to $\dim(2)=3$ and $\dim(3)=11$, respectively. Therefore the
density operator can be written as ${\mathfrak
	D}_m(\lambda_1,\dots,\lambda_m)=\sum_{k=1}^{\dim(m)}
\rho_k^{[m]}(\lambda_1,\dots,\lambda_m) P_{k}^{[m]}$, for conveniently chosen
operators $P_{k}^{[m]}$. In addition, the mixed density operator (Figure
\ref{mixDmatrix}) has also the advantage of simpler reduction properties,
since under partial anti-symmetrization of the semi-infinite lines it is
reduced directly to the physical density operator without the action of
$R$-matrices.

\begin{figure}
	\begin{center}
		\begin{minipage}{0.5\linewidth}
			\begin{tikzpicture}[scale=1.6]
			\draw (-0.,2.7) node {a)};
			
			\draw (-0.27,1.2) node {$\dots$};
			\draw (-0.27,1.5) node {$\dots$};
			\draw (-0.27,1.8) node {$\dots$};
			
			\draw (-0.22,0.5) node {$\dots$};
			\draw (-0.22,2.5) node {$\dots$};
			\draw (0,0) [<-,color=black, thick, rounded corners=7pt] +(-0.1,1.2)--(1.,1.2);
			\draw (0,0) [<-,color=black, thick, rounded corners=7pt] +(-0.1,1.5)--(1.,1.5);
			\draw (0,0) [<-,color=black, thick, rounded corners=7pt] +(-0.1,1.8)--(1.,1.8);
			\draw (1.,1.2)[fill=black]  circle (0.15ex);
			\draw (1.,1.5)[fill=black]  circle (0.15ex);
			\draw (1.,1.8)[fill=black]  circle (0.15ex);
			\draw (0.65,1.32) node {$\lambda_1+2$};
			\draw (0.65,1.62) node {$\lambda_1+1$};
			\draw (0.48,1.95) node {$\lambda_1$};
			
			\draw (0,0) [->,color=black, thick, rounded corners=7pt] +(0.25,0)--(0.25,2.75);
			
			\draw (0,0) [->,color=black, thick, rounded corners=7pt] +(1.25,1.65)--(1.25,2.75);
			\draw (0,0) [-,color=black, thick, rounded corners=7pt] +(1.25,0)--(1.25,1.35);
			
			\draw (0,0) [->,color=black, thick, rounded corners=7pt] +(1.75,1.65)--(1.75,2.75);
			\draw (0,0) [-,color=black, thick, rounded corners=7pt] +(1.75,0)--(1.75,1.35);
			
			\draw (0,0) [->,color=black, thick, rounded corners=7pt] +(2.25,0)--(2.25,2.75);
			
			\draw (0.25,-0.25) node {$0$};
			
			\draw (1.2,-0.25) node {$\lambda_2$};
			\draw (1.52	,-0.25) node {$\dots$};
			\draw (1.52	,0.25) node {$\dots$};
			\draw (1.8,-0.25) node {$\lambda_m$};
			\draw (2.25,-0.25) node {$0$};

			\draw (1.25,1.35)[fill=black]  circle (0.15ex);
			\draw (1.25,1.65)[fill=black]  circle (0.15ex);
			
			\draw (1.75,1.35)[fill=black]  circle (0.15ex);
			\draw (1.75,1.65)[fill=black]  circle (0.15ex);
			
			\draw (0,0) [->,color=black, thick, rounded corners=7pt] +(-0.05,0.5)--(2.75,0.5);
			
			\draw (0,0) [->,color=black, thick, rounded corners=7pt] +(-0.05,2.5)--(2.75,2.5);

			\draw (2.5,0.6) node {$u_1$};
			\draw (2.5,0.9) node {$\vdots$};
			\draw (2.5,2.6) node {$u_{N}$};
			
			\end{tikzpicture}
		\end{minipage}%
		\begin{minipage}{0.5\linewidth}
			\begin{tikzpicture}[scale=1.6]
			\draw (-0.27,1.2) node {$\dots$};
			\draw (-0.27,1.5) node {$\dots$};
			\draw (-0.27,0.9) node {$\dots$};
			
			\draw (-0.22,0.5) node {$\dots$};
			\draw (-0.22,2.5) node {$\dots$};
			\draw (0,0) [<-,color=black, thick, rounded corners=7pt] +(-0.1,1.2)--(1.,1.2);
			\draw (0,0) [<-,color=black, thick, rounded corners=7pt] +(-0.1,1.5)--(1.,1.5);
			\draw (0,0) [->|,color=black, thick, rounded corners=7pt] +(1.,1.8)--(1.,2)--(2,2)--(2,1.5);
			\draw (0,0) [->,color=black, thick, rounded corners=7pt] +(2.,1.5)--(2.,0.9)--(-0.1,0.9);

			\draw (1.,1.2)[fill=black]  circle (0.15ex);
			\draw (1.,1.5)[fill=black]  circle (0.15ex);
			\draw (1.,1.8)[fill=black]  circle (0.15ex);
			\draw (0.65,1.02) node {$\lambda_1+3$};			
			\draw (0.65,1.32) node {$\lambda_1+2$};
			\draw (0.65,1.62) node {$\lambda_1+1$};
			\draw (0.48,1.95) node {$\lambda_1$};
			
			\draw (0,0) [->,color=black, thick, rounded corners=7pt] +(0.25,0)--(0.25,2.75);
			
			\draw (0,0) [->,color=black, thick, rounded corners=7pt] +(1.25,1.65)--(1.25,2.75);
			\draw (0,0) [-,color=black, thick, rounded corners=7pt] +(1.25,0)--(1.25,1.35);
			
			\draw (0,0) [->,color=black, thick, rounded corners=7pt] +(1.75,1.65)--(1.75,2.75);
			\draw (0,0) [-,color=black, thick, rounded corners=7pt] +(1.75,0)--(1.75,1.35);
			
			\draw (0,0) [->,color=black, thick, rounded corners=7pt] +(2.25,0)--(2.25,2.75);
			
			\draw (0.25,-0.25) node {$0$};
			
			\draw (1.2,-0.25) node {$\lambda_2$};
			\draw (1.52	,-0.25) node {$\dots$};
			\draw (1.52	,0.25) node {$\dots$};
			\draw (1.8,-0.25) node {$\lambda_m$};
			\draw (2.25,-0.25) node {$0$};

			\draw (1.25,1.35)[fill=black]  circle (0.15ex);
			\draw (1.25,1.65)[fill=black]  circle (0.15ex);
			
			\draw (1.75,1.35)[fill=black]  circle (0.15ex);
			\draw (1.75,1.65)[fill=black]  circle (0.15ex);

			\draw (0,0) [->,color=black, thick, rounded corners=7pt] +(-0.05,0.5)--(2.75,0.5);
			
			\draw (0,0) [->,color=black, thick, rounded corners=7pt] +(-0.05,2.5)--(2.75,2.5);

			\draw (2.5,0.6) node {$u_1$};
			\draw (2.5,0.9) node {$\vdots$};
			\draw (2.5,2.6) node {$u_{N}$};

		\draw (-1.,1.5) node {$\stackrel{\lambda_1=u_i}{=}$};
			
			\end{tikzpicture}
		\end{minipage}
		\begin{minipage}{0.5\linewidth}
			\begin{tikzpicture}[scale=1.6]
			\draw (-0.,2.7) node {b)};			
			
			\draw (-0.27,1.2) node {$\dots$};
			\draw (-0.27,1.5) node {$\dots$};
			\draw (-0.27,1.8) node {$\dots$};

			\draw (-0.22,0.5) node {$\dots$};
			\draw (-0.22,2.5) node {$\dots$};

			\draw (0,0) [<-,color=black, thick, rounded corners=7pt] +(-0.1,1.2)--(1.,1.2);
			\draw (0,0) [<-,color=black, thick, rounded corners=7pt] +(-0.1,1.5)--(1.,1.5);
			\draw (0,0) [<-,color=black, thick, rounded corners=7pt] +(-0.1,1.8)--(1.,1.8);
			\draw (1.,1.2)[fill=black]  circle (0.15ex);
			\draw (1.,1.5)[fill=black]  circle (0.15ex);
			\draw (1.,1.8)[fill=black]  circle (0.15ex);
			\draw (0.65,1.32) node {$\lambda_1+2$};
			\draw (0.65,1.62) node {$\lambda_1+1$};
			\draw (0.48,1.95) node {$\lambda_1$};

			\draw (1.15,1.2) node {$i_1$};
			\draw (1.15,1.5) node {$i_2$};
			\draw (1.15,1.8) node {$i_3$};
			
			\draw (0,0) [->,color=black, thick, rounded corners=7pt] +(0.25,0)--(0.25,2.75);
			
			\draw (0,0) [->,color=black, thick, rounded corners=7pt] +(1.45,1.8)--(1.45,2.75);
			\draw (0,0) [-,color=black, thick, rounded corners=7pt] +(1.45,0)--(1.45,1.2);
			
			\draw (0,0) [->,color=black, thick, rounded corners=7pt] +(1.85,1.8)--(1.85,2.75);
			\draw (0,0) [-,color=black, thick, rounded corners=7pt] +(1.85,0)--(1.85,1.2);

			\draw (1.45,-0.25) node {$\lambda_2$};
			\draw (1.9,-0.25) node {$\lambda_3$};

			\draw (1.45,1.35) node {$s_1$};
			\draw (1.45,1.65) node {$r_1$};
			
			\draw (1.85,1.35) node {$s_2$};
			\draw (1.85,1.65) node {$r_2$};

			\draw (0,0) [->,color=black, thick, rounded corners=7pt] +(2.25,0)--(2.25,2.75);

			\draw (1.45,1.2)[fill=black]  circle (0.15ex);
			\draw (1.45,1.8)[fill=black]  circle (0.15ex);
			
			\draw (1.85,1.2)[fill=black]  circle (0.15ex);
			\draw (1.85,1.8)[fill=black]  circle (0.15ex);

			\draw (0.25,-0.25) node {$0$};
			
			\draw (2.25,-0.25) node {$0$};

			\draw (0,0) [->,color=black, thick, rounded corners=7pt] +(-0.05,0.5)--(2.75,0.5);
			
			\draw (0,0) [->,color=black, thick, rounded corners=7pt] +(-0.05,2.5)--(2.75,2.5);

			\draw (2.5,0.6) node {$u_1$};
			\draw (2.5,0.9) node {$\vdots$};
			\draw (2.5,2.6) node {$u_{N}$};
			
			\end{tikzpicture}
		\end{minipage}%
		\begin{minipage}{0.5\linewidth}
			\begin{tikzpicture}[scale=1.6]
			\draw (-0.77,1.2) node {$\dots$};
			\draw (-0.77,1.5) node {$\dots$};
			\draw (-0.77,0.9) node {$\dots$};

			\draw (-0.72,0.5) node {$\dots$};
			\draw (-0.72,2.5) node {$\dots$};
			\draw (0,0) [<-,color=black, thick, rounded corners=7pt] +(-0.6,1.2)--(0.7,1.2);
			\draw (0,0) [<-,color=black, thick, rounded corners=7pt] +(-0.6,1.5)--(0.7,1.5);
			\draw (0,0) [->|,color=black, thick, rounded corners=7pt] +(0.7,2)--(2.1,2)--(2.1,1.5);
			\draw (0,0) [->,color=black, thick, rounded corners=7pt] +(2.1,1.5)--(2.1,0.9)--(-0.6,0.9);
			
			\draw (0.7,1.2)[fill=black]  circle (0.15ex);
			\draw (0.7,1.5)[fill=black]  circle (0.15ex);
			\draw (0.7,2)[fill=black]  circle (0.15ex);
			\draw (-0.05,1.02) node {$\lambda_1+3$};			
			\draw (-0.05,1.32) node {$\lambda_1+2$};
			\draw (-0.05,1.62) node {$\lambda_1+1$};
			\draw (0.95,2.15) node {$\lambda_1$};
			\draw (0.75,1.34) node {$\bar i_2=i_1$};
			\draw (0.75,1.65) node {$\bar i_3=i_2$};
			\small
			\draw (1.2,1.85) node {$i_3$};
			
			\draw (0,0) [->,color=black, thick, rounded corners=7pt] +(-0.4,0)--(-0.4,2.75);
			
			\draw (0,0) [->,color=black, thick, rounded corners=7pt] +(1.45,1.8)--(1.45,2.75);
			\draw (0,0) [-,color=black, thick, rounded corners=7pt] +(1.45,0)--(1.45,1.2);
			
			\draw (0,0) [->,color=black, thick, rounded corners=7pt] +(1.85,1.8)--(1.85,2.75);
			\draw (0,0) [-,color=black, thick, rounded corners=7pt] +(1.85,0)--(1.85,1.2);

			\draw (1.25,1.05) node {$\bar i_1$};

			\draw (1.45,-0.25) node {$\lambda_2$};
			\draw (1.9,-0.25) node {$\lambda_3$};
			
			\draw (1.35,0.65) node {$\bar s_1$};
			\draw (2.0,0.65) node {$\bar s_2$};
			
			\draw (1.35,2.2) node {$\bar r_1$};
			\draw (2.,2.2) node {$\bar r_2$};

			\draw (1.65,2.1) node {$\alpha_1$};
			\draw (2.25,1.75) node {$\alpha_2$};

			\draw (1.65,1.05) node {$\beta_2$};
			\draw (2.25,1.15) node {$\beta_1$};
			
			\draw (1.45,1.35) node {$s_1$};
			\draw (1.45,1.65) node {$r_1$};
			
			\draw (1.85,1.35) node {$s_2$};
			\draw (1.85,1.65) node {$r_2$};

			\draw (0,0) [->,color=black, thick, rounded corners=7pt] +(2.4,0)--(2.4,2.75);

			\draw (1.45,1.2)[fill=black]  circle (0.15ex);
			\draw (1.45,1.8)[fill=black]  circle (0.15ex);
			
			\draw (1.85,1.2)[fill=black]  circle (0.15ex);
			\draw (1.85,1.8)[fill=black]  circle (0.15ex);

			\draw (-0.4,-0.25) node {$0$};
			\draw (2.4,-0.25) node {$0$};

			\draw (0,0) [->,color=black, thick, rounded corners=7pt] +(-0.6,0.5)--(2.85,0.5);
			
			\draw (0,0) [->,color=black, thick, rounded corners=7pt] +(-0.6,2.5)--(2.85,2.5);

			\draw (2.6,0.6) node {$u_1$};
			\draw (2.6,0.9) node {$\vdots$};
			\draw (2.6,2.6) node {$u_{N}$};

			\draw (-1.4,1.5) node {$\stackrel{\lambda_1=u_i}{=}$};
			
			\end{tikzpicture}
			
		\end{minipage}
	\end{center}
		\caption{Graphical depiction of the functional equations for the mixed density operator ${\mathfrak D}_m(\lambda_1,\dots,\lambda_m)$ valid for $\lambda_1=u_i$: for: a) $m$-sites; b) $3$-sites matrix elements. }
	\label{mixed-funeqs}
\end{figure}
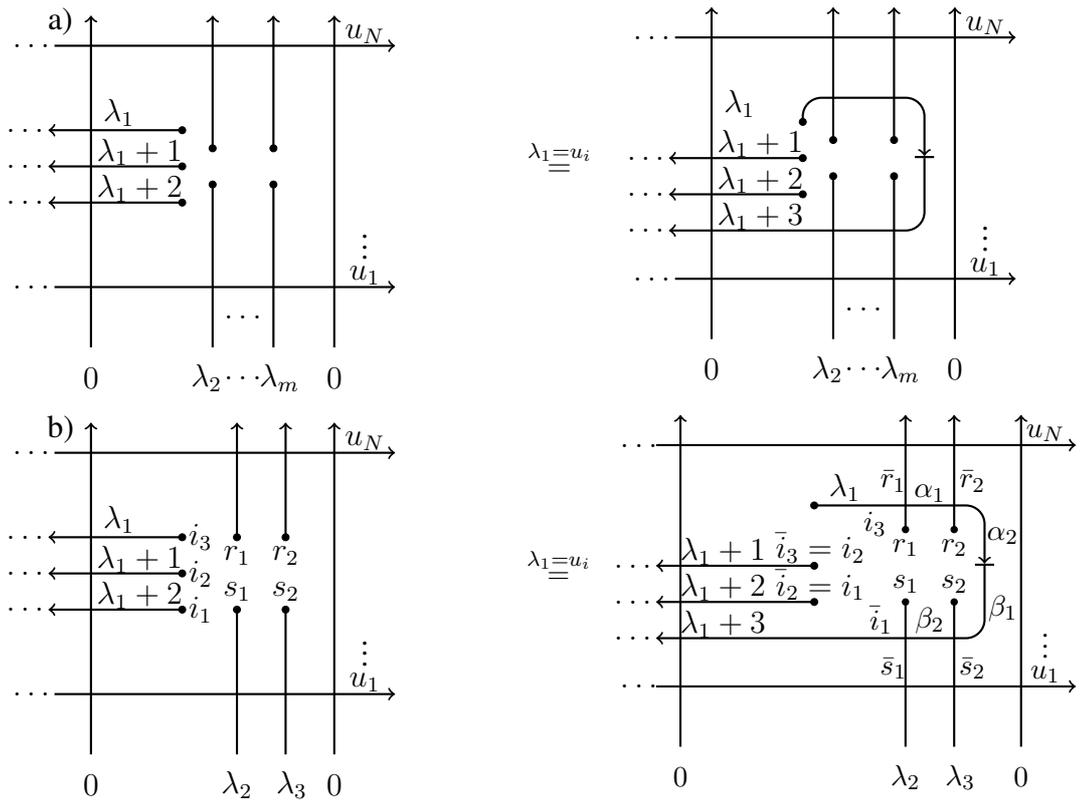

The mixed density operator also fulfills a functional equation
  of the form (\ref{qkzsun}). More specifically the equation satisfied by
  $\mathfrak D_m(\lambda_1,\dots,\lambda_m)$ is shown for the $SU(3)$ case in
  Figure \ref{mixed-funeqs}a, which again is derived for $\lambda_1=u_i$ by use
  of unitarity, Yang-Baxter and special unitarity condition.
\eq
{\mathfrak A}_m(\lambda_1,\dots,\lambda_m)[{\mathfrak D}_m(\lambda_1+1,\lambda_2,\dots,\lambda_m)] = {\mathfrak D}_m(\lambda_1,\lambda_2,\dots,\lambda_m), \qquad \lambda_1 = u_i,
\label{qkzsun-frak}
\en
where ${\mathfrak A}_m(\lambda_1,\dots,\lambda_m)$ is a linear operator
which consists of a product of $2(m-1)$ $R$-matrices 
depicted in Figure \ref{mixed-funeqs} for the case $SU(3)$,
 whose expression is given by (see Figure \ref{mixed-funeqs}b),
\bear
&&{{\mathfrak A}_m(\lambda_1,\dots,\lambda_m)}^{i_1 i_2 i_3 r_1 \cdots r_{m-1} s_1 \cdots s_{m-1}}_{\bar i_1 \bar i_2 \bar i_3 \bar r_1 \cdots \bar r_{m-1} \bar s_1 \cdots \bar s_{m-1}}= \label{frakA}\\
&=& [\check{R}^{(3,3)}(\lambda_1-\lambda_2)]_{ i_3  r_1}^{ \bar r_1 \alpha_1} [\check{R}^{(3,3)}(\lambda_1-\lambda_3)]_{ \alpha_1  r_2}^{ \bar r_2 \alpha_2}\cdots [\check{R}^{(3,3)}(\lambda_1-\lambda_m)]_{\alpha_{m-2}   r_{m-1} }^{\bar r_{m-1} \alpha_{m-1}} \delta_{\beta_1}^{\alpha_{m-1}} \delta_{ i_1}^{\bar i_2} \delta_{ i_2}^{\bar i_3} \nonumber \\
&\times& [\check{R}^{(3,\bar 3)}(\lambda_1+3-\lambda_m)]_{ \beta_1 s_{m-1}}^{ \bar s_{m-1} \beta_2 }\cdots [\check{R}^{(3,\bar 3)}(\lambda_1+3-\lambda_3)]_{\beta_{m-2} s_2}^{ \bar s_2 \beta_{m-1}}[\check{R}^{( 3,\bar 3)}(\lambda_1+3-\lambda_2)]_{ \beta_{m-1} s_1}^{ \bar s_1 \bar i_1}. \nonumber 
\ear

\begin{figure}
	\begin{center}
		\begin{minipage}{0.5\linewidth}
		\begin{tikzpicture}[scale=1.7]
		

		\draw (0,0) [->,color=black, thick , rounded corners=7pt] +(-0.5,1.2)--(-0.5,0);
		\draw (0,0) [-,color=black, directed, thick , rounded corners=7pt] +(-1.1,0)--(-1.1,1.5);
		\draw (0,0) [-,color=black, reverse directed, thick, rounded corners=7pt] +(-1.1,1.5)--(-0.5,1.5);
		\draw (0,0) [->,color=black, thick, rounded corners=7pt] +(-0.5,1.8)--(-1.1,2.1)--(-1.1,3);
		\draw (0,0) [-,color=black,  directed, thick, rounded corners=7pt] +(-0.5,3)--(-0.5,2.1)--(-1.1,1.5);

		\draw (-0.27,1.2) node {$\dots$};
		\draw (-0.27,1.5) node {$\dots$};
		\draw (-0.27,1.8) node {$\dots$};

		\draw (0,0) [->,color=black, thick, rounded corners=7pt] +(-1.2,0.5)--(-0.4,0.5);
		\draw (0,0) [->,color=black, thick, rounded corners=7pt] +(-1.2,2.5)--(-0.4,2.5);

		\draw (-0.22,0.5) node {$\dots$};
		\draw (-0.22,2.5) node {$\dots$};

		\draw (-0.45,-0.25) node {$\lambda_1+2$};
		\draw (-0.85,1.35) node {{\tiny $\lambda_1+1$}};
		\draw (-1.1,-0.25) node {$\lambda_1$};

		\draw (0,0) [<-,color=black, thick, rounded corners=7pt] +(-0.1,1.2)--(1.,1.2);
		\draw (0,0) [<-,color=black, thick, rounded corners=7pt] +(-0.1,1.5)--(1.,1.5);
		\draw (0,0) [<-,color=black, thick, rounded corners=7pt] +(-0.1,1.8)--(1.,1.8);
		\draw (1.,1.2)[fill=black]  circle (0.15ex);
		\draw (1.,1.5)[fill=black]  circle (0.15ex);
		\draw (1.,1.8)[fill=black]  circle (0.15ex);
		\draw (0.65,1.32) node {$\lambda_1+2$};
		\draw (0.65,1.62) node {$\lambda_1+1$};
		\draw (0.48,1.95) node {$\lambda_1$};
		
		\draw (0,0) [->,color=black, thick, rounded corners=7pt] +(0.25,0)--(0.25,3);
		
		\draw (0,0) [->,color=black, thick, rounded corners=7pt] +(1.25,1.65)--(1.25,3);
		\draw (0,0) [-,color=black, thick, rounded corners=7pt] +(1.25,0)--(1.25,1.35);
		
		\draw (0,0) [->,color=black, thick, rounded corners=7pt] +(1.75,1.65)--(1.75,3);
		\draw (0,0) [-,color=black, thick, rounded corners=7pt] +(1.75,0)--(1.75,1.35);
		
		\draw (0,0) [->,color=black, thick, rounded corners=7pt] +(2.25,0)--(2.25,3);
		
		\draw (0.25,-0.25) node {$0$};
		
		\draw (1.2,-0.25) node {$\lambda_2$};
		\draw (1.52	,-0.25) node {$\dots$};
		\draw (1.8,-0.25) node {$\lambda_m$};
		\draw (2.25,-0.25) node {$0$};

		\draw (1.25,1.35)[fill=black]  circle (0.15ex);
		\draw (1.25,1.65)[fill=black]  circle (0.15ex);
		
		\draw (1.75,1.35)[fill=black]  circle (0.15ex);
		\draw (1.75,1.65)[fill=black]  circle (0.15ex);

		\draw (0,0) [->,color=black, thick, rounded corners=7pt] +(-0.05,0.5)--(2.75,0.5);
		
		\draw (0,0) [->,color=black, thick, rounded corners=7pt] +(-0.05,2.5)--(2.75,2.5);

		\draw (2.5,0.6) node {$u_1$};
		\draw (2.5,0.9) node {$\vdots$};
		\draw (2.5,2.6) node {$u_{N}$};
				
		\end{tikzpicture}
		\end{minipage}%
		\begin{minipage}{0.5\linewidth}
		\begin{tikzpicture}[scale=1.7]
		\draw (0,0) [->,color=black, thick , rounded corners=7pt] +(1.25,1.3)--(1.25,0);
		\draw (0,0) [-,color=black, directed, thick , rounded corners=7pt] +(0.65,0)--(0.65,1.5);
		\draw (0,0) [-,color=black, directed, thick, rounded corners=7pt] +(1.25,1.5)--(0.65,1.5);
		\draw (0,0) [->,color=black, thick, rounded corners=7pt] +(1.25,1.7)--(0.65,2.1)--(0.65,3);
		\draw (0,0) [-,color=black,  directed, thick, rounded corners=7pt] +(1.25,3)--(1.25,2.1)--(0.65,1.5);
		\draw (1.25,1.3)[fill=black]  circle (0.15ex);
		\draw (1.25,1.5)[fill=black]  circle (0.15ex);
		\draw (1.25,1.7)[fill=black]  circle (0.15ex);
		\draw (0.65,1.5)[fill=black]  circle (0.15ex);
		\draw (1.15,-0.25) node {$\lambda_1+2$};
		\draw (0.95,1.4) node {{\tiny $\lambda_1+1$}};
		\draw (0.65,-0.25) node {$\lambda_1$};
		
		\draw (0,0) [->,color=black, thick, rounded corners=7pt] +(0.25,0)--(0.25,3);
		
		\draw (0,0) [->,color=black, thick, rounded corners=7pt] +(1.6,1.65)--(1.6,3);
		\draw (0,0) [-,color=black, thick, rounded corners=7pt] +(1.6,0)--(1.6,1.35);
		
		\draw (0,0) [->,color=black, thick, rounded corners=7pt] +(2.15,1.65)--(2.15,3);
		\draw (0,0) [-,color=black, thick, rounded corners=7pt] +(2.15,0)--(2.15,1.35);
		
		\draw (0,0) [->,color=black, thick, rounded corners=7pt] +(2.65,0)--(2.65,3);
		
		\draw (0.25,-0.25) node {$0$};
		
		\draw (1.65,-0.25) node {$\lambda_2$};
		\draw (1.95	,-0.25) node {$\dots$};
		\draw (2.25,-0.25) node {$\lambda_m$};
		\draw (2.65,-0.25) node {$0$};

		\draw (1.6,1.35)[fill=black]  circle (0.15ex);
		\draw (1.6,1.65)[fill=black]  circle (0.15ex);
		
		\draw (2.15,1.35)[fill=black]  circle (0.15ex);
		\draw (2.15,1.65)[fill=black]  circle (0.15ex);

		\draw (-0.25,0) [->,color=black, thick, rounded corners=7pt] +(0.25,0.5)--(3.25,0.5);
		
		\draw (-0.25,0) [->,color=black, thick, rounded corners=7pt] +(0.25,2.5)--(3.25,2.5);
		
		\draw (-0.5,1.5) node {$=$};
		
		\draw (2.9,0.6) node {$u_1$};
		\draw (2.9,0.9) node {$\vdots$};
		\draw (2.9,2.6) node {$u_{N}$};
		
		\end{tikzpicture}
		
		\end{minipage}
	\end{center}
\caption{The mixed density operator ${\mathfrak
    D}_m(\lambda_1,\dots,\lambda_m)$ for the case $SU(3)$ and $m=3$, i.e. one
  bunch of semi-infinite horizontal lines and two open vertical lines. The
  semi-infinite lines can be rearranged without changing the correlations in
  the form of two vertical lines with spectral parameters $\lambda_1$ and
  $\lambda_1+2$, however different arrow directions resp.~representations.
}
\label{analyticitymixed}
\end{figure}
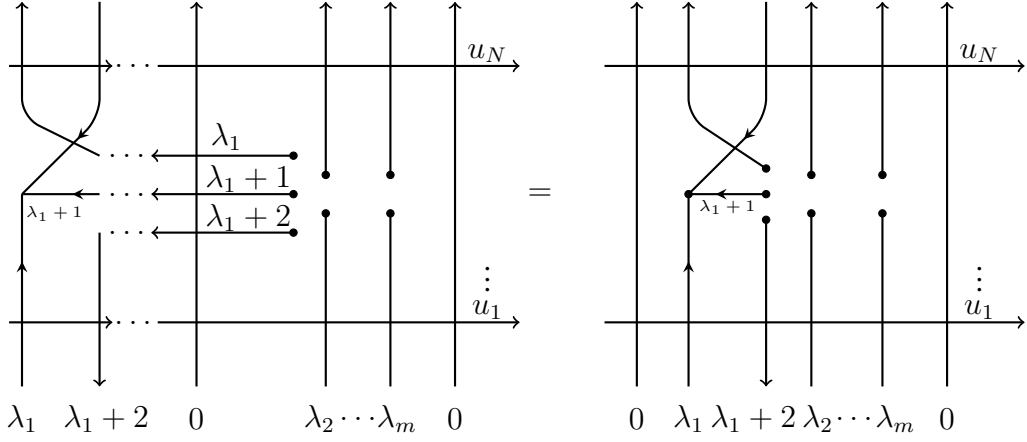

As for the generalized density operator, also for the mixed density operator
$\mathfrak D_m(\lambda_1,\dots,\lambda_m)$ the analytical properties with
regard to the dependence on $\lambda_1$ are not obvious from the definition of
the operator. However, by exploiting the properties
(\ref{sym-property1}-\ref{sym-property2}) as above, we transform the semi-infinite
horizontal lines carrying the spectral parameter $\lambda_1$ into two vertical
lines, see Figure \ref{analyticitymixed}.

This makes the analytical structure of the mixed density operator ${\mathfrak
  D}_m(\lambda_1,\dots,\lambda_m)$ clear, since for the $SU(3)$ case the numerator of the matrix elements must be
a multivariate polynomial of degree $2N$ in the variable $\lambda_1$ and of
degree $N$ for the remaining $\lambda_i$, for $i=2,3,\dots,m$,
\eq
\frac{P(\lambda_1,\dots,\lambda_m)}
{\Lambda_0^{(3)}(\lambda_1)\Lambda_0^{(\bar 3)}(\lambda_1+2)\prod_{i=2}^m\Lambda_0^{(3)}(\lambda_i)
},
\label{analyt-prop}
\en
where again $\Lambda_0^{(3)}(\lambda)$ is the leading
eigenvalue of the quantum transfer matrix with fundamental representation in
the auxiliary space and $\Lambda_0^{(\bar 3)}(\lambda)$ is the leading
eigenvalue of the quantum transfer matrix with anti-fundamental representation
in the auxiliary space.

\subsection{Computation of the two-sites density operator}

Due to $SU(n)$ symmetry the usual (normalized) two-point density operator for the
fundamental-funda\-mental and also for anti-fundamental--fundamental
representations can be written as follows,
\bear
D_2^{(nn)}(\lambda_1,\lambda_2)&=&\left(\frac{1}{n^2}-\frac{\alpha_{nn}(\lambda_1,\lambda_2)}{n} \right) I + \alpha_{nn}(\lambda_1,\lambda_2) P, \label{qq}\\
D_2^{(\bar{n}n)}(\lambda_1,\lambda_2)&=&\left(\frac{1}{n^2}-\frac{\alpha_{\bar{n}n}(\lambda_1,\lambda_2)}{n} \right) I + \alpha_{\bar{n}n}(\lambda_1,\lambda_2)E.
\ear
It is convenient to define some simple two-point correlation functions to work with,
\bear 
\omega_{nn}(\lambda_1,\lambda_2)&=&\tr[P D_2^{(nn)}] =\frac{1}{n}+ (n^2-1) \alpha_{nn}(\lambda_1,\lambda_2), \label{ome1}\\
\omega_{\bar{n}n}(\lambda_1,\lambda_2)&=&\tr[E D_2^{(\bar{n}n)}]=\frac{1}{n}+ (n^2-1) \alpha_{\bar{n}n}(\lambda_1,\lambda_2). 
\ear
The above correlation functions will be useful in the coming computations.

The operator (\ref{qq}) represents the full density matrix whose non-trivial
matrix elements are
${D_2^{(nn)}}_{ii}^{ii}(\lambda_1,\lambda_2)=\frac{1}{n^2}+\frac{(n-1)}{n}
\alpha_{nn}(\lambda_1,\lambda_2)$,
${D_2^{(nn)}}_{ij}^{ji}(\lambda_1,\lambda_2)=\alpha_{nn}(\lambda_1,\lambda_2)$,
${D_2^{(nn)}}_{ij}^{ij}(\lambda_1,\lambda_2)=\frac{1}{n^2}-\frac{1}{n}\alpha_{nn}(\lambda_1,\lambda_2)$
for $i,j=1,\cdots,n$, $i\neq j$. Therefore, in order to fully determine the
two-sites density operator (\ref{qq}), one only needs to compute
$\alpha_{nn}(\lambda_1,\lambda_2)$ or equivalently
$\omega_{nn}(\lambda_1,\lambda_2)$ above.

As discussed above, for deriving a closed set of functional equations we have
to use the mixed density operator ${\mathfrak D}_2$ and due to the $SU(3)$
symmetry this operator can be explicitly written as a superposition of $3$ linearly
independent operators as follows,
\eq
{\mathfrak D}_2(\lambda_1,\lambda_2)= \rho_1^{[2]}(\lambda_1,\lambda_2) P_{1}^{[2]} + \rho_2^{[2]}(\lambda_1,\lambda_2) P_{2}^{[2]} + \rho_3^{[2]}(\lambda_1,\lambda_2) P_{3}^{[2]},
\label{D2frak}
\en
where the operators $P_{k}^{[2]}$ are chosen as,
\eq
\begin{minipage}{0.33\linewidth}
\begin{tikzpicture}[scale=1]
\draw (-1.,0) node {$P_1^{[2]}=$};
\draw (0,0) [-,color=black, thick,directed, rounded corners=8pt]+(0,0.5) -- +(0.4,0.4) -- +(0.5,0.);
\draw (0,0) [-,color=black, thick, directed, rounded corners=8pt]+(0,0.) --  +(0.5,0.);
\draw (0,0) [-,color=black, thick,directed, rounded corners=8pt]+(0,-0.5) -- +(0.4,-0.4) -- +(0.5,0.);
		\draw (0.,0.5)[fill=black]  circle (0.3ex);
		\draw (0,0)[fill=black]  circle (0.3ex);
		\draw (0.,-0.5)[fill=black]  circle (0.3ex);
\draw (0,0) [color=black, directed, thick, rounded corners=8pt]+(0.75,0.5) -- +(0.75,-0.5);
		\draw (0.75,-0.5)[fill=black]  circle (0.3ex);
		\draw (0.75,0.5,0)[fill=black]  circle (0.3ex);
\draw (1.25,0) node {,};
\end{tikzpicture}
\end{minipage}%
\begin{minipage}{0.33\linewidth}
\begin{tikzpicture}[scale=1]
\draw (-1.,0) node {$P_2^{[2]}=$};
\draw (0,0) [-,color=black, thick,directed, rounded corners=8pt]+(0,-0.5) -- +(0.25,-0.45) -- +(0.5,-0.25);
\draw (0,0) [-,color=black, thick, directed, rounded corners=8pt]+(0,0.) --  +(0.25,-0.05)-- +(0.5,-0.25);

\tikzstyle directed=[postaction={decorate,decoration={markings, mark=at position .45 with {\arrow[arrowstyle]{stealth}}}}]
\draw (0,0) [color=black, directed, thick, rounded corners=8pt]+(0.75,0.5) -- +(0.5,-0.25);
\tikzstyle reverse directed=[postaction={decorate,decoration={markings, mark=at position .65 with {\arrowreversed[arrowstyle]{stealth};}}}]
\draw (0,0) [-,color=black, thick,directed, rounded corners=8pt]+(0,0.5) -- +(0.25,0.4) -- +(0.55,0.1);
\draw (0,0) [-,color=black, thick, rounded corners=8pt]+(0.65,-0.05) -- +(0.75,-0.3) -- +(0.75,-0.5);

\draw (0.,0.5)[fill=black]  circle (0.3ex);
\draw (0,0)[fill=black]  circle (0.3ex);
\draw (0.,-0.5)[fill=black]  circle (0.3ex);
\draw (0.75,-0.5)[fill=black]  circle (0.3ex);
\draw (0.75,0.5,0)[fill=black]  circle (0.3ex);
\draw (1.25,0) node {,};
\end{tikzpicture}
\end{minipage}%
\begin{minipage}{0.33\linewidth}
\begin{tikzpicture}[scale=1]
\draw (-1.,0) node {$P_3^{[2]}=$};
\draw (0,0) [-,color=black, thick,directed, rounded corners=8pt]+(0,-0.5) -- +(0.4,-0.35) -- +(0.75,-0.5);
\draw (0,0) [-,color=black, thick, directed, rounded corners=8pt]+(0,0.) --  +(0.25,0.05)-- +(0.45,0.25);
\draw (0,0) [-,color=black, thick,directed, rounded corners=8pt]+(0,0.5) -- +(0.25,0.45) -- +(0.45,0.25);

\draw (0.,0.5)[fill=black]  circle (0.3ex);
\draw (0,0)[fill=black]  circle (0.3ex);
\draw (0.,-0.5)[fill=black]  circle (0.3ex);
\draw (0,0) [color=black, directed, thick, rounded corners=8pt]+(0.75,0.5) -- +(0.45,0.25);
\draw (0.75,-0.5)[fill=black]  circle (0.3ex);
\draw (0.75,0.5,0)[fill=black]  circle (0.3ex);
\draw (1.25,0) node {,};
\end{tikzpicture}
\end{minipage}
\en
and the functions $\rho_{k}^{[2]}(\lambda_1,\lambda_2)$ are to be
determined. It is worth to note that we only need three linearly independent
operators/states which can be seen as tensor anti-symmetric in the three
connected indices (black dots) and symmetric in the other two. Alternatively,
we can consider the operators $P_j^{[2]}$ as vector in $[\bar 3]^{\otimes 4} \otimes
[3]$, with the indices $i_1,i_2,i_3,r_1,s_1=1,2,3$ assigned to the dots, such that e.g.
for $P_1^{[2]}$ we have
\eq
\	\begin{tikzpicture}[scale=1]
	\draw (-5,0) node
              {$\left(P_1^{[2]}\right)_{i_1,i_2,i_3,r_1}^{s_1}
:=\quad\epsilon_{i_1,i_2,i_3}\cdot\delta_{r_1}^{s_1}\quad
\equiv$};
	\draw (0,0) [-,color=black, thick,directed, rounded corners=8pt]+(0,0.5) -- +(0.4,0.4) -- +(0.5,0.);
	\draw (0,0) [-,color=black, thick, directed, rounded corners=8pt]+(0,0.) --  +(0.5,0.);
	\draw (0,0) [-,color=black, thick,directed, rounded corners=8pt]+(0,-0.5) -- +(0.4,-0.4) -- +(0.5,0.);
	\draw (0.,0.5)[fill=black]  circle (0.3ex);
	\draw (0,0)[fill=black]  circle (0.3ex);
	\draw (0.,-0.5)[fill=black]  circle (0.3ex);
	\draw (-0.25,0.5) node {$i_1$};
	\draw (-0.25,0) node {$i_2$};
	\draw (-0.25,-0.5) node {$i_3$};
	\draw (0.75,0.75) node {$r_1$};
	\draw (0.75,-0.75) node {$s_1$};

	\draw (0,0) [color=black, directed, thick, rounded corners=8pt]+(0.75,0.5) -- +(0.75,-0.5);
	\draw (0.75,-0.5)[fill=black]  circle (0.3ex);
	\draw (0.75,0.5,0)[fill=black]  circle (0.3ex);
	\draw (1.25,0) node {.};
	\end{tikzpicture}
\en
Note that upper indices refer to states from $[3]$ and lower indices refer to
states from $[\bar 3]$.
And furthermore, this object has the right transformation properties, namely
invariance under $SU(3)$, because for arbitrary $g\in SU(3)$ we have
$g^*\otimes g^*\otimes g^*\cdot\epsilon = \mbox{det}(g^*)\epsilon=\epsilon$
and $g^*\otimes g\cdot \mbox{id} = \mbox{id}$.
For convenience we have chosen to work with operators which resemble the
usual identity, permutation and Temperley-Lieb defined before and that after
partial anti-symmetrization reduce to the combination of identity,
permutation and Temperley-Lieb. This way, the density operator (\ref{D2frak})
would nicely reduce to the usual density operator (\ref{qq}) as
$D_2(\lambda_1,\lambda_2)=(2 \rho^{[2]}_{1} + \rho^{[2]}_{3}) I + (2
\rho^{[2]}_{2} - \rho^{[2]}_{3}) P_{12}$ thanks to the properties
(\ref{other2}).

Using the above representation of the density operator (\ref{D2frak}), 
equation (\ref{qkzsun-frak}) implies the following set of functional equations
\eq
\left(\begin{array}{c}
	\rho_{1}^{[2]}(\lambda_1,\lambda_2) \\
	\rho_{2}^{[2]}(\lambda_1,\lambda_2) \\
	\rho_{3}^{[2]}(\lambda_1,\lambda_2)
\end{array}\right)=
A^{[2]}(\lambda) \cdot\left(\begin{array}{c}
	\rho_{1}^{[2]}(\lambda_1+1,\lambda_2) \\
	\rho_{2}^{[2]}(\lambda_1+1,\lambda_2) \\
	\rho_{3}^{[2]}(\lambda_1+1,\lambda_2)
\end{array}\right), \qquad \lambda_1=u_i,
\label{matrixL-qKZ}
\en
where $\lambda=\lambda_1-\lambda_2$ and the matrix $A^{[2]}(\lambda)$ is given by,
\eq
A^{[2]}(\lambda)=\left(\begin{array}{ccc}
\frac{(-1 + 3\lambda + \lambda^2)}{\lambda( \lambda + 3)} & \frac{(-2 + 2\lambda + \lambda^2)}{\lambda( \lambda+3)} & \frac{1}{ \lambda + 3} \\
\frac{3}{\lambda( \lambda + 3)} & \frac{-(-3 + \lambda + \lambda^2)}{\lambda( \lambda + 3)} & \frac{\lambda}{ \lambda + 3} \\ 
0 & \frac{-(-1 + \lambda)}{\lambda} & 0  
\end{array}\right).
\nonumber
\en

These equations can be disentangled by the transformation matrix
\eq
\left(\begin{array}{c}
	1 \\
	\omega_{33}(\lambda_1,\lambda_2) \\
	\omega_{\bar{3}3}(\lambda_1-1,\lambda_2)
\end{array}\right)
=\left(\begin{array}{ccc}
18 &  6  & 6\\
6 & 18  &-6 \\
6 & -6 & 18
\end{array}\right) \cdot \left(\begin{array}{c}
\rho_{1}^{[2]}(\lambda_1,\lambda_2) \\
\rho_{2}^{[2]}(\lambda_1,\lambda_2) \\
\rho_{3}^{[2]}(\lambda_1,\lambda_2)
\end{array}\right),
\en
which expresses the normalization condition and the partial antisymmetrization
in the lower and upper two semi-infinite lines of the density operator
represented in Figure \ref{mixDmatrix}. Here, the properties (\ref{fus1}-\ref{fus2}) were used.

In terms of the above functions, the functional equation becomes,
\bear
\omega_{33}(\lambda_1,\lambda_2)&=& \frac{(\lambda-1)(\lambda+1)}{\lambda(\lambda+3)} \omega_{\bar{3}3}(\lambda_1,\lambda_2) +\frac{1}{\lambda},  \label{qKZ1}\\
\omega_{\bar{3}3}(\lambda_1-1,\lambda_2)&=&-\frac{(\lambda-1)(\lambda+3)}{\lambda(\lambda+3)}
\omega_{33}(\lambda_1+1,\lambda_2)
-\frac{(\lambda-1)(\lambda+2)}{\lambda(\lambda+3)}
\omega_{\bar{3}3}(\lambda_1,\lambda_2)\nonumber\\
&&+\frac{\lambda-1}{\lambda}, \label{qKZ2} 
\ear
for $\lambda_1=u_i$ and arbitrary $\lambda_2$.

\subsubsection{Zero temperature solution}

At zero temperature, the above functional equations hold for arbitrary
$\lambda_1$. This is because at zero temperature one has to take the Trotter
limit ($N\to\infty$) and therefore the horizontal  
spectral parameters $u_i$ can take an infinite number of continuous values.

Therefore, we solve Eq.(\ref{qKZ1}) for
$\omega_{\bar{3}3}(\lambda_1,\lambda_2)$ and  
insert this into
  Eq.~(\ref{qKZ2}), resulting in
\bear
\frac{\omega_{33}(\lambda_1+1,\lambda_2)}{\lambda(\lambda+2)} + \frac{\omega_{33}(\lambda_1,\lambda_2)}{(\lambda-1)(\lambda+1)} + \frac{\omega_{33}(\lambda_1-1,\lambda_2)}{\lambda(\lambda-2)} = \frac{\lambda^2+2}{(\lambda^2-4)(\lambda^2-1)}.
\label{eqq}
\ear
In addition, at zero temperature the bi-variate function
$\omega_{33}(\lambda_1,\lambda_2)$ turns into a single-variable function
$\omega_{33}(\lambda_1-\lambda_2)$
depending only on the difference of the arguments, which allows us to define
\eq
\sigma(\lambda)=\frac{\omega_{33}(\lambda)}{(\lambda-1)(\lambda+1)}.
\en
Therefore, equation (\ref{eqq}) can be written as
\bear
\sigma(\lambda+1) + \sigma(\lambda) + \sigma(\lambda-1) = \frac{\lambda^2+2}{(\lambda-2)(\lambda-1)(\lambda+1)(\lambda+2)},
\label{eqqq}
\ear
whose solution obtained via Fourier transform is given by,
\bear
\sigma(\lambda)=- \frac{d}{d\lambda}\log\left\{  \frac{\Gamma(1+\frac{1}{3}+\frac{\lambda}{3} ))\Gamma(1-\frac{\lambda}{3})}{\Gamma(1+\frac{1}{3}-\frac{\lambda}{3})  \Gamma(1+\frac{\lambda}{3})  } \right\}- \frac{1}{\lambda^2-1}.
\label{sol-lam}
\ear

Having this solution we obtain $\omega_{33}(\lambda_1,\lambda_2)$. Taking the homogeneous limit ($\lambda_k \to 0$), we obtain
\bear
\omega_{33}(0,0)=-\sigma(0)=1-\frac{\pi}{3\sqrt3}- \log3  \approx -0.70321207674618 ,
\label{sol}
\ear
which is precisely the ground state energy, as expected
\cite{UIMIN,SUTHERLAND}, which is a special correlation
function. The $\alpha_{33}$
coefficient in the density operator (\ref{qq}) is obtained from (\ref{ome1}) and (\ref{sol}) such that,
\bear
\alpha_{33}(0,0)=\frac{1}{24}\left[2-\frac{\pi}{\sqrt{3}} -3 \log3\right]\approx -0.12956817625994.
\label{soldens}
\ear

\subsubsection{Properties of the two-point function}

Differently from $SU(2)$, in the higher-rank case of $SU(3)$ the function $\omega_{33}(\lambda)$
is a generating function of special combinations of modified $\zeta$ functions (Hurwitz' zeta function).

We define a function $G(\lambda)$ as follows
\bear
G(\lambda)&=& \frac{\omega_{33}(\lambda)+1 }{\lambda^2-1}   \\
&=& \frac{1}{3}\left[\psi_0\left(1-\frac{\lambda}{3}\right)-\psi_0\left(1+\frac{1}{3}-\frac{\lambda}{3}\right) +\psi_0\left(1+\frac{\lambda}{3}\right)-\psi_0\left(1+\frac{1}{3}+\frac{\lambda}{3}\right)\right],
\nonumber
\ear
where $\psi_0(z)$ is the digamma function $\psi_0(z)=\frac{d}{dz}\log \Gamma(z)$.

Expanding $G(\lambda)$ in a power series we obtain
\bear
3 G(\lambda)=2 \sum_{k=0}^{\infty} \frac{1}{2 k!} \left[\psi_{2k}\left( 1 \right) - \psi_{2k}\left(1+\frac{1}{3}\right) \right] \left(\frac{\lambda}{3}\right)^{2k},
\label{func}
\ear 
where now $\psi_m(\lambda)$ is the polygamma function. 

We can use the fact that $\psi_{m}(z)=(-1)^{m+1}(m)!\zeta(m+1,z)$, where $\zeta(m,z)$ is the modified zeta function defined as
\eq
\zeta(m,a)=\sum_{k=0}^{\infty} \frac{1}{(k+a)^m},
\en
and re-write expression (\ref{func}) as
\eq
3 G(\lambda)=2 \sum_{k=1}^{\infty} \left[\zeta\left(2k+1,1\right) - \zeta\left(2k+1,1+\frac{1}{3}\right) \right] \left(\frac{\lambda}{3}\right)^{2k} + 2\left[\psi_0\left(1\right)-\psi_0\left(1+\frac{1}{3}\right)\right].
\en  
Then we see that $G(\lambda)$ is a generating function of
differences of the modified zeta function $\zeta\left(2k+1,1\right) -\zeta\left(2k+1,1+\frac{1}{3}\right)$, albeit not of the modified zeta function itself.

\subsection{Computation of the three-sites density operator}

In the three-sites case, the density operator can be written as a superposition of $11$ linearly independent operators as follows,
\eq
{\mathfrak D}_3(\lambda_1,\lambda_2,\lambda_3)=\sum_{k=1}^{11} \rho_k^{[3]}(\lambda_1,\lambda_2,\lambda_3) P_{k}^{[3]}
\label{D3frak}
\en
where the operators $P_{k}^{[3]}$ are chosen as,
\eq
\begin{minipage}{0.25\linewidth}
	\begin{tikzpicture}[scale=1] 
	\draw (-1.,0) node {$P_1^{[3]}=$};
	\draw (0,0) [-,color=black, thick,directed, rounded corners=8pt]+(0,0.5) -- +(0.4,0.4) -- +(0.5,0.);
	\draw (0,0) [-,color=black, thick, directed, rounded corners=8pt]+(0,0.) --  +(0.5,0.);
	\draw (0,0) [-,color=black, thick,directed, rounded corners=8pt]+(0,-0.5) -- +(0.4,-0.4) -- +(0.5,0.);
	\draw (0.,0.5)[fill=black]  circle (0.3ex);
	\draw (0,0)[fill=black]  circle (0.3ex);
	\draw (0.,-0.5)[fill=black]  circle (0.3ex);
	\draw (0,0) [color=black, directed, thick, rounded corners=8pt]+(0.75,0.5) -- +(0.75,-0.5);
	\draw (0.75,-0.5)[fill=black]  circle (0.3ex);
	\draw (0.75,0.5,0)[fill=black]  circle (0.3ex);
	\draw (0,0) [color=black, directed, thick, rounded corners=8pt]+(1.,0.5) -- +(1,-0.5);
	\draw (1.,-0.5)[fill=black]  circle (0.3ex);
	\draw (1.,0.5,0)[fill=black]  circle (0.3ex);
	\draw (1.25,0) node {,};
	\end{tikzpicture}
\end{minipage}%
\begin{minipage}{0.25\linewidth}
	\begin{tikzpicture}[scale=1]
	\draw (-1.,0) node {$P_2^{[3]}=$};
	\draw (0,0) [-,color=black, thick,directed, rounded corners=8pt]+(0,-0.5) -- +(0.25,-0.45) -- +(0.5,-0.25);
	\draw (0,0) [-,color=black, thick, directed, rounded corners=8pt]+(0,0.) --  +(0.25,-0.05)-- +(0.5,-0.25);

\tikzstyle directed=[postaction={decorate,decoration={markings, mark=at position .45 with {\arrow[arrowstyle]{stealth}}}}]
	\draw (0,0) [color=black, directed, thick, rounded corners=8pt]+(0.75,0.5) -- +(0.5,-0.25);
\tikzstyle directed=[postaction={decorate,decoration={markings, mark=at position .65 with {\arrow[arrowstyle]{stealth}}}}]

	\draw (0,0) [-,color=black, thick,directed, rounded corners=8pt]+(0,0.5) -- +(0.25,0.4) -- +(0.55,0.1);
	\draw (0,0) [-,color=black, thick, rounded corners=8pt]+(0.65,-0.05) -- +(0.75,-0.3) -- +(0.75,-0.5);
	
	\draw (0.,0.5)[fill=black]  circle (0.3ex);
	\draw (0,0)[fill=black]  circle (0.3ex);
	\draw (0.,-0.5)[fill=black]  circle (0.3ex);
	\draw (0.75,-0.5)[fill=black]  circle (0.3ex);
	\draw (0.75,0.5,0)[fill=black]  circle (0.3ex);
	\draw (0,0) [color=black, directed, thick, rounded corners=8pt]+(1.,0.5) -- +(1,-0.5);
	\draw (1.,-0.5)[fill=black]  circle (0.3ex);
	\draw (1.,0.5,0)[fill=black]  circle (0.3ex);
	\draw (1.25,0) node {,};
	\end{tikzpicture}
\end{minipage}%
\begin{minipage}{0.25\linewidth}
	\begin{tikzpicture}[scale=1]
	\draw (-1.,0) node {$P_3^{[3]}=$};
	\draw (0,0) [-,color=black, thick,directed, rounded corners=8pt]+(0,0.5) -- +(0.4,0.4) -- +(0.5,0.);
	\draw (0,0) [-,color=black, thick, directed, rounded corners=8pt]+(0,0.) --  +(0.5,0.);
	\draw (0,0) [-,color=black, thick,directed, rounded corners=8pt]+(0,-0.5) -- +(0.4,-0.4) -- +(0.5,0.);
	\draw (0.,0.5)[fill=black]  circle (0.3ex);
	\draw (0,0)[fill=black]  circle (0.3ex);
	\draw (0.,-0.5)[fill=black]  circle (0.3ex);
\tikzstyle directed=[postaction={decorate,decoration={markings, mark=at position .65 with {\arrow[arrowstyle]{stealth}}}}]
	\draw (0,0) [-,color=black,  directed, thick,  rounded corners=8pt]+(0.75,0.5) -- +(0.85,0.05);
	\draw (0,0) [color=black,  thick,  rounded corners=8pt]+(0.9,-0.05) -- +(1.,-0.5);

	\draw (0.75,-0.5)[fill=black]  circle (0.3ex);
	\draw (0.75,0.5,0)[fill=black]  circle (0.3ex);
\tikzstyle directed=[postaction={decorate,decoration={markings, mark=at position .3 with {\arrow[arrowstyle]{stealth}}}}]
	\draw (0,0) [color=black, directed, thick, rounded corners=8pt]+(1.,0.5) -- +(0.75,-0.5);
\tikzstyle directed=[postaction={decorate,decoration={markings, mark=at position .65 with {\arrow[arrowstyle]{stealth}}}}]
	\draw (1.,-0.5)[fill=black]  circle (0.3ex);
	\draw (1.,0.5,0)[fill=black]  circle (0.3ex);
	\draw (1.25,0) node {,};
	\end{tikzpicture}
\end{minipage}%
\begin{minipage}{0.25\linewidth}
	\begin{tikzpicture}[scale=1]
	\draw (-1.,0) node {$P_4^{[3]}=$};
	\draw (0,0) [-,color=black, thick,directed, rounded corners=8pt]+(0,-0.5) -- +(0.25,-0.45) -- +(0.5,-0.25);
	\draw (0,0) [-,color=black, thick, directed, rounded corners=8pt]+(0,0.) --  +(0.25,-0.05)-- +(0.5,-0.25);
	
\tikzstyle directed=[postaction={decorate,decoration={markings, mark=at position .85 with {\arrow[arrowstyle]{stealth}}}}]
	\draw (0,0) [color=black, directed, thick, rounded corners=8pt]+(1.,0.5) -- +(0.5,-0.25);
\tikzstyle reverse directed=[postaction={decorate,decoration={markings, mark=at position .65 with {\arrowreversed[arrowstyle]{stealth};}}}]
	\draw (0,0) [-,color=black, thick,directed, rounded corners=8pt]+(0,0.5) -- +(0.25,0.4) -- +(0.67,0.16);
	\draw (0,0) [-,color=black, thick, rounded corners=8pt]+(0.8,0.08) -- +(1,-0.3) -- +(1.,-0.5);
	
	\draw (0.,0.5)[fill=black]  circle (0.3ex);
	\draw (0,0)[fill=black]  circle (0.3ex);
	\draw (0.,-0.5)[fill=black]  circle (0.3ex);
	\draw (1.,-0.5)[fill=black]  circle (0.3ex);
	\draw (1.,0.5)[fill=black]  circle (0.3ex);
\tikzstyle directed=[postaction={decorate,decoration={markings, mark=at position .75 with {\arrow[arrowstyle]{stealth}}}}]
	\draw (0,0) [color=black, directed, thick, rounded corners=8pt]+(0.75,0.5) -- +(0.75,-0.5);
\tikzstyle reverse directed=[postaction={decorate,decoration={markings, mark=at position .65 with {\arrowreversed[arrowstyle]{stealth};}}}]
	\draw (0.75,-0.5)[fill=black]  circle (0.3ex);
	\draw (0.75,0.5,0)[fill=black]  circle (0.3ex);
	\draw (1.25,0) node {,};
	\end{tikzpicture}
\end{minipage}
\nonumber
\en
\eq
\begin{minipage}{0.25\linewidth}
	\begin{tikzpicture}[scale=1]
	\draw (-1.,0) node {$P_5^{[3]}=$};
	\draw (0,0) [-,color=black, thick,directed, rounded corners=8pt]+(0,-0.5) -- +(0.25,-0.45) -- +(0.5,-0.25);
	\draw (0,0) [-,color=black, thick, directed, rounded corners=8pt]+(0,0.) --  +(0.25,-0.05)-- +(0.5,-0.25);
	
\tikzstyle directed=[postaction={decorate,decoration={markings, mark=at position .55 with {\arrow[arrowstyle]{stealth}}}}]
	\draw (0,0) [color=black, directed, thick, rounded corners=8pt]+(0.75,0.5) -- +(0.5,-0.25);
	\draw (0,0) [-,color=black, thick,directed, rounded corners=8pt]+(0,0.5) -- +(0.25,0.4) -- +(0.55,0.1);
\tikzstyle reverse directed=[postaction={decorate,decoration={markings, mark=at position .65 with {\arrowreversed[arrowstyle]{stealth};}}}]
	\draw (0,0) [-,color=black, thick, rounded corners=8pt]+(0.7,-0.05) -- +(0.85,-0.2) -- +(1.,-0.5);
	
	\draw (0.,0.5)[fill=black]  circle (0.3ex);
	\draw (0,0)[fill=black]  circle (0.3ex);
	\draw (0.,-0.5)[fill=black]  circle (0.3ex);
	\draw (0.75,-0.5)[fill=black]  circle (0.3ex);
	\draw (0.75,0.5,0)[fill=black]  circle (0.3ex);
	\draw (0,0) [color=black, directed, thick, rounded corners=8pt]+(1.,0.5) -- +(0.75,-0.5);
	\draw (1.,-0.5)[fill=black]  circle (0.3ex);
	\draw (1.,0.5,0)[fill=black]  circle (0.3ex);
	\draw (1.25,0) node {,};
	\end{tikzpicture}
\end{minipage}%
\begin{minipage}{0.25\linewidth}
	\begin{tikzpicture}[scale=1]
	\draw (-1.,0) node {$P_6^{[3]}=$};
	\draw (0,0) [-,color=black, thick,directed, rounded corners=8pt]+(0,-0.5) -- +(0.25,-0.45) -- +(0.5,-0.25);
	\draw (0,0) [-,color=black, thick, directed, rounded corners=8pt]+(0,0.) --  +(0.25,-0.05)-- +(0.5,-0.25);
	
	\draw (0,0) [color=black, directed, thick, rounded corners=8pt]+(1.,0.5) -- +(0.5,-0.25);
	\draw (0,0) [-,color=black, thick,directed, rounded corners=8pt]+(0,0.5) -- +(0.25,0.4) -- +(0.6,0.05);
	\draw (0,0) [-,color=black, thick, rounded corners=8pt]+(0.71,-0.16) -- +(0.75,-0.3) -- +(0.75,-0.5);
	
	\draw (0.,0.5)[fill=black]  circle (0.3ex);
	\draw (0,0)[fill=black]  circle (0.3ex);
	\draw (0.,-0.5)[fill=black]  circle (0.3ex);
	\draw (0.75,-0.5)[fill=black]  circle (0.3ex);
	\draw (0.75,0.5,0)[fill=black]  circle (0.3ex);
	\draw (0,0) [color=black, directed, thick, rounded corners=8pt]+(0.75,0.5) -- +(1,-0.5);
	\draw (1.,-0.5)[fill=black]  circle (0.3ex);
	\draw (1.,0.5,0)[fill=black]  circle (0.3ex);
	\draw (1.25,0) node {,};
\end{tikzpicture}
\end{minipage}%
\begin{minipage}{0.25\linewidth}
	\begin{tikzpicture}[scale=1]
	\draw (-1.,0) node {$P_7^{[3]}=$};
	\draw (0,0) [-,color=black, thick,directed, rounded corners=8pt]+(0,-0.5) -- +(0.4,-0.35) -- +(0.75,-0.5);
	\draw (0,0) [-,color=black, thick, directed, rounded corners=8pt]+(0,0.) --  +(0.25,0.05)-- +(0.45,0.25);
	\draw (0,0) [-,color=black, thick,directed, rounded corners=8pt]+(0,0.5) -- +(0.25,0.45) -- +(0.45,0.25);
	
	\draw (0.,0.5)[fill=black]  circle (0.3ex);
	\draw (0,0)[fill=black]  circle (0.3ex);
	\draw (0.,-0.5)[fill=black]  circle (0.3ex);
	\draw (0,0) [color=black, directed, thick, rounded corners=8pt]+(0.75,0.5) -- +(0.45,0.25);
	\draw (0.75,-0.5)[fill=black]  circle (0.3ex);
	\draw (0.75,0.5,0)[fill=black]  circle (0.3ex);
	\draw (1.25,0) node {,};
	\draw (0,0) [color=black, directed, thick, rounded corners=8pt]+(1.,0.5) -- +(1,-0.5);
	\draw (1.,-0.5)[fill=black]  circle (0.3ex);
	\draw (1.,0.5,0)[fill=black]  circle (0.3ex);
	\draw (1.25,0) node {,};
	\end{tikzpicture}
\end{minipage}%
\begin{minipage}{0.25\linewidth}
	\begin{tikzpicture}[scale=1]
	\draw (-1.,0) node {$P_8^{[3]}=$};
	\draw (0,0) [-,color=black, thick, reverse directed, rounded corners=8pt]+(0.75,-0.5) -- +(0.45,-0.25) --+(0.,0.);
	\draw (0,0) [-,color=black, thick, reverse directed, rounded corners=8pt]+(1,-0.5) --  +(0.45,0.15)-- +(0.,0.5);
	\draw (0,0) [-,color=black, thick, directed, rounded corners=8pt]+(0,-0.5) -- +(0.35,-0.45) -- +(0.87,0.15);
	
	\draw (0.,0.5)[fill=black]  circle (0.3ex);
	\draw (0,0)[fill=black]  circle (0.3ex);
	\draw (0.,-0.5)[fill=black]  circle (0.3ex);
	\draw (0.75,-0.5)[fill=black]  circle (0.3ex);
	\draw (0.75,0.5,0)[fill=black]  circle (0.3ex);
	\draw (1.25,0) node {,};
	\draw (0,0) [color=black, directed, thick, rounded corners=8pt]+(1,0.5) -- +(0.87,0.15);
	\draw (0,0) [color=black, directed, thick, rounded corners=8pt]+(0.75,0.5) -- +(0.87,0.15);
	\draw (1.,-0.5)[fill=black]  circle (0.3ex);
	\draw (1.,0.5,0)[fill=black]  circle (0.3ex);
	\draw (1.25,0) node {,};
	\end{tikzpicture}
\end{minipage}
\nonumber
\en
\eq
\begin{minipage}{0.25\linewidth}
	\begin{tikzpicture}[scale=1]
	\draw (-1.,0) node {$P_9^{[3]}=$};
	\draw (0,0) [-,color=black, thick,directed, rounded corners=8pt]+(0,-0.5) -- +(0.5,-0.25) -- +(1.,-0.5);
	\draw (0,0) [-,color=black, thick, directed, rounded corners=8pt]+(0,0.) --  +(0.25,0.05)-- +(0.45,0.25);
	\draw (0,0) [-,color=black, thick,directed, rounded corners=8pt]+(0,0.5) -- +(0.25,0.45) -- +(0.45,0.25);
	
	\draw (0.,0.5)[fill=black]  circle (0.3ex);
	\draw (0,0)[fill=black]  circle (0.3ex);
	\draw (0.,-0.5)[fill=black]  circle (0.3ex);
\tikzstyle directed=[postaction={decorate,decoration={markings, mark=at position .85 with {\arrow[arrowstyle]{stealth}}}}]
	\draw (0,0) [color=black, directed, thick, rounded corners=8pt]+(1.,0.5) --+(0.75,0.15) -- +(0.45,0.25);
	\draw (1.,-0.5)[fill=black]  circle (0.3ex);
	\draw (1.,0.5,0)[fill=black]  circle (0.3ex);
	\draw (1.25,0) node {,};
\tikzstyle directed=[postaction={decorate,decoration={markings, mark=at position .6 with {\arrow[arrowstyle]{stealth}}}}]
	\draw (0,0) [color=black, directed, thick, rounded corners=8pt]+(0.75,0.5) -- +(0.75,-0.5);
\tikzstyle reverse directed=[postaction={decorate,decoration={markings, mark=at position .65 with {\arrowreversed[arrowstyle]{stealth};}}}]
	\draw (0.75,-0.5)[fill=black]  circle (0.3ex);
	\draw (0.75,0.5,0)[fill=black]  circle (0.3ex);
	\draw (1.25,0) node {,};
	\end{tikzpicture}
\end{minipage}%
\begin{minipage}{0.25\linewidth}
	\begin{tikzpicture}[scale=1]
	\draw (-1.,0) node {$P_{10}^{[3]}=$};
	\draw (0,0) [-,color=black, thick,directed, rounded corners=8pt]+(0,-0.5) -- +(0.5,-0.25) -- +(1.,-0.5);
	\draw (0,0) [-,color=black, thick, directed, rounded corners=8pt]+(0,0.) --  +(0.25,0.05)-- +(0.45,0.25);
	\draw (0,0) [-,color=black, thick,directed, rounded corners=8pt]+(0,0.5) -- +(0.25,0.45) -- +(0.45,0.25);
	
	\draw (0.,0.5)[fill=black]  circle (0.3ex);
	\draw (0,0)[fill=black]  circle (0.3ex);
	\draw (0.,-0.5)[fill=black]  circle (0.3ex);
	\draw (0,0) [color=black, directed, thick, rounded corners=8pt]+(0.75,0.5) -- +(0.45,0.25);
	\draw (0.75,-0.5)[fill=black]  circle (0.3ex);
	\draw (0.75,0.5,0)[fill=black]  circle (0.3ex);
	\draw (1.25,0) node {,};
	\draw (0,0) [color=black, directed, thick, rounded corners=8pt]+(1.,0.5) -- +(0.75,-0.5);
	\draw (1.,-0.5)[fill=black]  circle (0.3ex);
	\draw (1.,0.5,0)[fill=black]  circle (0.3ex);
	\draw (1.25,0) node {,};
	\end{tikzpicture}
\end{minipage}%
\begin{minipage}{0.25\linewidth}
	\begin{tikzpicture}[scale=1]
	\draw (-1.,0) node {$P_{11}^{[3]}=$};
	\draw (0,0) [-,color=black, thick,directed, rounded corners=8pt]+(0,-0.5) -- +(0.5,-0.25) -- +(0.75,-0.5);
	\draw (0,0) [-,color=black, thick, directed, rounded corners=8pt]+(0,0.) --  +(0.25,0.05)-- +(0.45,0.25);
	\draw (0,0) [-,color=black, thick,directed, rounded corners=8pt]+(0,0.5) -- +(0.25,0.45) -- +(0.45,0.25);
	
	\draw (0.,0.5)[fill=black]  circle (0.3ex);
	\draw (0,0)[fill=black]  circle (0.3ex);
	\draw (0.,-0.5)[fill=black]  circle (0.3ex);
\tikzstyle directed=[postaction={decorate,decoration={markings, mark=at position .85 with {\arrow[arrowstyle]{stealth}}}}]
	\draw (0,0) [color=black, directed, thick, rounded corners=8pt]+(1.,0.5) --+(0.75,0.15) -- +(0.45,0.25);
\tikzstyle reverse directed=[postaction={decorate,decoration={markings, mark=at position .65 with {\arrowreversed[arrowstyle]{stealth};}}}]
	\draw (1.,-0.5)[fill=black]  circle (0.3ex);
	\draw (1.,0.5,0)[fill=black]  circle (0.3ex);
	\draw (1.25,0) node {,};
	\draw (0,0) [color=black, directed, thick, rounded corners=8pt]+(0.75,0.5) -- +(1.,-0.5);
	\draw (0.75,-0.5)[fill=black]  circle (0.3ex);
	\draw (0.75,0.5,0)[fill=black]  circle (0.3ex);
	\draw (1.25,0) node {,};
	\end{tikzpicture}
\end{minipage}%
\nonumber
\en
and the functions $\rho_{k}^{[3]}(\lambda_1,\lambda_2)$ are to be
determined.  Like in the two-point case, we have conveniently chosen the
operators as tensor products of the totally anti-symmetric tensor and the
identity/Kronecker symbol
to resemble identity, permutation and Temperley-Lieb acting in
different spaces as $I, P_{12}, P_{23}, P_{13}, P_{12} P_{23},$ $ P_{23} P_{12},
E_{12}, E_{13}$, $ E_{12}P_{23}, P_{23}E_{12}$ plus one operator which due to
symmetry must allow for a symmetric combination of the indices in the first column
of dots/indices as given in $P_8^{[3]}$. 
 We have checked that the above chosen operators are indeed linearly independent.
Again, after partial
anti-symmetrization, the density operator (\ref{D3frak}) can be reduced to
the usual density operator for three-sites
$D_3(\lambda_1,\lambda_2,\lambda_3)=(2\rho^{[3]}_1 -\rho^{[3]}_7-\rho^{[3]}_9) I +(2\rho^{[3]}_2+\rho^{[3]}_7) P_{12} + (2\rho^{[3]}_3-\rho^{[3]}_{10}-\rho^{[3]}_{11}) P_{23} + (2\rho^{[3]}_4-\rho^{[3]}_8+\rho^{[3]}_9) P_{13} + (2\rho^{[3]}_5 +\rho^{[3]}_8 +\rho^{[3]}_{10}) P_{12} P_{23} + (2\rho^{[3]}_6+\rho^{[3]}_{11}) P_{23} P_{12}$ by means of the use of the properties (\ref{other2}).

Analogously to the two-sites case, the operator $P_j^{[3]}$ can be seen as vector with the indices $i_1,i_2,i_3,r_1,r_2,s_1,s_2=1,2,3$ assigned to the dots, such that e.g.
for $P_1^{[3]}$ we have
\eq
\	\begin{tikzpicture}[scale=1]
\draw (-6.25,0) node {$\left(P_1^{[3]}\right)_{i_1,i_2,i_3,r_1,r_2}^{s_1,s_2}
:=\quad\epsilon_{i_1,i_2,i_3}\delta_{r_1}^{s_1}\delta_{r_2}^{s_2} \quad
\equiv$};
\draw (0,0) [-,color=black, thick,directed, rounded corners=8pt]+(0,0.5) -- +(0.4,0.4) -- +(0.5,0.);
\draw (0,0) [-,color=black, thick, directed, rounded corners=8pt]+(0,0.) --  +(0.5,0.);
\draw (0,0) [-,color=black, thick,directed, rounded corners=8pt]+(0,-0.5) -- +(0.4,-0.4) -- +(0.5,0.);
\draw (0.,0.5)[fill=black]  circle (0.3ex);
\draw (0,0)[fill=black]  circle (0.3ex);
\draw (0.,-0.5)[fill=black]  circle (0.3ex);
\draw (-0.25,0.5) node {$i_1$};
\draw (-0.25,0) node {$i_2$};
\draw (-0.25,-0.5) node {$i_3$};
\draw (0.75,0.75) node {$r_1$};
\draw (0.75,-0.75) node {$s_1$};
\draw (1.25,0.75) node {$r_2$};
\draw (1.25,-0.75) node {$s_2$};

\draw (0,0) [color=black, directed, thick, rounded corners=8pt]+(0.75,0.5) -- +(0.75,-0.5);
\draw (0.75,-0.5)[fill=black]  circle (0.3ex);
\draw (0.75,0.5,0)[fill=black]  circle (0.3ex);
\draw (0,0) [color=black, directed, thick, rounded corners=8pt]+(1.25,0.5) -- +(1.25,-0.5);
\draw (1.25,-0.5)[fill=black]  circle (0.3ex);
\draw (1.25,0.5,0)[fill=black]  circle (0.3ex);
\draw (1.55,0) node {,};
\end{tikzpicture}
\en
which shows that the computation for three-sites goes along the same lines as in the two-sites case, we just have to deal with a large number  of extended operators $P_j^{[3]}$.
 
Inserting the above expansion of the density operator (\ref{D3frak}) into equation
(\ref{qkzsun-frak}) yields the set of functional equations
\eq
\vec{\rho}(\lambda_1,\lambda_2,\lambda_3)= A^{[3]}(\lambda_1,\lambda_2)\cdot\vec{\rho}(\lambda_1+1,\lambda_2,\lambda_3),
\label{funct11}
\en 
where $\vec{\rho}(\lambda_1,\lambda_2,\lambda_3)$ is a $11$ dimensional
vector whose entries are the expansion coefficients
$\rho_k^{[3]}(\lambda_1,\lambda_2,\lambda_3)$ for $k=1,\dots,11$ and the
matrix $A^{[3]}(\lambda_1,\lambda_2,\lambda_3)$,
which is obtained from the action of the linear operator ${\mathfrak A}_3(\lambda_1,\lambda_2,\lambda_3)$ (\ref{frakA}) on the mixed density operator (\ref{D3frak}), is given in appendix B.

For convenience of the presentation of the results,
we define the intermediate auxiliary functions
$f_k(\lambda_1,\lambda_2,\lambda_3)= ({P}_k^{[3]})^t\cdot {\mathfrak D}_3(\lambda_1,\lambda_2,\lambda_3)$ 
by 
\eq
\vec{f}(\lambda_1,\lambda_2,\lambda_3)= M \cdot \vec{\rho}(\lambda_1,\lambda_2,\lambda_3),
\label{aux}
\en
where the matrix $M$ is given by
\eq
M =\left(\begin{array}{ccccccccccc} 
54 & 18 & 18 & 18 & 6 & 6 & 18 & 6 & 18 & 6 & 6 \\
18 & 54 & 6 & 6 & 18 & 18 & -18 & -6 & 6 & -6 & -6 \\
18 & 6 & 54 & 6 & 18 & 18 & 6 & -6 & 6 & 18 & 18 \\
18 & 6 & 6 & 54 & 18 & 18 & 6 & 18 & -18 & -6 & -6 \\
6 & 18 & 18 & 18 & 54 & 6 & -6 & -18 & -6 & -18 & 6 \\
6 & 18 & 18 & 18 & 6 & 54 & -6 & 6 & -6 & 6 & -18 \\
18 & -18 & 6 & 6 & -6 & -6 & 54 & -6 & 6 & 18 & 18 \\
6 & -6 & -6 & 18 & -18 & 6 & -6 & 54 & -6 & 6 & 6 \\
18 & 6 & 6 & -18 & -6 & -6 & 6 & -6 & 54 & 18 & 18 \\
6 & -6 & 18 & -6 & -18 & 6 & 18 & 6 & 18 & 54 & 6 \\
6 & -6 & 18 & -6 & 6 & -18 & 18 & 6 & 18 & 6 & 54
\end{array}\right).
\en 

The equations (\ref{funct11}) can be disentangled by making a suitable
transformation. This can be done by using the reduction properties like the
intertwining symmetry, the partial trace and so on in order to 
identify 8 linearly independent combinations of the functions
$f_k(\lambda_1,\lambda_2,\lambda_3)$ as two-site functions (or simpler) and
3 remaining combinations as true three-site functions.  Therefore, the
suitable functions can be written in terms of the auxiliary functions as follows,
\bear
&&1=f_1, \nonumber\\
&&\omega_{33}(\lambda_1,\lambda_2)=f_2, \nonumber\\
&&\omega_{\bar{3}3}(\lambda_1-1,\lambda_2)=f_7, \nonumber\\
&&\omega_{33}(\lambda_1,\lambda_3) (1-y^2)= f_3 + y f_6 - y f_5 -y^2 f_4, \nonumber\\
&&\omega_{\bar{3}3}(\lambda_1-1,\lambda_3) (1-y)(2+y)=  f_7 -(y+2) f_{10} + (y-1)f_{11} - (y-1)(y+2)f_9,  \nonumber \\ 
&&\omega_{33}(\lambda_1,\lambda_2) (1-x^2) (1-(x-y)^2) =(1 - (x - y)^2) f_3 + x(x - 2y) f_4  \nonumber\\
&&+ x(-1 + xy - y^2) f_5 + x(1 - xy + y^2) f_6 + x(x - y)(-2 + x^2 - xy) f_2, \nonumber\\
&&\omega_{\bar{3}3}(\lambda_1-1,\lambda_2) (1-x)(2+x)(1-(x-y)^2)= \nonumber\\
&&(1 - (-1 + x)(x - y) - (2 + x)(x - y) + (-1 + x)(2 + x)(x - y)^2) f_7 \nonumber\\
&&+ (2 - y - y^2) f_9 + (2 + y)(-1 + x^2 + y - x(1 + y)) f_{10} \nonumber\\
&& + (-1 + y)(1 - x^2 + x(-2 + y) + 2y)  f_{11}, \nonumber\\
&&\omega_{33}(\lambda_2,\lambda_3)=f_3, \nonumber \\
&&F_1(\lambda_1,\lambda_2,\lambda_3)=2 x (2 + x) y (2 + y) f_1 + 2 x (2 + x) (2 + y) f_2 \nonumber \\
&& + 2 (2 + x) y (2 + y) f_4  + 2 (2 + x) (2 + y) f_5, \label{inter}\\ 
&&F_2(\lambda_1,\lambda_2,\lambda_3)= 2 (-2 - y - x (2 + y) + x (2 + x) y (2 + y)) f_1 \nonumber\\
&& - 2 (-1 + x + x^2) (2 + y) f_2 + 2 (1 + x) (2 + y) f_3  \nonumber\\
&& +  2 (1 + x + (2 + x) y - (2 + x) y (2 + y)) f_4 -  2 (1 + x + 2 y + x y) f_5 \nonumber\\  
&& + 2 (2 + y) f_6 - 2 (-2 - y + x y + x^2 (1 + y)) f_7 + 2 (1 + x - y) f_8  \nonumber\\
&& - 2 (1 + x) (-2 + y + y^2) f_9 - 2 (1 + x) (2 + y) f_{10} -  2 x f_{11}, \nonumber \\
&&F_3(\lambda_1,\lambda_2,\lambda_3)= 2 (x^2-1) (y^2-1)f_1 +  2 (x^2-1) (1 + y) f_7 \nonumber \\
&&+ 2 (1 + x) (y^2-1) f_9  + 2 (1 + x) (1 + y) f_{10}, \nonumber 
\ear
where $x=\lambda_1-\lambda_3$ and $y=\lambda_1-\lambda_2$ and the combination $f_1(\lambda_1,\lambda_2,\lambda_3)$ is the normalization condition (analogue of the total trace). 

Substituting 
Eq. (\ref{aux}) in Eqs. (\ref{inter}) and solving the 11 linear
equations for $\rho_1^{[3]},...,\rho_{11}^{[3]}$ in terms of
the known two-site functions and the yet unknown three-site functions $F_1,
F_2, F_3$, inserting these expressions into the functional equations
(\ref{funct11}) and using those for the two-point functions, 
reduces these to just three equations for the unknown functions.
So the only remaining unknown functions are $F_1, F_2, F_3$ and satisfy a set
of linear functional equations. By a suitable rescaling the coefficients of
the occurring matrix take a very simple form
\eq
\left(\begin{array}{c}
	G_1(\lambda_1,\lambda_2,\lambda_3) \\
	G_2(\lambda_1,\lambda_2,\lambda_3) \\
	G_3(\lambda_1,\lambda_2,\lambda_3)
\end{array}\right)
=\left(\begin{array}{ccc}
	0 & 0  & 1\\
	1 & 0  & 0 \\
	0 & 1  & 0
\end{array}\right) \cdot \left(\begin{array}{c}
	G_1(\lambda_1+1,\lambda_2,\lambda_3) \\
	G_2(\lambda_1+1,\lambda_2,\lambda_3) \\
	G_3(\lambda_1+1,\lambda_2,\lambda_3)
\end{array}\right)
+
\left(\begin{array}{c}
	r(\lambda_1,\lambda_2,\lambda_3) \\
	0 \\
	0
\end{array}\right),
\label{3ptseq}
\en
where $\lambda_1=u_i$, and we have introduced for convenience the $G_k$-functions as,
\bear
	G_1(\lambda_1,\lambda_2,\lambda_3)&=& \frac{x y}{(x^2-1)(y^2-1) (x+2)(y+2)} F_1(\lambda_1,\lambda_2,\lambda_3), \nonumber \\
	G_2(\lambda_1,\lambda_2,\lambda_3)&=& \frac{(x+1)(y+1)}{(x^2-1)(y^2-1)(x+2)(y+2)} F_2(\lambda_1,\lambda_2,\lambda_3), \\
	G_3(\lambda_1,\lambda_2,\lambda_3)&=& \frac{1}{(x^2-1)(y^2-1)} F_3(\lambda_1,\lambda_2,\lambda_3), \nonumber
\ear
 and
\bear
r(\lambda_1,\lambda_2,\lambda_3)& =& \frac{2 (-1 + 2 x^2 + 2 y^2)}{(x^2-1)(y^2-1)} + \frac{2(x + y)}{(x^2-1)(y^2-1)} \omega_{33}(\lambda_2, \lambda_3)  \\
&+&\frac{2 (-1 + 3 x + x^2 - 3 y - 2 x y + y^2 - 3 x y^2 + 3 y^3)}{x ( x+3)(x - y)(y^2-1)} \omega_{\bar{3}3}(\lambda_1, \lambda_3)  \nonumber \\
&-& \frac{2(-1 - 3 x + x^2 + 3 x^3 + 3 y - 2 x y - 3 x^2 y + y^2)}
{(x^2-1)(x - y) y ( y+3)} \omega_{\bar{3}3}(\lambda_1, \lambda_2). \nonumber
\ear

\subsubsection{Zero temperature solution}

At zero temperature, again, the above functional equations hold
for arbitrary $\lambda_1$. Since we have already obtained the solution for
the two-site functions from Eq.~(\ref{qKZ1}-\ref{qKZ2}), it only remains to
solve equation (\ref{3ptseq}).

However, equations (\ref{3ptseq}) are more complicated to deal
with, since one of the equations contains the inhomogeneity
$r(\lambda_1,\lambda_2,\lambda_3)$ with a more complicated pole structure
than in the two-site case.  The inhomogeneity can be written in terms of
digamma functions. This increases the complexity to obtain a closed solution
for (\ref{3ptseq}).

In order to avoid carrying out, for the time being, the Fourier
transform of rational functions times digamma functions, we have chosen to
write the solution in terms of convolutions. Eventually, the convolution
integrals can be evaluated numerically as we will describe in what follows.

The problem can be significantly simplified by  partially taking the homogeneous limit $\lambda_2=\lambda_3=0$ and decoupling the equations (\ref{3ptseq}) by the following transformation
\bear
g_0(\lambda_1) &=& G_1(\lambda_1,0,0) + G_2(\lambda_1,0,0) +G_3(\lambda_1,0,0), \nonumber\\
g_{1}(\lambda_1) &=& G_1(\lambda_1,0,0) +w G_2(\lambda_1,0,0) + w^2 G_3(\lambda_1,0,0), \\
g_{-1}(\lambda_1) &=& G_1(\lambda_1,0,0) +w^{-1} G_2(\lambda_1,0,0) + w^{-2} G_3(\lambda_1,0,0), \nonumber
\ear
where $w=e^{\frac{2\pi\im}{3}}$. 

Therefore, the resulting equations become (we now set
  $\lambda_1=\lambda$)
\bear
g_l(\lambda) &=&w^l g_l(\lambda+1)  + \varphi(\lambda), 
\ear
where $l=0,1,-1$ and  $\varphi(\lambda)=\lim_{\lambda_2,\lambda_3\to
  0}r(\lambda,\lambda_2,\lambda_3)$,
\bear
\varphi(\lambda)&=&-\frac{12}{(\lambda^2-1)}
\omega_{33}(\lambda,0)-\frac{2}{(\lambda^2-1)^2} \omega_{33}'(\lambda,0)
+\frac{4\lambda}{(\lambda^2-1)^2}\omega_{33}(0,0)\nonumber\\
&& +\frac{2(4 \lambda^4 +6 \lambda^3-\lambda^2-6 \lambda-1)}{\lambda^2 (\lambda^2-1)^2},
\ear
and the prime denotes the derivative with respect to the argument $\lambda$.
Naturally the zero temperature solution for three-sites correlation must depend on the two-sites function $\omega_{33}(\lambda,0)$ and its derivative $\omega_{33}'(\lambda,0)$ via the $\varphi(\lambda)$, therefore the modified zeta function would also appear explicitly in case of an analytic solution for the three-site correlation.

We use analyticity in the variable $\lambda$ and Fourier
transform the above equations. The resulting equations are algebraically
solved for the Fourier coefficients and yield product expressions. Then, we
Fourier transform back and find integrals of convolution type
\bear
g_l(\lambda) &=& \int_{-\infty}^{\infty}  h_l(\lambda-\mu) \varphi(\mu) \frac{d\mu}{2\pi}, 
\label{gconv}
\ear
where 
\eq
h_l(z)= \int_{{\mathbb R} +\im 0} \frac{ e^{\im k z}}{1-w^{l}e^k} dk.
\en 
The integral expression can be evaluated numerically at the
homogeneous point $\lambda=0$. This allows us to obtain the functions
$G_i(0,0,0)$ (and derivatives of $G_i$ at $(0,0,0)$) from which we compute
$F_i(0,0,0)$ directly in the homogeneous limit. The function $F_1(0,0,0)$
is related to a simple three point correlation function $F_1(0,0,0)=8\langle
P_{12} P_{23} \rangle$. Using the result of the numerical evaluation of
the integral equation (\ref{gconv}), we obtain
\eq
\langle P_{12} P_{23} \rangle = 0.191368820116674 
\en
\begin{table}[h]
	\begin{center}
	\begin{tabular}{|l|l|l|}
		\hline  
		Length & $\omega_{33}(0,0)$  & $\langle P_{12} P_{23} \rangle$  \\ 
		\hline
		$L=3$ & $-1.000000000000000$ &  $1.000000000000000$ \\
		\hline
		$L=6$ & $-0.767591879243998$ &  $0.309579305659537$ \\
		\hline
		$L=9$ & $-0.731082881703061$ & $0.239661721591669$ \\
		\hline
		$L\rightarrow\infty$ & $-0.703212076746182$ & $0.191368820116674
		$ \\
		\hline 
	\end{tabular}
	\end{center}
\caption{Comparison of numerical results from exact
diagonalization for $L=3,6$ and Lanczos calculations for $L=9$
sites with the analytical result in the thermodynamic limit.}
\end{table}

The numerical data for finite lattices indicates an agreement with the
infinite lattice result obtained from the solution of the functional
equations, see Table 1. Although the numerical evaluation of the integral
equations (\ref{gconv}) is not computationally demanding, it would be
desirable to have an exact analytical expression. As indicated
above, the analytical calculation of the convolution integral requires the
computation of the Fourier transform of a product of a rational function
with a digamma functions which for the moment we leave as an exercise.

\section{Lack of factorization of the correlation functions}\label{lack}

In the case of $SU(2)$ spin chains, it is well known that the correlation
functions factorize in terms of two-site correlations \cite{BOKO01,BGKS}. This property was useful in
obtaining the correlation functions for the spin-1/2 system and also for
higher-spin cases \cite{KNS013,RK2016}.

Unfortunately, in the case of $SU(3)$ our attempts of solving
Eq.~(\ref{3ptseq}) in terms of some naive factorized ansatz failed. 

Besides, we have also investigated the factorization for finite Trotter number
and (at first sight) surprisingly we realized that the three-point correlation
functions are
expressed in terms of the two-sites ($m=2$) and also a three-sites ($m=3$)
emptiness formation probability (EFP). For instance, the following correlation
function is given by
\bear 
&&\tr{\left[P_{singlet} D_3(\lambda_1,\lambda_2,\lambda_3) \right]}=
\frac{Q_3^{(s)}(\lambda_1,\lambda_2,\lambda_3)}{\Lambda_0^{(n)}(\lambda_1)\Lambda_0^{(n)}(\lambda_2)\Lambda_0^{(n)}(\lambda_3)}
\nonumber \\ 
&=&6-24 \Big[ (1 +\frac{1}{\lambda_{13}\lambda_{23}})
  P_2(\lambda_1,\lambda_2) + (1 +\frac{1}{\lambda_{12}\lambda_{32}})
  P_2(\lambda_1,\lambda_3) \\
&+&(1 +\frac{1}{\lambda_{21}\lambda_{31}})
  P_2(\lambda_2,\lambda_3) \Big] + 60 P_3(\lambda_1,\lambda_2,\lambda_3), \nonumber 
\ear 
where $P_{singlet} $ is the $SU(3)$ singlet projector and 
\bear 
P_2(\lambda_1,\lambda_2)&=&
{\left[D_2^{(3,3)}(\lambda_1,\lambda_2)\right]}_{11}^{11}=\frac{Q_2(\lambda_1,\lambda_2)}{\Lambda_0^{(n)}(\lambda_1)\Lambda_0^{(n)}(\lambda_2)},
\\ P_3(\lambda_1,\lambda_2,\lambda_3)&=&{\left[D_3^{(3,3)}(\lambda_1,\lambda_2,\lambda_3))\right]}_{111}^{111}=\frac{Q_3(\lambda_1,\lambda_2,\lambda_3)}{\Lambda_0^{(n)}(\lambda_1)\Lambda_0^{(n)}(\lambda_2)\Lambda_0^{(n)}(\lambda_3)},
\ear 
and $Q_m(\lambda_1,\dots,\lambda_m)$ are polynomials of known degree in
each variable and $\Lambda_0^{(n)}(\lambda)$ is again the leading eigenvalue of the
quantum transfer matrix but with finite Trotter number $N$. 
It is worth to emphasize that $P_3(\lambda_1,\lambda_2,\lambda_3)$ cannot
be written only in terms of the $P_2(\lambda_i,\lambda_j)$, so it does not
factorize in terms of the two-point emptiness formation probability. The
correlations for the case $m=4$ 
also do not factorize only in terms of $m=2,3$ correlators, but require one
four-point function, e.g. the emptiness formation probability of four-sites
$P_4(\lambda_1,\lambda_2,\lambda_3,\lambda_4)$. 

Therefore, this might indicate that the correlations of high rank spin chains
cannot be factorized only in terms of two-point functions, which brings an
additional degree of difficulty in order to push the calculation to correlations at
longer distances.

\section{Conclusions}\label{conclusion}

We have formulated a consistent approach to deal with short-distance
correlation functions of the $SU(n)$ spin chains for $n>2$ at zero and
finite-temperature. The fact that the model does not have crossing symmetry
turned the derivation of functional equations for the correlation functions
much more challenging than in the $SU(2)$ case. The difficulties which arise
were circumvented by working with generalized density
operators of two types. These operators not only contain the physically
interesting correlation functions, but many other correlations with mixed
representations. In this sense, this approach exploits the full $SU(n)$ structure.

We considered in detail the special case of $SU(3)$ for two- and three-site
correlations ($m=2,3$). We used the discrete functional equations to obtain
the equations which fix the two-point correlation functions. Besides, we have
solved the equations via Fourier transform at zero temperature and its
solutions are explicitly given in terms of digamma functions.
The correlation function for the local Hamiltonian gives the ground state energy
as expected.

In addition, we considered the case of three-point correlations. The
computation is much more involved in this case, since the dimension of the
singlet space of the generalized mixed density operator is $11$. 
Therefore, we had to obtain this large number of equations and
by appropriate identification of two and three-site functions, we reduced
these $11$ equations to just 3 decoupled functional equations. We derived an
integral expression of convolution type for the remaining
three-point functions. The integrals were evaluated numerically giving the
result for three-point correlation functions in the thermodynamical limit. We
compared the infinite system size result with the result obtained for very
finite lattices $L=3,6$ and $9$, which shows the correct trend.

Moreover, we have also investigated the possibility of the three-point
function to be factorized in terms of two-point correlations. Our attempts
were based on the proposition of an ansatz for the solution of the functional
equations for the three-point functions (\ref{3ptseq}),
which always led us to contradictions of the proposed ansatz, indicating that the
factorized ansatz does not apply to the model. Additionally, we
investigated the possibility of factorization at finite Trotter number. At
finite Trotter number the correlators are rational functions, which allowed us
to realize that three-point correlations can be decomposed in terms of
two-point function and an additional three-point function. Therefore, this is
also another indication of lack of factorization.

In this work, we have obtained the first results about the
correlation functions of Yang-Baxter integrable $SU(n)$ quantum spin chains.  We
obtained analytical and numerical results for nearest ($m=2$) and
next-nearest ($m=3$) correlators. We would still like to obtain an
analytical evaluation of the convolution integrals for the $m=3$ case. It is
completely open, how to solve the functional equation for the cases of $m\ge 4$.
In the general case of $SU(n)$ ($n>3$), we have  only the solution  for $m=2$.
Of course, here it would also be desirable to obtain the explicit functional equations for $m=3$ and its solution, at least for $SU(4)$.  
Another interesting goal is the explicit evaluation of the correlations at finite temperature. In
order to do that, we have to derive the non-linear integral equation for
the generalized quantum transfer matrix and more challenging we have to
devise a way to evaluate the three-point functions at finite
temperature. The standard trick for the evaluation of the two-point
function at finite temperature comprises the derivative of the leading
eigenvalue with respect to some inhomogeneity parameter \cite{AuKl12},
however this trick cannot be applied to three-point correlations. The
above mentioned issues are currently under investigation.

\section*{Acknowledgments}

G.A.P. Ribeiro thanks the S\~ao Paulo Research Foundation (FAPESP) for financial support through the grants 2017/16535-1 and 2015/01643-8. He also acknowledges the hospitality of Bergische Universit\"at Wuppertal.
This  work  has  been  carried  out  within  the  DFG research  unit Correlations  in  Integrable  Quantum Many-Body Systems (FOR2316).

Note added: After our paper appeared on the arXiv, we became aware of the related preprint \cite{BOOS18}.

\newpage

\section*{\bf Appendix A: Reduction of ${\mathbb D}_m$ to ${\mathbb D}_{m-1}$}
\setcounter{equation}{0}
\renewcommand{\theequation}{A.\arabic{equation}}
\setcounter{figure}{0}
\renewcommand{\thefigure}{A.\arabic{figure}}

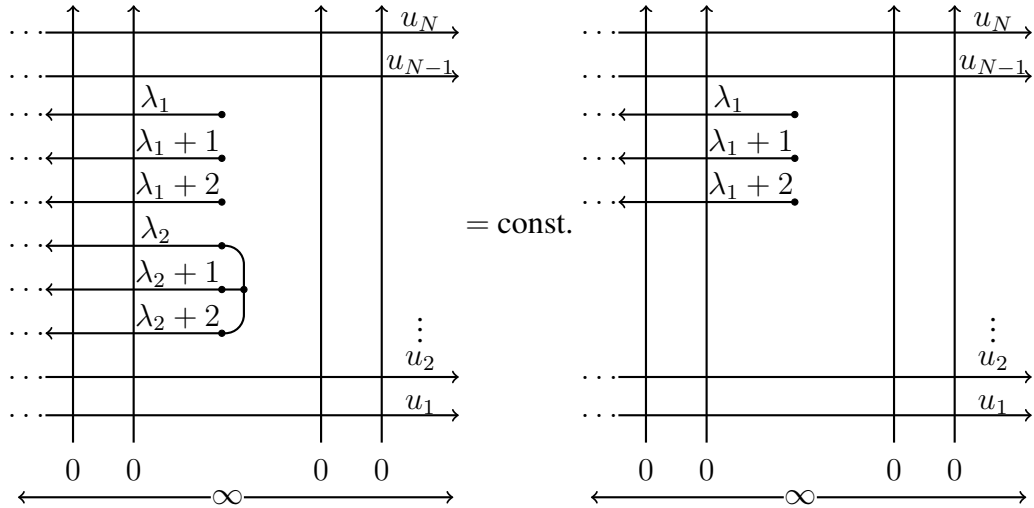
\begin{figure}[h]
	\begin{center}
\begin{minipage}{0.55\linewidth}
		\begin{tikzpicture}[scale=1.45]
		\draw (0,0) [->,color=black, thick, rounded corners=7pt] +(0.25,0)--(0.25,4);
		\draw (0,0) [->,color=black, thick, rounded corners=7pt] +(0.8,0)--(0.8,4);
		
		\draw (0,0) [<-,color=black, thick, rounded corners=7pt] +(0.,1.0)--(1.6,1.0);
		\draw (0,0) [<-,color=black, thick, rounded corners=7pt] +(0.,1.4)--(1.6,1.4);
		\draw (0,0) [<-,color=black, thick, rounded corners=7pt] +(0.,1.8)--(1.6,1.8);
		
		\draw (0,0) [<-,color=black, thick, rounded corners=7pt] +(0.,2.2)--(1.6,2.2);
		\draw (0,0) [<-,color=black, thick, rounded corners=7pt] +(0.,2.6)--(1.6,2.6);
		\draw (0,0) [<-,color=black, thick, rounded corners=7pt] +(0.,3.0)--(1.6,3.0);

		\draw (-0.15,0.25) node {$\dots$};
		\draw (-0.15,0.6) node {$\dots$};
		\draw (-0.15,3.35) node {$\dots$};
		\draw (-0.15,3.75) node {$\dots$};
\draw (-0.15,1) node {$\dots$};
\draw (-0.15,1.4) node {$\dots$};
\draw (-0.15,1.8) node {$\dots$};
\draw (-0.15,2.2) node {$\dots$};
\draw (-0.15,2.6) node {$\dots$};
\draw (-0.15,3.0) node {$\dots$};
		
	\draw (0,0) [-,color=black, thick, rounded corners=7pt] +(1.8,1.4)--(1.6,1.4);
	\draw (0,0) [-,color=black, thick, rounded corners=7pt]+(1.6,1.0) -- +(1.8,1.0) -- +(1.8,1.8)-- +(1.6,1.8);
		\draw (1.8,1.4)[fill=black]  circle (0.15ex);

		\draw (0,0) [->,color=black, thick, rounded corners=7pt] +(2.5,0)--(2.5,4);
		\draw (0,0) [->,color=black, thick, rounded corners=7pt] +(3.05,0)--(3.05,4);
		
		\draw (1.2,1.15) node {$\lambda_2+2$};
		\draw (1.2,1.55) node {$\lambda_2+1$};
		\draw (1.,1.95) node {$\lambda_2$};

		\draw (1.2,2.35) node {$\lambda_1+2$};
		\draw (1.2,2.75) node {$\lambda_1+1$};
		\draw (1.,3.15) node {$\lambda_1$};		
		
		\draw (0.25,-0.25) node {$0$};
		\draw (0.8,-0.25) node {$0$};
		\draw (2.5,-0.25) node {$0$};
		\draw (3.05,-0.25) node {$0$};
		
		\draw (0,0) [->,color=black, thick, rounded corners=7pt] +(1.8,-0.5)--(3.7,-0.5);
		\draw (1.65,-0.5) node {$\infty$};
		\draw (0,0) [<-,color=black, thick, rounded corners=7pt] +(-0.25,-0.5)--(1.5,-0.5);
		
		\draw (-0.25,0) [->,color=black, thick, rounded corners=7pt] +(0.25,0.25)--(3.75,0.25);
		\draw (-0.25,0) [->,color=black, thick, rounded corners=7pt] +(0.25,0.6)--(3.75,0.6);
		
		\draw (-0.25,0) [->,color=black, thick, rounded corners=7pt] +(0.25,3.35)--(3.75,3.35);
		\draw (-0.25,0) [->,color=black, thick, rounded corners=7pt] +(0.25,3.75)--(3.75,3.75);

		\draw (3.4,0.35) node {$u_1$};
		\draw (3.4,0.75) node {$u_2$};
		\draw (3.4,1.1) node {$\vdots$};
		\draw (3.4,3.85) node {$u_{N}$};
		\draw (3.4,3.45) node {$u_{N-1}$};
		\draw (1.6,1)[fill=black]  circle (0.15ex);
		\draw (1.6,1.4)[fill=black]  circle (0.15ex);
		\draw (1.6,1.8)[fill=black]  circle (0.15ex);
		
		\draw (1.6,2.2)[fill=black]  circle (0.15ex);
		\draw (1.6,2.6)[fill=black]  circle (0.15ex);
		\draw (1.6,3)[fill=black]  circle (0.15ex);

		\draw (4.3,2) node {$=$ const.};
		
		\end{tikzpicture}
\end{minipage}%
\begin{minipage}{0.45\linewidth}
		\begin{tikzpicture}[scale=1.45]
		\draw (0,0) [->,color=black, thick, rounded corners=7pt] +(0.25,0)--(0.25,4);
		\draw (0,0) [->,color=black, thick, rounded corners=7pt] +(0.8,0)--(0.8,4);
				
		\draw (0,0) [<-,color=black, thick, rounded corners=7pt] +(-0,2.2)--(1.6,2.2);
		\draw (0,0) [<-,color=black, thick, rounded corners=7pt] +(-0,2.6)--(1.6,2.6);
		\draw (0,0) [<-,color=black, thick, rounded corners=7pt] +(-0,3.0)--(1.6,3.0);
				
		\draw (-0.15,0.25) node {$\dots$};
		\draw (-0.15,0.6) node {$\dots$};
		\draw (-0.15,3.35) node {$\dots$};
		\draw (-0.15,3.75) node {$\dots$};
		\draw (-0.15,2.2) node {$\dots$};
		\draw (-0.15,2.6) node {$\dots$};
		\draw (-0.15,3.0) node {$\dots$};

		\draw (0,0) [->,color=black, thick, rounded corners=7pt] +(2.5,0)--(2.5,4);
		\draw (0,0) [->,color=black, thick, rounded corners=7pt] +(3.05,0)--(3.05,4);

		\draw (1.2,2.35) node {$\lambda_1+2$};
		\draw (1.2,2.75) node {$\lambda_1+1$};
		\draw (1.,3.15) node {$\lambda_1$};		
		
		\draw (0.25,-0.25) node {$0$};
		\draw (0.8,-0.25) node {$0$};
		\draw (2.5,-0.25) node {$0$};
		\draw (3.05,-0.25) node {$0$};
		
		\draw (0,0) [->,color=black, thick, rounded corners=7pt] +(1.8,-0.5)--(3.7,-0.5);
		\draw (1.65,-0.5) node {$\infty$};
		\draw (0,0) [<-,color=black, thick, rounded corners=7pt] +(-0.25,-0.5)--(1.5,-0.5);
		
		\draw (-0.25,0) [->,color=black, thick, rounded corners=7pt] +(0.25,0.25)--(3.75,0.25);
		\draw (-0.25,0) [->,color=black, thick, rounded corners=7pt] +(0.25,0.6)--(3.75,0.6);
		
		\draw (-0.25,0) [->,color=black, thick, rounded corners=7pt] +(0.25,3.35)--(3.75,3.35);
		\draw (-0.25,0) [->,color=black, thick, rounded corners=7pt] +(0.25,3.75)--(3.75,3.75);

		\draw (3.4,0.35) node {$u_1$};
		\draw (3.4,0.75) node {$u_2$};
		\draw (3.4,1.1) node {$\vdots$};
		\draw (3.4,3.85) node {$u_{N}$};
		\draw (3.4,3.45) node {$u_{N-1}$};

		\draw (1.6,2.2)[fill=black]  circle (0.15ex);
		\draw (1.6,2.6)[fill=black]  circle (0.15ex);
		\draw (1.6,3)[fill=black]  circle (0.15ex);

		\end{tikzpicture}
\end{minipage}
		\caption{Graphical illustration of the reduction property of two-sites to one site correlation. This property can be iterated until we reach the normalization condition of the generalized density operator ${\mathbb D}_2(\lambda_1,\lambda_2)$.}
		\label{reduction}
	\end{center}
\end{figure}

We may apply the completely anti-symmetric tensor $\epsilon$ to any of the
bunches of $n$ semi-infinite lines of ${\mathbb D}_m$. Then by use of the
properties (\ref{sym-property1}) the anti-symmetrizer can be moved towards the
far left resulting in ${\mathbb D}_{m-1}$ times some proportionality factor.
This is illustrated in Figure \ref{reduction} for the case $m=2$ and $SU(3)$.

We like to point out that repeated applications of anti-symmetrizers to 
${\mathbb D}_m$ yield ${\mathbb D}_{\widetilde m}$ with arbitrary $\widetilde m\ (\le m)$.
Note, the application of $m$ times the anti-symmetric tensor $\epsilon$
removes all degrees of freedom and serves as the normalization of
${\mathbb D}_m$.

\section*{\bf Appendix B: The matrix $A^{[3]}$ for the three point case.}
\setcounter{equation}{0}
\renewcommand{\theequation}{B.\arabic{equation}}

Here we give the  $11\times 11$ matrix $A^{[3]}$ which defines the system of
functional equations for the functions
$\rho^{[3]}_k(\lambda_1,\lambda_2,\lambda_3)$. This was obtained by replacing (\ref{D3frak}) into equation
(\ref{qkzsun-frak}), along the same lines as in the two-sites case.
\eq
A^{[3]}=\left(\begin{array}{ccccccccccc}
	a_{1,1} & a_{1,2} & a_{1,3} & a_{1,4} & a_{1,5} & a_{1,6} & a_{1,7} & a_{1,8} & a_{1,9} & a_{1,10} & a_{1,11} \\
	a_{2,1} & a_{2,2} & a_{2,3} & a_{2,4} & a_{2,5} & a_{2,6} & a_{2,7} & a_{2,8} & a_{2,9} & a_{2,10} & a_{2,11} \\
	a_{3,1} & a_{3,2} & a_{3,3} & a_{3,4} & a_{3,5} & a_{3,6} & a_{3,7} & a_{3,8} & a_{3,9} & a_{3,10} & a_{3,11} \\
	a_{4,1} & 0 & a_{4,3} & a_{4,4} & a_{4,5} & a_{4,6} & 0 & a_{4,8} & a_{4,9} & 0 & a_{4,11} \\
	
	a_{5,1} & a_{5,2} & a_{5,3} & a_{5,4} & a_{5,5} & a_{5,6} & a_{5,7} & a_{5,8} & a_{5,9} & a_{5,10} & a_{5,11} \\
	a_{6,1} & a_{6,2} & a_{6,3} & a_{6,4} & a_{6,5} & a_{6,6} & 0 & a_{6,8} & a_{6,9} & 0 & a_{6,11} \\
	0 & a_{7,2} & 0 & a_{7,4} & a_{7,5} & a_{7,6} & 0 & a_{7,8} & 0 & 0 & 0 \\
	0 & a_{8,2} & 0 & a_{8,4} & a_{8,5} & a_{8,6} & 0 & a_{8,8} & 0 & 0 & 0 \\
	0 & 0 & 0 & a_{9,4} & a_{9,5} & 0 & 0 & a_{9,8} & 0 & 0 & 0 \\
	0 & 0 & 0 & a_{10,4} & a_{10,5} & 0 & 0 & a_{10,8} & 0 & 0 & 0 \\
	0 & a_{11,2} & 0 & a_{11,4} & a_{11,5} & a_{11,6} & 0 & a_{11,8} & 0 & 0 & 0 \\
\end{array}\right)
\label{matrixL3-qKZ}
\en
where the non-trivial matrix elements are written as follows,
\begin{align}
&a_{1,1}=\frac{(-1+3 x+x^2) (-1+3 y+y^2)}{x (3+x) y (3+y)},   
a_{1,2}=\frac{(-1+3 x+x^2) (-2+2 y+y^2)}{x (3+x) y (3+y)}, \nonumber \\
&a_{1,3}=-\frac{3}{x (3+x) y (3+y)},    
a_{1,4}=\frac{(1+y) (-8+3 x+2 x^2-2 y+2 x y+x^2 y)}{x (3+x) y (3+y)}, \nonumber \\
&a_{1,5}=\frac{(1+y) (-7+x^2-3 y-x y)}{x (3+x) y (3+y)},    
a_{1,6}=\frac{-3+y+y^2}{x (3+x) y (3+y)}, \nonumber \\
&a_{1,7}=\frac{-1+3 x+x^2}{x (3+x) (3+y)},   a_{1,8}=\frac{-1+3 x+x^2+y+3 x y+x^2 y}{x (3+x) y (3+y)}, \nonumber \\
\end{align}
\begin{align}
&a_{1,9}=\frac{1+3 x+x y}{x (3+x) (3+y)},    
a_{1,10}=-\frac{-1+x^2-x y}{x (3+x) y (3+y)}, a_{1,11}=-\frac{y}{x (3+x) (3+y)}, \nonumber \\
&a_{2,1}=\frac{3 (-1+3 x+x^2)}{x (3+x) y (3+y)},
a_{2,2}=-\frac{(-1+3 x+x^2) (-3+y+y^2)}{x (3+x) y (3+y)}, \nonumber \\
&a_{2,3}=-\frac{-1+3 y+y^2}{x (3+x) y (3+y)},
a_{2,4}=\frac{2 (1+y)}{x (3+x) y (3+y)}, 
a_{2,5}=\frac{(1+y)^2}{x (3+x) y (3+y)} \nonumber\\
&a_{2,6}=-\frac{-2+2 y+y^2}{x (3+x) y (3+y)},
a_{2,7}=\frac{(-1+3 x+x^2) y}{x (3+x) (3+y)},
a_{2,8}=-\frac{-1+y}{x (3+x) y (3+y)},\nonumber \\
&a_{2,9}=\frac{1+3 x+x y}{x (3+x) y (3+y)},
a_{2,10}=-\frac{-1+x^2-x y}{x (3+x) (3+y)},
a_{2,11}=-\frac{1}{x (3+x) (3+y)}, \nonumber \\
&a_{3,1}=-\frac{3}{x (3+x) y (3+y)},
a_{3,2}=-\frac{3 (2+y)}{x (3+x) y (3+y)}, \nonumber\\
&a_{3,3}=\frac{-3 x-3 y+7 x y+3 x^2 y+3 x y^2+x^2 y^2}{x (3+x) y (3+y)},
a_{3,4}=-\frac{-6+6 x+3 x^2-6 y-2 x y}{x (3+x) y (3+y)}, \nonumber\\
&a_{3,5}=\frac{3-6 x-3 x^2+6 y+6 x y+x^2 y+3 y^2+4 x y^2+x^2 y^2}{x (3+x) y (3+y)}, \nonumber\\
&a_{3,6}=\frac{-3 x-6 y+6 x y+3 x^2 y-3 y^2+2 x y^2+x^2 y^2}{x (3+x) y (3+y)}, \nonumber\\
&a_{3,7}=\frac{3}{x (3+x) (3+y)},
a_{3,8}=-\frac{-3+3 y+4 x y+x^2 y}{x (3+x) y (3+y)},
a_{3,9}=-\frac{1}{x (3+y)},\nonumber\\
&a_{3,10}=\frac{-3+3 x^2+x^2 y}{x (3+x) y (3+y)},
a_{3,11}=\frac{y}{x (3+y)},\nonumber \\
&a_{4,1}=\frac{3}{x (3+x)},
a_{4,3}=-\frac{-9+x^2-3 y-2 x y}{x (3+x) y (3+y)}, \nonumber\\
&a_{4,4}=-\frac{-3 x-x^2-9 y+3 x y+3 x^2 y-3 y^2+x y^2+x^2 y^2}{x (3+x) y (3+y)},\nonumber \\
&a_{4,5}=-\frac{-3+x-y}{x y (3+y)},
a_{4,6}=\frac{3+x-y}{(3+x) y (3+y)},\nonumber\\
&a_{4,8}=-\frac{9-3 x-2 x^2-6 y+x y+2 x^2 y-3 y^2+x y^2+x^2 y^2}{x (3+x) y (3+y)},\nonumber\\
&a_{4,9}=\frac{-3-x+y+3 x y+x y^2}{(3+x) y (3+y)},
a_{4,11}=\frac{3+x-y}{(3+x) (3+y)},\nonumber \\
&a_{5,1}=-\frac{3}{x (3+x) (3+y)}, \qquad
a_{5,2}=\frac{3}{x (3+x) (3+y)},\nonumber\\
&a_{5,3}=\frac{-3+8 x+3 x^2+x^2 y-x y^2}{x (3+x) y (3+y)},
a_{5,4}=-\frac{-1+y}{x y (3+y)},\nonumber\\
\end{align}
\begin{align}
&a_{5,5}=-\frac{3-8 x-3 x^2-3 y+2 x y+2 x^2 y+2 x y^2+x^2 y^2}{x (3+x) y (3+y)},\nonumber\\
&a_{5,6}=\frac{1}{x y (3+y)},
a_{5,7}=\frac{3 y}{x (3+x) (3+y)},\nonumber\\
&a_{5,8}=\frac{6-7 x-3 x^2-3 y+2 x y+2 x^2 y-3 y^2+x y^2+x^2 y^2}{x (3+x) y (3+y)},\nonumber\\
&a_{5,9}=-\frac{1}{x y (3+y)},
a_{5,10}=\frac{-3+3 x^2+x^2 y}{x (3+x) (3+y)},
a_{5,11}=\frac{1}{x (3+y)},\nonumber \\
&a_{6,1}=\frac{3}{x (3+x) y},
a_{6,2}=\frac{3}{x (3+x) y},
a_{6,3}=-\frac{-x-9 y+x^2 y-3 y^2-x y^2}{x (3+x) y (3+y)},\nonumber\\
&a_{6,4}=\frac{2+y}{(3+x) y (3+y)},
a_{6,5}=\frac{1}{(3+x) y (3+y)},\nonumber\\
&a_{6,6}=-\frac{-2 x-9 y+2 x y+2 x^2 y-3 y^2+2 x y^2+x^2 y^2}{x (3+x) y (3+y)},\nonumber\\
&a_{6,8}=\frac{1}{(3+x) y (3+y)},
a_{6,9}=\frac{1+3 x-3 y}{(3+x) y (3+y)},
a_{6,11}=\frac{-1+3 y+x y}{(3+x) (3+y)},\nonumber\\
&a_{7,2}=-\frac{(-1+3 x+x^2)(-1+y)}{x (3+x) y},
a_{7,4}=-\frac{-8+6 x+3 x^2-4 y-x y}{x (3+x) y (3+y)},\nonumber\\
&a_{7,5}=-\frac{-1-8 y+x^2 y-3 y^2-x y^2}{x (3+x) y (3+y)},
a_{7,6}=-\frac{-1+y}{x (3+x) y},\nonumber\\
&a_{7,8}=\frac{7-6 x-3 x^2-5 y+x y+x^2 y-2 y^2+2 x y^2+x^2 y^2}{x (3+x) y (3+y)},\nonumber \\
&a_{8,2}=\frac{3}{x (3+x)},
a_{8,4}=-\frac{3 (-3+2 x+x^2-y)}{x (3+x) y (3+y)}, \nonumber\\
&a_{8,5}=-\frac{-9-x+x^2-3 y-x y}{x (3+x) (3+y)},
a_{8,6}=-\frac{-3+2 x+x^2-x y}{x (3+x) y},\nonumber \\
&a_{8,8}=\frac{9-6 x-3 x^2-6 y-x y+x^2 y-3 y^2+x y^2+x^2 y^2}{x (3+x) y (3+y)},\nonumber \\
&a_{9,4}=\frac{(1+y) (3-2 x+y-x y)}{x y (3+y)},
a_{9,5}=\frac{(3-x+y) (1+y)}{x y (3+y)},
a_{9,8}=-\frac{1+y}{y (3+y)},\nonumber\\
&a_{10,4}=\frac{-2+3 x-2 y}{x y (3+y)},
a_{10,5}=-\frac{1-3 x+2 y+x y+y^2+x y^2}{x y (3+y)},
a_{10,8}=\frac{-1+y+x y}{x y (3+y)}, \nonumber \\
&a_{11,2}=\frac{3}{x (3+x) y},
a_{11,4}=\frac{-9+5 x+3 x^2-3 y-x y}{x (3+x) y (3+y)}, \nonumber\\
&a_{11,5}=\frac{x-9 y+x^2 y-3 y^2-x y^2}{x (3+x) y (3+y)},
a_{11,6}=-\frac{-x-3 y+2 x y+x^2 y}{x (3+x) y},\nonumber\\
&a_{11,8}=-\frac{9-4 x-3 x^2-6 y+2 x y+x^2 y-3 y^2+2 x y^2+x^2 y^2}{x (3+x) y (3+y)},\nonumber
\end{align}
where $x=\lambda_1-\lambda_3$ and $y=\lambda_1-\lambda_2$.



\begin{thebibliography}{100}
\bibitem{BOOK-KBI} V.E. Korepin, N.M. Bogoliubov, and A.G. Izergin \textit{Quantum inverse scattering method and correlation functions} (CUP, Cambridge, 1993).
\bibitem{BOOK-JM} M. Jimbo and T. Miwa, \textit{Algebraic Analysis of Solvable Lattice Models} (AMS, Rhode Island, 1995).
\bibitem{TAKA} M. Takahashi, J. Phys. C: Solid State Phys. \textbf{10} (1977) 1289.
\bibitem{JMMN92} M.~Jimbo, K.~Miki, T.~Miwa and A.~Nakayashiki, Phys. Lett. A \textbf{168} (1992) 256.
\bibitem{JM96} M.~Jimbo and T.~Miwa, J. Phys. A \textbf{29} (1996) 2923.
\bibitem{KMT00} N.~Kitanine, J.~M. Maillet and V.~Terras, Nucl. Phys. B \textbf{567} (2000) 554.
\bibitem{GAS05} F.~G\"ohmann, A.~Kl\"umper and A.~Seel, J. Phys. A \textbf{38} (2005) 1833.
\bibitem{BOKO01} H.~E. Boos and V.~E. Korepin, J. Phys. A \textbf{34} (2001) 5311.
\bibitem{BOOS05} H.~Boos, M.~Jimbo, T.~Miwa, F.~Smirnov and Y.~Takeyama, St Pertersburg Math. J. \textbf{17} (2006) 85. %
\bibitem{BOOS2}  H. Boos, M. Jimbo, T. Miwa, F. Smirnov, Y. Takeyama, Comm. Math. Phys. 261 (2006) 245. %
\bibitem{BGKS06} H.~Boos, F.~G\"ohmann, A.~Kl\"umper and J.~Suzuki, J. Stat. Mech.  (2006) P04001.
\bibitem{DGHK07} J.~Damerau, F.~G\"ohmann, N.~P. Hasenclever and A.~Kl\"umper, J. Phys. A \textbf{40} (2007) 4439.
\bibitem{KKMST09} N.~Kitanine, K.~Kozlowski, J.~M. Maillet, N.~A. Slavnov and V.~Terras, J. Stat. Mech. (2009) P04003.
\bibitem{AuKl12} Britta Aufgebauer and Andreas Kl\"umper, J.Phys. A: Math. Theor. 45 (2012) 345203.
\bibitem{BoWe94} A.~H. Bougourzi and R.~A. Weston,  Nucl. Phys. B \textbf{417} (1994) 439.
\bibitem{Idzumi94} M.~Idzumi,  Int. J. Mod. Phys. A \textbf{9} (1994) 4449.
\bibitem{Kitanine01} N.~Kitanine, J. Phys. A \textbf{34} (2001) 8151.
\bibitem{DeMa10} T.~Deguchi and C.~Matsui, Nucl. Phys. B \textbf{831} (2010) 359.
\bibitem{GSS10} F.~G\"ohmann, A.~Seel and J.~Suzuki, J. Stat. Mech. (2010) P11011.
\bibitem{KNS013} A.~Kl\"umper, D. Nawrath and J.~Suzuki, J. Stat. Mech. (2013) P08009
\bibitem{RK2016} G.A.P. Ribeiro and A. Kl\"umper, J. Phys. A: Math. Theor. {\bf 49} (2016) 254001.
\bibitem{UIMIN} G. V. Uimin, JETP Lett.{\bf 12} (1970) 225.
\bibitem{SUTHERLAND} B. Sutherland, Phys. Rev. B {\bf 12} (1975) 3795.
\bibitem{GoKlSe04} F.~G\"ohmann, A.~Kl\"umper and A.~Seel, J. Phys. A \textbf{37} (2004) 7625.
\bibitem{EFP} H.E. Boos, V.E. Korepin, F.A. Smirnov, Nucl.Phys. B \textbf{658} (2003) 417.
\bibitem{BGKS} H.E.~Boos, F.~G\"ohmann, A.~Kl\"umper, J.~Suzuki, J. Phys. A \textbf{40} (2007) 10699.
\bibitem{BOOS18} H.E. Boos, A. Hutsalyuk, K. Nirov, J. Phys. A: Math. Theor. 51 (2018)
445202 (arXiv:1804.09756 [hep-th]).
\end{thebibliography}
\end{document}